\documentclass[longauth]{aa} 

\def\simgt{\lower.5ex\hbox{$\; \buildrel > \over \sim \;$}}
\def\simlt{\lower.5ex\hbox{$\; \buildrel < \over \sim \;$}}
\def\logM {log ${\cal M}/{{\cal M}_{\odot}}\ $}

\usepackage{graphicx}
\usepackage{txfonts}
\begin{document}

\title{The zCOSMOS survey:
\thanks{based on data obtained with the European Southern Observatory 
Very Large Telescope, Paranal, Chile, program 175.A-0839}}
\subtitle{the role of the environment in the evolution 
\\
of the luminosity function of different galaxy types}

\author{
E.Zucca\inst{1}
\and
S.Bardelli\inst{1}
\and
M.Bolzonella\inst{1}
\and 
G.Zamorani\inst{1}
\and
O.Ilbert\inst{2}
\and
L.Pozzetti\inst{1}
\and
M.Mignoli\inst{1}
\and
K.Kova\v{c}\inst{3}
\and
S.Lilly\inst{3}
\and
L.Tresse\inst{2}
\and 
L.Tasca\inst{4}
\and
P.Cassata\inst{5}
\and
C.Halliday\inst{6}
\and
D.Vergani\inst{1}
\and
K.Caputi\inst{3}
\and  
C.M.Carollo\inst{3}
\and
T.Contini\inst{7}
\and
J.-P.Kneib\inst{2}
\and
O.Le~F\`{e}vre\inst{2}
\and
V.Mainieri\inst{8}
\and
A.Renzini\inst{9}
\and
M.Scodeggio\inst{4}
\and  
A.Bongiorno\inst{10}
\and
G.Coppa\inst{1,14}
\and
O.Cucciati\inst{2}
\and
S.de~la~Torre\inst{11,4}
\and
L.de~Ravel\inst{2}
\and
P.Franzetti\inst{4}
\and
B.Garilli\inst{4}
\and
A.Iovino\inst{11}
\and
P.Kampczyk\inst{3}
\and
C.Knobel\inst{3}
\and
F.Lamareille\inst{7}
\and
J.-F.Le~Borgne\inst{7}
\and
V.Le~Brun\inst{2}
\and
C.Maier\inst{3}
\and
R.Pell\`o\inst{7}
\and
Y.Peng\inst{3}
\and
E.Perez-Montero\inst{7,26}
\and
E.Ricciardelli\inst{12}
\and
J.D.Silverman\inst{3}
\and
M.Tanaka\inst{8}
\and  
U.Abbas\inst{13}
\and
D.Bottini\inst{4}
\and
A.Cappi\inst{1}
\and
A.Cimatti\inst{14}
\and 
L.Guzzo\inst{11}
\and 
A.M.Koekemoer\inst{15}
\and 
A.Leauthaud\inst{16}
\and 
D.Maccagni\inst{4}
\and 
C.Marinoni\inst{17}
\and 
H.J.McCracken\inst{18}
\and 
P.Memeo\inst{4}
\and 
B.Meneux\inst{10,27}
\and
M.Moresco\inst{14}
\and 
P.Oesch\inst{3}
\and 
C.Porciani\inst{3,28}
\and 
R.Scaramella\inst{19}
\and  
S.Arnouts\inst{20}
\and
H.Aussel\inst{21}
\and 
P.Capak\inst{22}
\and 
J.Kartaltepe\inst{23}
\and 
M.Salvato\inst{22}
\and
D.Sanders\inst{23}
\and
N.Scoville\inst{22}
\and 
Y.Taniguchi\inst{24}
\and 
D.Thompson\inst{25}
}

\offprints{Elena Zucca (elena.zucca@oabo.inaf.it)}

\institute{
INAF - Osservatorio Astronomico di Bologna, via Ranzani 1, I-40127 Bologna, Italy \\
\email{elena.zucca@oabo.inaf.it}
\and    
{Laboratoire d'Astrophysique de Marseille, Universit\'{e} d'Aix-Marseille, CNRS, 38 rue Frederic Joliot-Curie, F-13388 Marseille Cedex 13, France}
\and    
{Institute of Astronomy, Swiss Federal Institute of Technology (ETH H\"onggerberg), CH-8093, Z\"urich, Switzerland}
\and    
{INAF - IASF Milano, via Bassini 15, I-20133 Milano, Italy}
\and    
{Department of Astronomy, University of Massachusetts, 710 North Pleasant Street, Amherst, MA 01003, USA }
\and    
{INAF - Osservatorio Astrofisico di Arcetri, Largo Enrico Fermi 5, I-50125 Firenze, Italy }
\and    
{Laboratoire d'Astrophysique de Toulouse-Tarbes, Universit\'{e} de Toulouse, CNRS, 14 avenue Edouard Belin, F-31400 Toulouse, France}
\and    
{European Southern Observatory, Karl-Schwarzschild-Strasse 2, Garching, D-85748, Germany}
\and    
{INAF - Osservatorio Astronomico di Padova, vicolo Osservatorio 5, I-35122 Padova, Italy}
\and    
{Max-Planck-Institut f\"ur extraterrestrische Physik, D-84571 Garching, Germany}
\and    
{INAF - Osservatorio Astronomico di Brera, via Brera 28, I-20121 Milano, Italy}
\and    
{Dipartimento di Astronomia, Universit\`a di Padova, vicolo Osservatorio 3, I-35122 Padova, Italy}
\and    
{INAF - Osservatorio Astronomico di Torino, strada Osservatorio 20, I-10025 Pino Torinese, Italy}
\and    
{Dipartimento di Astronomia, Universit\`a di Bologna, via Ranzani 1, I-40127, Bologna, Italy}
\and    
{Space Telescope Science Institute, 3700 San Martin Drive, Baltimore, MS 21218, USA}
\and    
{LBNL \& BCCP, University of California, Berkeley, CA 94720, USA}
\and    
{Centre de Physique Theorique, Marseille, France}
\and    
{Institut d'Astrophysique de Paris, UMR 7095 CNRS, Universit\'e Pierre et Marie Curie, 98 bis Boulevard Arago, F-75014 Paris, France}
\and    
{INAF, Osservatorio Astronomico di Roma, via di Frascati 33, I-00040 Monteporzio Catone, Italy}
\and    
{Canada-France-Hawaii Telescope Corporation, 65-1238 Mamalahoa Hwy, Kamuela, HI 96743, USA} 
\and    
{AIM Unit\'e Mixte de Recherche CEA CNRS, Université Paris VII UMR n158, Paris, France} 
\and    
{California Institute of Technology, MC 105-24, 1200 East California Boulevard, Pasadena, CA 91125, USA}
\and    
{Institute for Astronomy, University of Hawaii, 2680 Woodlawn Drive, Honolulu, HI, 96822}
\and    
{Research Center for Space and Cosmic Evolution, Ehime University, Bunkyo-cho, Matsuyama 790-8577, Japan}
\and    
{Large Binocular Telescope Observatory, University of Arizona, 933 N. Cherry Ave., Tucson, AZ 85721-0065, USA }
\and    
{Instituto de Astrofisica de Andalucia, CSIC, Apdo. 3004, 18080, Granada, Spain}
\and    
{Universit\"ats-Sternwarte, Scheinerstrasse 1, Munich D-81679, Germany}
\and    
{Argelander-Institut f\"{u}r Astronomie, Auf dem H\"{u}gel 71, D-53121 Bonn, Germany}
}

\date{Received -- -- ----; accepted -- -- ----}

\abstract
{}
{  
An unbiased and detailed characterization of the galaxy luminosity function (LF)
is a basic requirement in many astrophysical issues: it is of particular
interest in assessing the role of the environment in the evolution 
of the LF of different galaxy types.
} 
{
We studied the evolution in the $B$ band LF to
redshift $z\sim 1$ in the zCOSMOS 10k sample, for which both
accurate galaxy classifications (spectrophotometric and morphological) 
and a detailed description of the local density field are available.
}
{ 
The global $B$ band LF exhibits a brightening of $\sim 0.7$ mag
in $M^*$ from $z\sim 0.2 $ to $z\sim 0.9$.
At low redshifts ($z<0.35$), spectrophotometric late types dominate at 
faint magnitudes ($M_{B_{AB}} > -20$), while the bright end is populated 
mainly by spectrophotometric early types.
At higher redshift, spectrophotometric late-type galaxies evolve significantly and, at 
redshift $z\sim 1$,the contribution from the various types to the bright end of 
the LF is comparable. 
The evolution for spectrophotometric early-type galaxies is in both luminosity
and normalization: $M^*$ brightens by $\sim 0.6$ mag
but $\phi^*$ decreases by a factor $\sim 1.7$ between the first and the last 
redshift bin.
A similar behaviour is exhibited by spectrophotometric late-type galaxies, but with an
opposite trend for the normalization: a brightening of $\sim 0.5$ mag is
present in $M^*$, while $\phi^*$ increases by a factor $\sim 1.8$.
\\
Studying the role of the environment, we find that the global 
LF of galaxies in overdense regions has always a brighter $M^*$ and a 
flatter slope. In low density environments, the main contribution to the 
LF is from blue galaxies, while for high density
environments there is an important contribution from red galaxies
to the bright end.
\\
The differences between the global LF in the two
environments are not due to only a difference in the relative numbers
of red and blue galaxies, but also to their relative 
luminosity distributions: the value of $M^*$ for both types
in underdense regions is always fainter than in overdense environments. 
These results indicate that galaxies of the same type in
different environments have different properties.
\\
We also detect a differential evolution in blue galaxies
in different environments: the evolution in their LF 
is similar in underdense and overdense regions between
$z\sim 0.25$ and $z\sim 0.55$, and is mainly in luminosity.
In contrast, between $z\sim 0.55$ and $z\sim 0.85$ 
there is little luminosity evolution but there is significant
evolution in $\phi^*$, that is, however, different between the two environments: 
in overdense regions $\phi^*$
increases by a factor $\sim 1.6$, while in underdense regions
this increase reaches a factor $\sim 2.8$.
Analyzing the blue galaxy population in more detail, 
we find that this evolution is driven mainly by the bluest types.
}
{
The ``specular" evolution of late- and early-type galaxies
is consistent with a scenario where a part of blue galaxies is transformed in
red galaxies with increasing cosmic time, without significant
changes in the fraction of intermediate-type galaxies.
The bulk of this tranformation in overdense regions probably
happened before $z\sim 1$, while it is still ongoing at lower
redshifts in underdense environments.
}

\authorrunning {E.Zucca et al.}

\titlerunning {The zCOSMOS luminosity function}


\maketitle


\section{Introduction}

The COSMOS project (Scoville et al. \cite{scoville07}) 
aims to identify the physical processes that drive the
evolution of galaxies. For example, dynamical processes are likely 
to play a major role in determining the galaxy morphology, whereas
dissipative phenomena affect the gas content and therefore the star 
formation, altering the galaxy spectral energy distribution.
The interplay between these two types of processes is not yet
completely understood, as well as the role of the environment
in the evolution of galaxy properties.
\\
From the observational point of view, the investigation of these
topics requires: 
{\it a)} high quality images, for deriving accurate morphological
classifications;
{\it b)} complete multiwavelength coverage, for determining 
spectral energy distributions;
{\it c)} galaxy spectra, for obtaining spectroscopic redshifts 
(necessary for a precise environment description) and measuring 
spectral features (to be used as diagnostics of the gas and stellar physics).
\\
The COSMOS multiwavelength imaging project provides data for points 
{\it a)} and {\it b)}, while
the zCOSMOS survey (Lilly et al. \cite{lilly07}) was developed  
to fullfill point {\it c)}.
The combination of these data allows us to study with unprecedented completeness
the properties and evolutionary histories of galaxies as a function of their 
type and environment at the same time.
The luminosity function is the first and most direct estimator in quantifying
this evolution. 
\\
Several works have already demonstrated that the global luminosity function
evolves. 
The Canadian Network for Observational Cosmology
field galaxy redshift survey (CNOC-2, Lin et al. \cite{lin99}) and the
ESO Sculptor Survey (ESS, deLapparent et al. \cite{delapparent03}) derived
the luminosity function up to $z\sim 0.5$ using $\sim 2000$ and $\sim 600$
redshifts, respectively. 
At higher redshift, 
the Canada France Redshift Survey (CFRS, Lilly et al. \cite{lilly95})
allowed to study the luminosity function up to $z\sim 1.1$ with a sample of
$\sim 600$ redshifts. 
\\
A major improvement in this field was obtained with the VIMOS VLT Deep Survey 
(VVDS, Le~F\`evre et al. \cite{vvds}), which detected a significant brightening of 
the $M^*$ parameter, amounting to $\sim 2$ mag in the $B$ band from $z\sim 0$ to 
$z\sim 2$ (Ilbert et al. \cite{vvdsLF}), using a sample of $\sim 11\ 000$ spectra. 
With the same sample, Zucca et al. 
(\cite{zucca06}) studied the evolution in the luminosity function for 
different spectrophotometric types, finding a strong type-dependent evolution
and identifying the latest types as being responsible for most of the evolution 
in the global luminosity function.
\\
Establishing the role of the environment is more difficult, because accurate
redshifts and surveys with high sampling rates are necessary for reliable
density estimates.
These constraints are satisfied by large local surveys, such as the
two-degree Field Galaxy Redshift Survey (2dFGRS, Colless et al. \cite{2dFGRS})
and the Sloan Digital Sky Survey (SDSS, York et al. \cite{SDSS}). 
From the 2dFGRS, Croton et al. (\cite{croton05}) measured the
dependence of the luminosity function on the density contrast (defined
in spheres of radius 8 h$^{-1}$ Mpc) and galaxy types, finding that the void
regions are dominated by faint late-type galaxies and that the cluster regions
exhibit an excess of very bright early-type galaxies. The parameter $M^*$ 
brightens in overdense regions for all galaxy types, while the slope
$\alpha$ steepens with increasing density for late-type galaxies 
and remains constant for early-type galaxies.
\\
At high redshift, the situation becomes more complicated: many surveys
are based on photometric redshifts, which do not allow an accurate reconstruction
of the density field, and/or lack galaxy classifications (morphological and/or
spectrophotometric). 
\\
Deep surveys based on spectroscopic redshifts, such as the VIMOS-VLT
Deep Survey (VVDS, Le~F\`evre et al. \cite{vvds}) and the DEEP2 Galaxy Redshift Survey
(Davis et al. \cite{deep2}), are able to study galaxy properties as a function
of the environment to $z \sim 1.5$.
Cucciati et al. (\cite{cucciati06}), using VVDS data, and Cooper et al. (\cite{cooper06},
\cite{cooper07}), using DEEP2 data, claimed that the color-density relation 
significantly evolves with redshift, and Cucciati et al. (\cite{cucciati06}) 
found that this relation is established at higher redshift for brighter galaxies.
To understand these results, it is necessary to study the evolution in the
luminosity function of different galaxy types in different environments.
Until now, the data required to complete this study as a function of galaxy type, 
environment, and redshift, has been unavailable.
\\
In this paper, we study the effect of the environment on the evolution in the 
luminosity function of different galaxy types for the zCOSMOS 10k bright sample.
Parallel studies discuss the evolution of the luminosity density (Tresse et al. \cite{tresse08}),
the mass function for different types (Pozzetti et al. \cite{pozzetti08}) 
and in various environments (Bolzonella et al. \cite{bolzonella08}), the 
spectrophotometric properties as a function of the environment (Cucciati et al.
\cite{cucciati08}; Iovino et al. \cite{iovino08}) and the morphology-density 
relation (Tasca et al. \cite{tasca08}, Kova\v{c} et al. \cite{kovac09}).
\\
The paper is organized as follows:
in Sect. \ref{sectdata}, we present the data on which this work is based
and in Sect. \ref{sectmeth} we describe the method used to estimate
luminosity functions.
The results concerning the luminosity function evolution are described
in Sect. \ref{sectLFtot} and the contribution of the different galaxy
types is presented in Sect. \ref{sectLFtype}, paying particular
attention to early-type galaxies in Sect. \ref{sect2types}.
The role of the environment is discussed in Sect. \ref{sectenv}, and
the results are summarized in Sect. \ref{sectconc}.
\\
Throughout this paper we adopt a flat $\Omega_m = 0.25$ and
$\Omega_\Lambda = 0.75$ cosmology, with H$_o = 70$ km s$^{-1}$
Mpc$^{-1}$. Magnitudes are given in the AB system. 


\section{Data}\label{sectdata}

The zCOSMOS project is a large redshift survey (Lilly et al. \cite{lilly07}) that was
undertaken to study the COSMOS field using 600 hours of observations with the ESO VLT.
COSMOS is an HST Treasury Project (Scoville et al. \cite{scoville07})
survey of a 2 square degree equatorial field with the Advanced Camera for Surveys (ACS). 
It is the largest survey ever completed by HST, utilizing 10\% (640 orbits) of its observing 
time over two years (HST Cycles 12 and 13), as described further in Koekemoer et al.
(\cite{koe07}). 
The primary goal of COSMOS is to understand how galaxies and AGN evolve over cosmic time 
in terms of their environment, on all scales from groups up to the large scale structure 
of filaments and voids. The COSMOS field is accessible to almost all astronomical facilities,
which has enabled the compilation of complete multiwavelength datasets (X-ray, UV, optical/IR, 
FIR/submillimeter to radio) to be used in combination with the high resolution HST images.


\subsection{Photometric data}

Photometric data in the COSMOS field are available for a wide range of wavelengths.
In the following, we make use of the magnitudes measured in the filters CFHT
$u^*$ and $K_s$, Subaru $B_J$, $V_J$, $g^+$, $r^+$, $i^+$ and $z^+$,  
and of the Spitzer IRAC magnitudes at 3.6 $\mu$m and 4.5 $\mu$m.
Details about Subaru observations are given in Taniguchi et al. (\cite{taniguchi07}),
while $K_s$ data are described in McCracken et al. (\cite{hjmcc09}).
Spitzer IRAC data are presented in Sanders et al. (\cite{sanders07}).
\\
Capak et al. (\cite{capak07}) provided a full description of the completed data 
reduction and characteristics of the optical observations. We recall that photometry
can be optimized by applying offsets to the observed magnitudes to reduce the 
differences between observed and reference magnitudes computed from a set of
template Spectral Energy Distributions (SEDs), as demonstrated by Capak et al.
(\cite{capak07}; see their Table 13). We adopted the same approach, but 
we derived our own offsets by using the set of SEDs that we used to compute 
absolute magnitudes (see Sect. \ref{sectmag}). In all cases, 
the offsets that we derived are similar to those of Capak et al. (\cite{capak07}).


\subsection{Morphologies from ACS images}\label{sectacs}

Morphological parameters for the galaxies are obtained from the HST ACS imaging
(Koekemoer et al. \cite{koe07}).
The COSMOS $I$ band ACS images have sufficient depth and resolution to allow
classical bulge-disk decomposition of $L^*$ galaxies at $z\le 2$.
The size of the COSMOS sample suggests the use of an automatic and objective 
morphological classification technique.
The method adopted for the morphological classification is described in detail
in Cassata et al. (\cite{cassata07}, \cite{cassata09}) and Tasca et al. 
(\cite{tasca08}), and here we summarize the main steps of the procedure.
\\
Using a training set of $\sim 500$ galaxies for which eye-ball morphological 
classification is available, the parameters describing the galaxy morphology 
were determined. 
The classification scheme was based on three non-parametric diagnostics of
galaxy structures (Abraham et al. \cite{abraham03}, Lotz et al. \cite{lotz04}),  
the concentration index $C$, 
the asymmetry parameter $A$ and the Gini coefficient $G$, and
the galaxy magnitude in the $I$ band (F814W HST/ACS band).  
These parameters were then converted into morphological classes:
this is done computing  in the multi-dimensional parameter space the distance 
of each galaxy from objects in the training set.
The morphological class was then assigned on the basis of the most frequent
class of the 11 nearest neighbours.
Given the fact that the classification is based on the observed F814W band,  
in the redshift range considered in this work ($z<1$) the effects of morphological 
K-correction are small (Cassata et al. \cite{cassata07}, Tasca et al. \cite{tasca08}). 
\\
In the following, we use this morphological classification, dividing
galaxies into early-types (including ellipticals and lenticulars), 
spirals, and irregulars.


\subsection{Spectroscopic data}

Spectroscopic redshifts in the zCOSMOS survey are obtained with
the VIMOS spectrograph (Le~F\`evre et al. \cite{vimos}) at the ESO Very Large
Telescope. 
The zCOSMOS survey consists of two parts. The first part (zCOSMOS--bright)
is a pure magnitude-limited survey with $15\le I \le 22.5$ (the
$I$ magnitude having been measured in the F814W HST/ACS band) and covers the entire
1.7 deg$^2$ COSMOS field. This magnitude limit will yield a sample of $\sim 20\ 000$ 
galaxies in the redshift range
$0.05 \simlt z \simlt 1.2$. The second part (zCOSMOS--deep) aims to observe 
$\sim 10\ 000$ galaxies in the redshift range $1.5 \simlt z \simlt 3.0$, selected
by clearly defined color criteria, and is restricted to the central
1 deg$^2$ of the COSMOS field. 
\\
Spectroscopic data were reduced with the VIMOS Interactive Pipeline
Graphical Interface (VIPGI, Scodeggio et al. \cite{vipgi}) and redshift 
measurements were derived using the EZ package (Garilli et al. \cite{ez}) 
and then visually checked.  Each redshift measurement was assigned a quality 
flag, between ``0'' (impossible to determine a redshift) and ``4''
(for which the measurement is 100\% certain); flag ``9'' indicates spectra with a single
emission line, for which multiple solutions are possible. 
Specific flags are used to denote Broad Line AGNs. 
A decimal digit indicates how closely the redshift agrees with its photometric
redshift (Feldmann et al. \cite{feldmann06}) computed from optical and
near IR photometry, using the code ZEBRA (Feldmann et al. \cite{feldmann08}).
Further details of the reduction procedure, redshift determination, and quality 
flags are given in Lilly et al. (\cite{lilly07}, \cite{lilly08}).


\subsection{Environment}

One of the main scientific objectives of the zCOSMOS survey is to study the role of the
environment on galaxy evolution to high redshift. 
For the zCOSMOS--bright survey, spectroscopic observations were performed with 
the medium resolution ($R\sim 600$) red grism, which provides a velocity accuracy 
of $\sim 100$ km/s. Repeated observations of $\sim 100$ galaxies enabled us to
estimate the rms velocity uncertainty in each measurement to be of the 
order of 110 km/s (Lilly et al. \cite{lilly08}). 
The measurement of redshifts to such precision allows us to define environments
of galaxies from the scale of galaxy groups to the larger scales of the
cosmic web.
\\
The density field of the COSMOS survey and the local environment of zCOSMOS
galaxies have been derived in Kova\v{c} et al. (\cite{kovac08}), where various 
estimators based on counts in fixed comoving apertures (cylindrical, spherical
and Gaussian) and the distance to the nearest neighbours are presented, 
using different tracers (flux-limited or volume-limited subsamples).
\\
In the following, we use overdensities derived by the 5th
nearest neighbour estimator and computed using volume-limited tracers.
This choice is a good compromise between the smallest accessible scales,
the robustness of the estimator, and the covered redshift range.
Bolzonella et al. (\cite{bolzonella08}) discuss in detail the
effects of different choices, in terms of estimators and
tracers, on the estimate of the galaxy stellar mass function.


\subsection{The zCOSMOS 10k bright sample}

The zCOSMOS survey is currently ongoing: the data used in the present paper
are the so-called ``10k bright sample" (Lilly et al. \cite{lilly08}), which 
consists of the first $10\ 644$ observed objects, over an area of 1.402 deg$^2$ with 
1 or 2 passes per VIMOS pointing and a mean sampling rate of $\sim 33\%$.
\\
For the present analysis we excluded Broad Line AGNs, stars, and objects 
that had not been included in the statistical sample defined in the magnitude range 
$15\le I \le 22.5$. We used only galaxies with reliable redshifts,
i.e. starting from flags 1.5 (see Lilly et al. \cite{lilly07} and 
\cite{lilly08} for details about the flag definition):
such a sample has an overall reliability of $\sim 99\%$.
These objects comprise $88\%$ of the overall sample and $95\%$ 
of the objects in the redshift range $0.5 < z < 0.8$. 
\\
To obtain a reliable SED fitting when deriving absolute 
magnitudes and spectrophotometric types (see Sect. \ref{sectmag}), we 
considered only objects 
with apparent magnitudes measured in more than 3 photometric bands.
\\
The final sample used in this paper consists of 8478 galaxies satisfying
the criteria described above.

\begin{figure*}
\centering
\includegraphics[width=0.8\hsize]{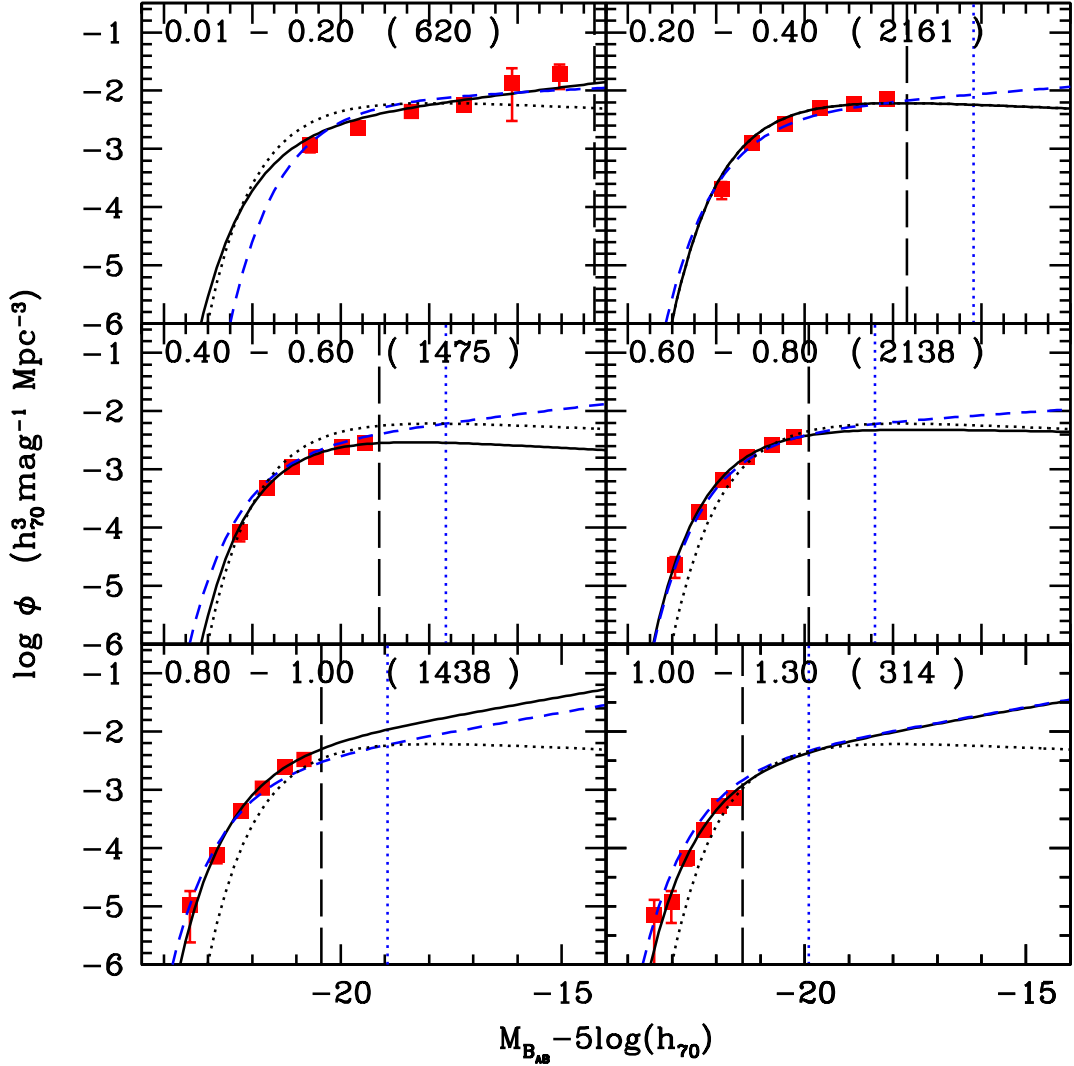}
\caption{ Evolution of the global luminosity function 
in the $B$ rest-frame band. 
Each panel refers to a different redshift bin, which is indicated
in the label, with also the number of galaxies. The vertical dashed and
dotted lines represent the faint absolute limit considered in the
$STY$ estimate for the zCOSMOS and VVDS sample, respectively.
The luminosity functions are estimated with different methods (see text 
for details) but for clarity we plot only the results from 
$C^+$ (squares) and $STY$ (solid line). The dotted line
represents the luminosity function estimated in the redshift range
$[0.2-0.4]$ and it is reported in each panel as a reference.
The $STY$ estimates from the VVDS sample are plotted with dashed lines.
In the redshift bins $[0.8-1.0]$ and $[1.0-1.3]$ for the zCOSMOS
sample we fixed $\alpha$ to the VVDS value.}
\label{LFtot}
\end{figure*}


\section{Luminosity function estimate}\label{sectmeth}


\subsection{Absolute magnitudes and spectrophotometric types}\label{sectmag}

Absolute magnitudes were computed following the method described in the
Appendix of Ilbert et al. (\cite{vvdsLF}). The K-correction was computed
using a set of templates and all available photometric information.  
However, to reduce the template dependency, the
rest-frame absolute magnitude in each band was derived using the
apparent magnitude from the closest observed band, shifted to the
redshift of the galaxy. With this method, the applied K-correction was
as small as possible.
\\
The spectrophotometric types are defined by matching the rest-frame magnitudes 
to the set of templates described in Ilbert et al. (\cite{ilbert06}):
the four locally observed CWW spectra (Coleman et al. \cite{cww}) and
two starburst SEDs from Kinney et al. (\cite{kinney96}),
extrapolated toward UV and mid-IR wavelengths, interpolated to obtain
62 SEDs, and optimized by using the VVDS spectroscopic data.
The SED fitting was performed by minimizing a $\chi^2$ variable on
these templates at the spectroscopic redshift of each galaxy,
providing as output the best-fit spectrum template. Galaxies were then
classified into four types, corresponding to colors of E/S0 (type 1),
early spirals (type 2), late spirals (type 3), and irregular and starburst
galaxies (type 4). 


\subsection{The ALF tool}\label{sectalf}
 
Luminosity functions were computed using the ``Algorithm for Luminosity
Function'' (ALF), a dedicated tool that implements several estimators:
the non-parametric $1/V_{max}$ (Schmidt \cite{schmidt68}), $C^+$
(Lynden Bell \cite{lyndenbell71}, in its modified version described by
Zucca et al. \cite{zucca97}), $SWML$ (Efstathiou et al.
\cite{swml}), and the parametric $STY$ (Sandage, Tammann \& Yahil
\cite{sty}), for which we assumed a Schechter function (Schechter
\cite{schechter76}).  The tool and these estimators were described in 
detail by Ilbert et al. (\cite{vvdsLF}).
\\
Ilbert et al. (\cite{ilbert04}) demonstrated that the estimate of the
global luminosity function can be biased, mainly at the faint end, when
the measurement band differs in wavelength considerably
from the rest-frame band in which galaxies are selected. 
This is caused by the fact that, because of the
K-corrections, different galaxy types are visible in different absolute
magnitude ranges at a given redshift and fixed apparent magnitude
limit. 
When computing the luminosity functions 
we avoided this bias by using in each redshift range (for the $C^+$,
$SWML$, and $STY$ estimates) only galaxies within the absolute magnitude 
range for which the entire wavelength range of their SEDs is potentially 
observable. 
We used the complete magnitude range for only the $1/V_{max}$ estimate.
Since this estimator underestimates the luminosity
function for absolute magnitudes fainter than the bias limit
(Ilbert et al. \cite{ilbert04}), it provides a lower limit of the
faint-end slope. 
\\
Even if this bias is less important when estimating the luminosity
function of galaxies divided by type, because the K-corrections
are more similar to each other, we have, however, taken it into
account.  The absolute magnitude limits for the $STY$ estimate are
indicated by vertical dashed lines in the figures, and in the tables
where the best-fit parameters are reported, we provide both the total number
of objects and the number of galaxies within this magnitude limit.

\begin{figure*}
\centering
\includegraphics[width=0.45\hsize]{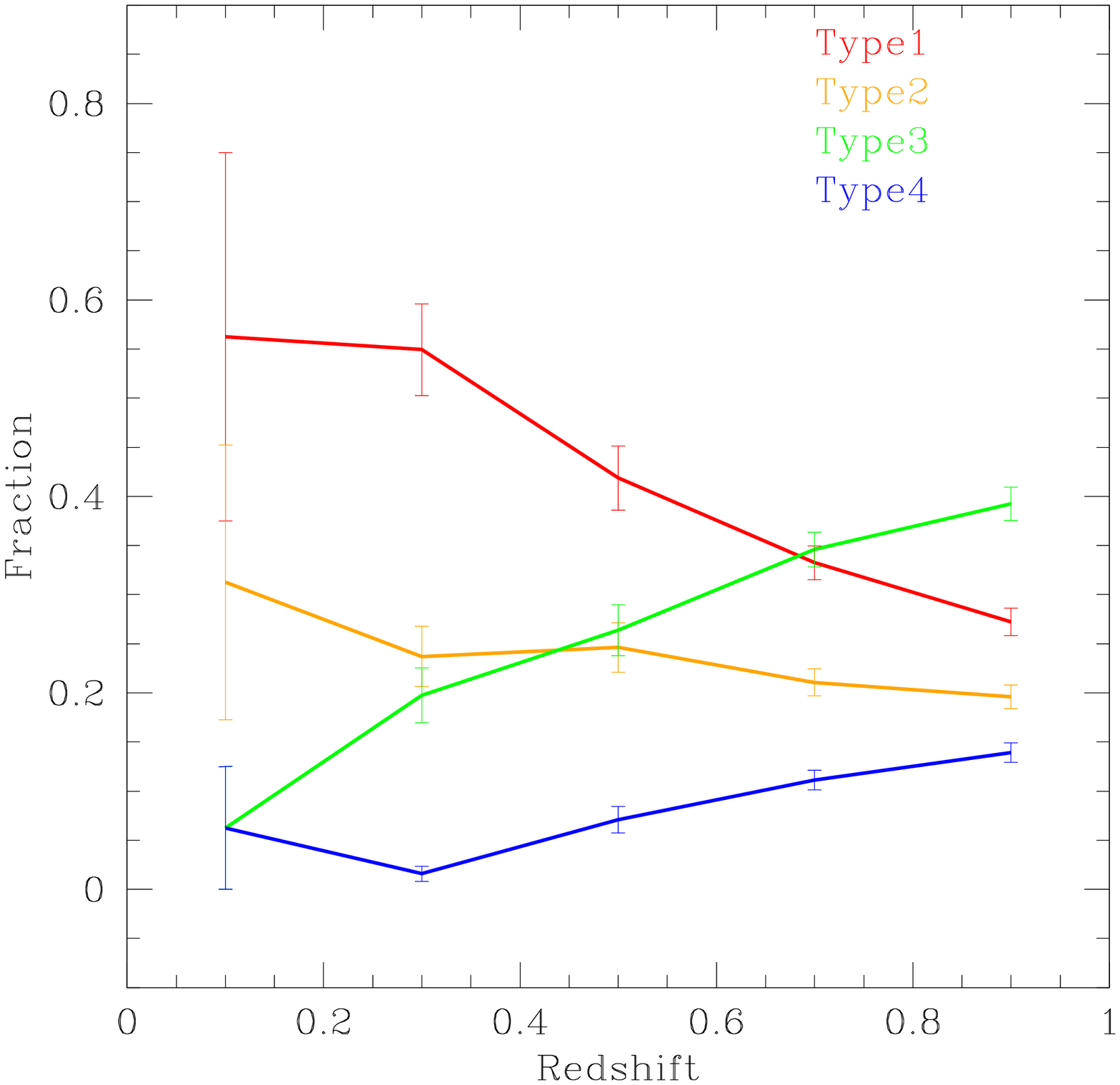}
\includegraphics[width=0.45\hsize]{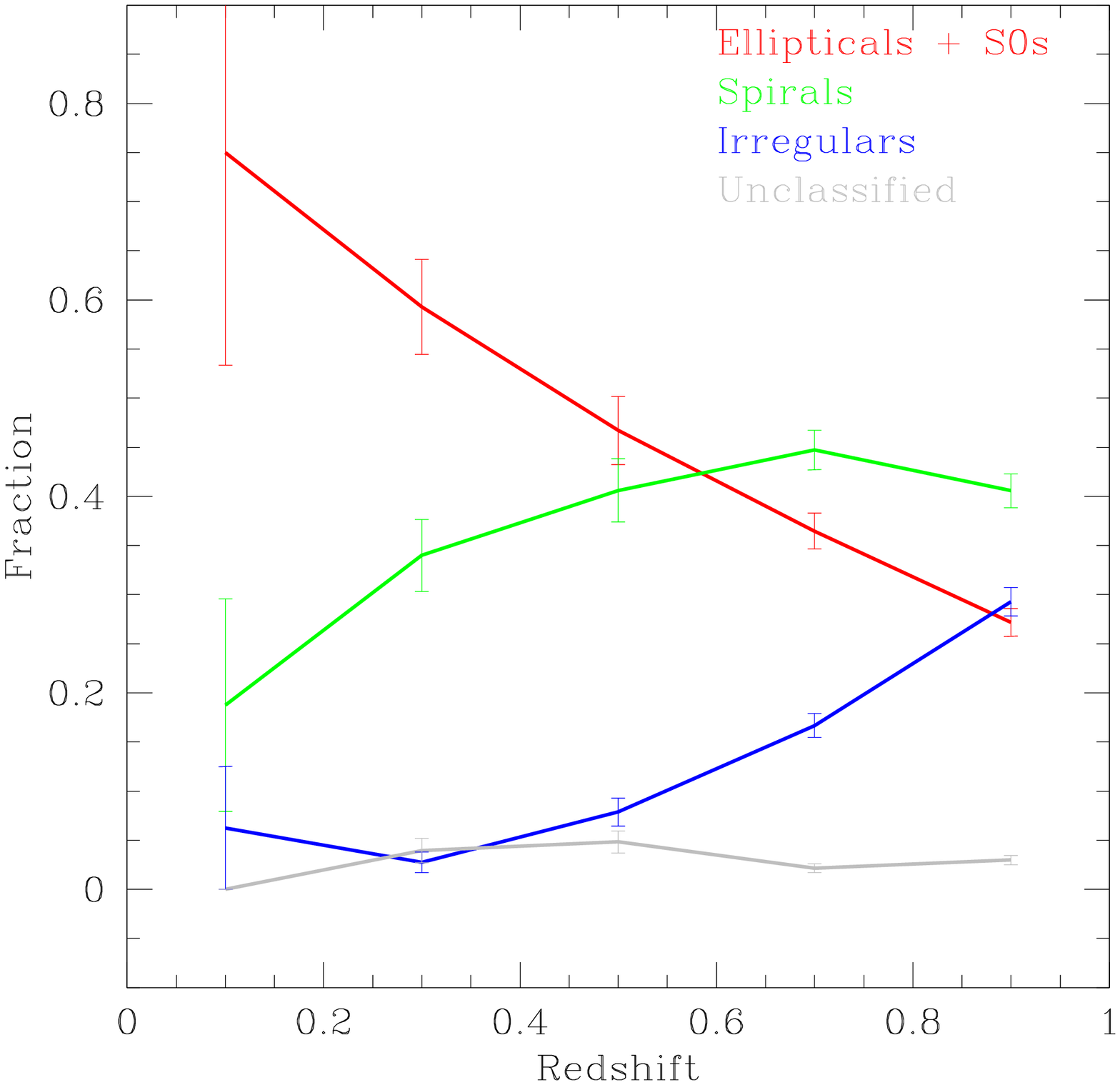}
\caption{Observed fraction of bright galaxies ($M_{B} < -20.77$) 
of different types as a function of redshift. Error bars are $1\sigma$
Poisson errors. Left panel: spectrophotometric type 1 (red), type 2
(orange), type 3 (green) and type 4 (blue) galaxies.
Right panel: morphological early-type (ellipticals and lenticulars)
(red), spiral (green) and
irregular (blue) galaxies; in gray unclassified galaxies.}
\label{fracTOT}
\end{figure*}

\begin{figure*}
\centering
\includegraphics[width=0.45\hsize]{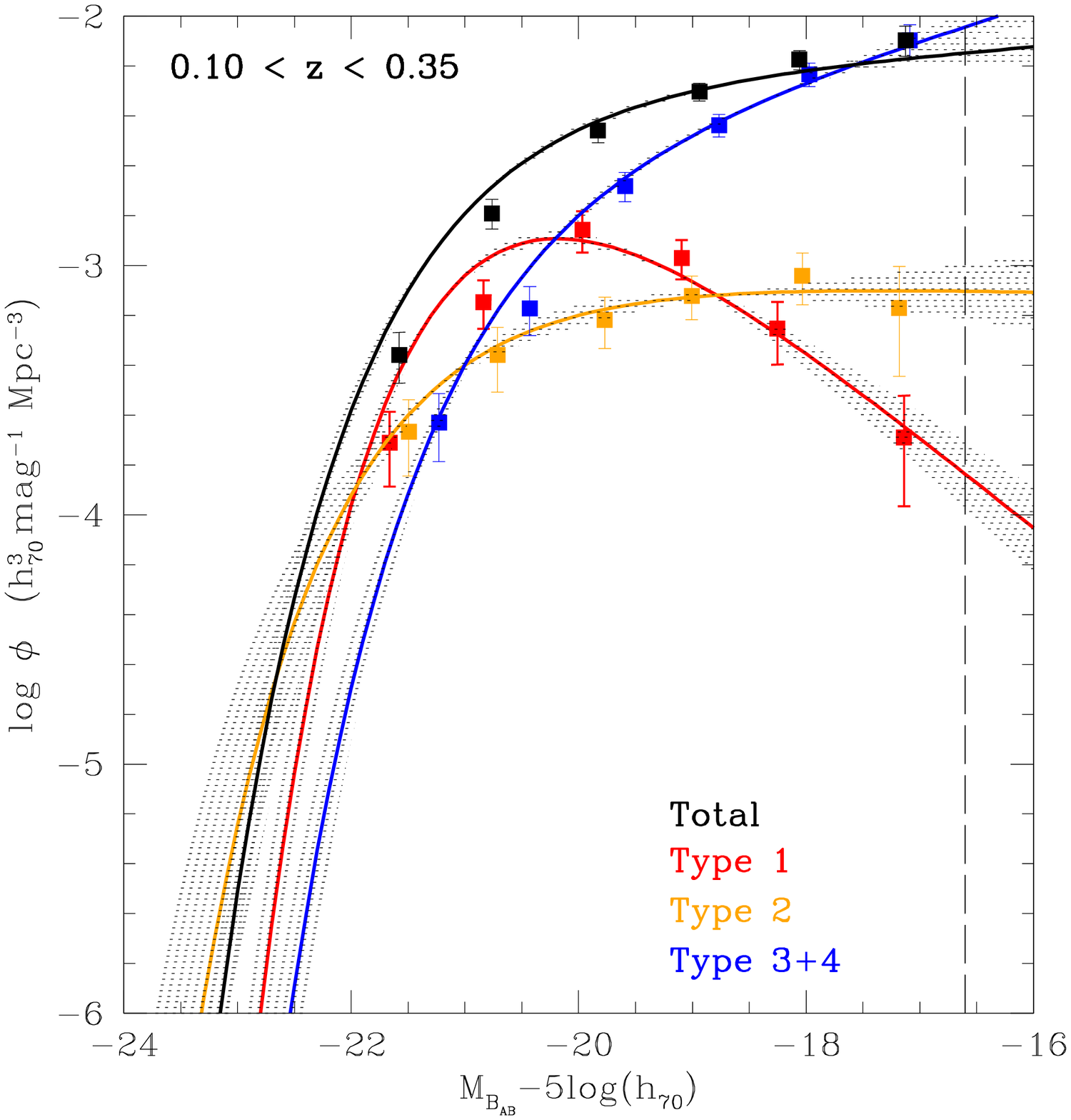}
\includegraphics[width=0.45\hsize]{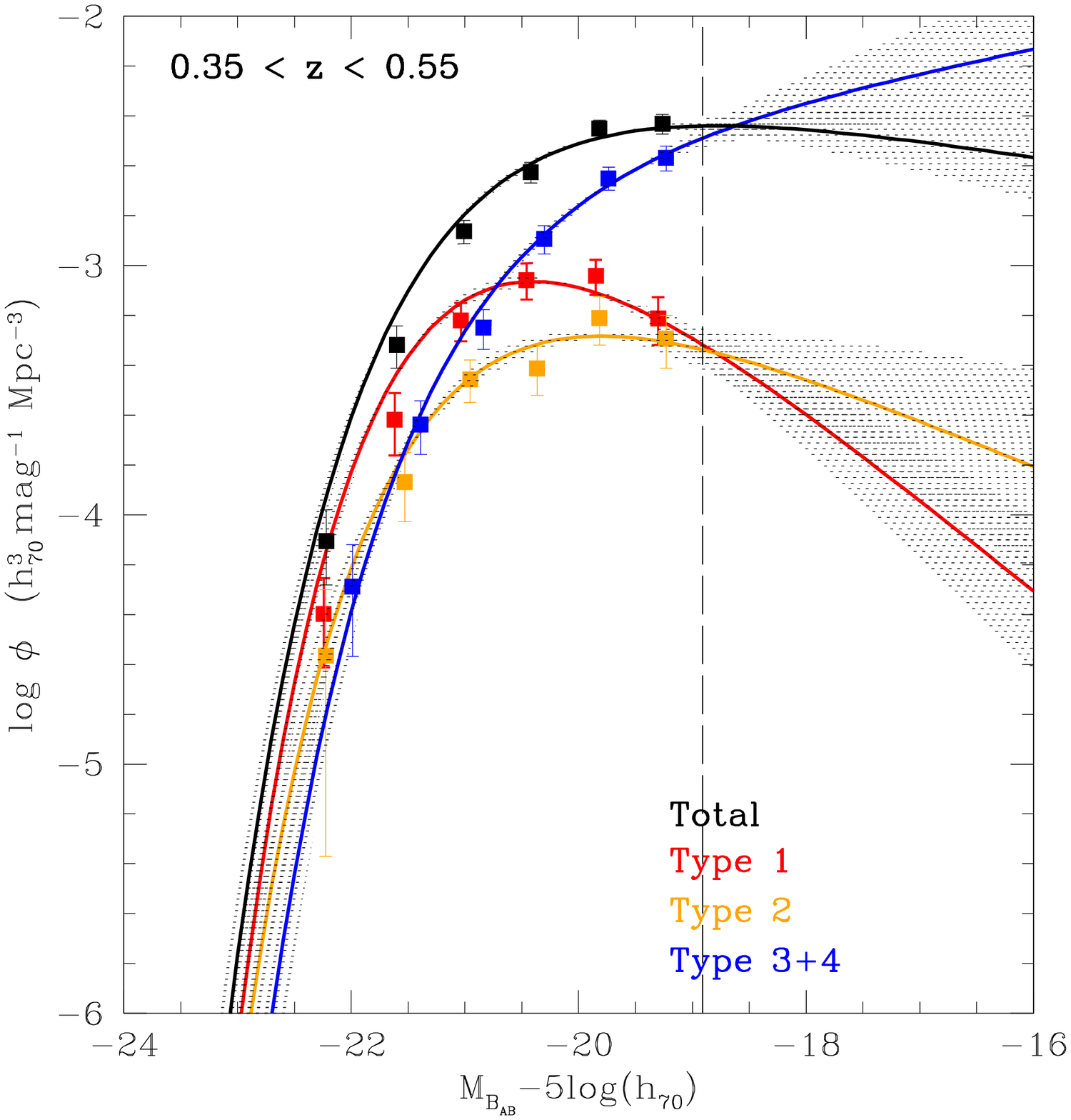}\\
\includegraphics[width=0.45\hsize]{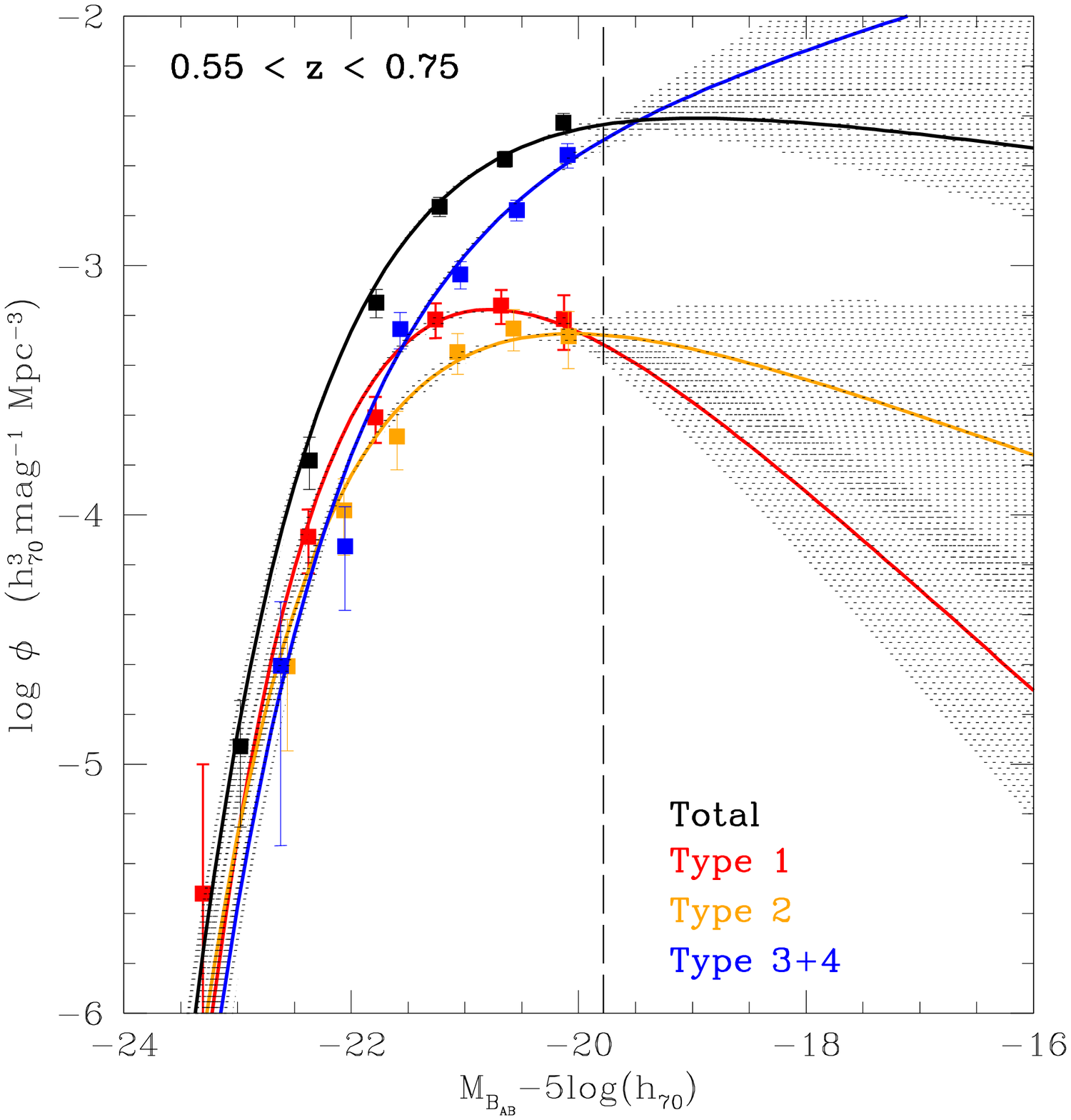}
\includegraphics[width=0.45\hsize]{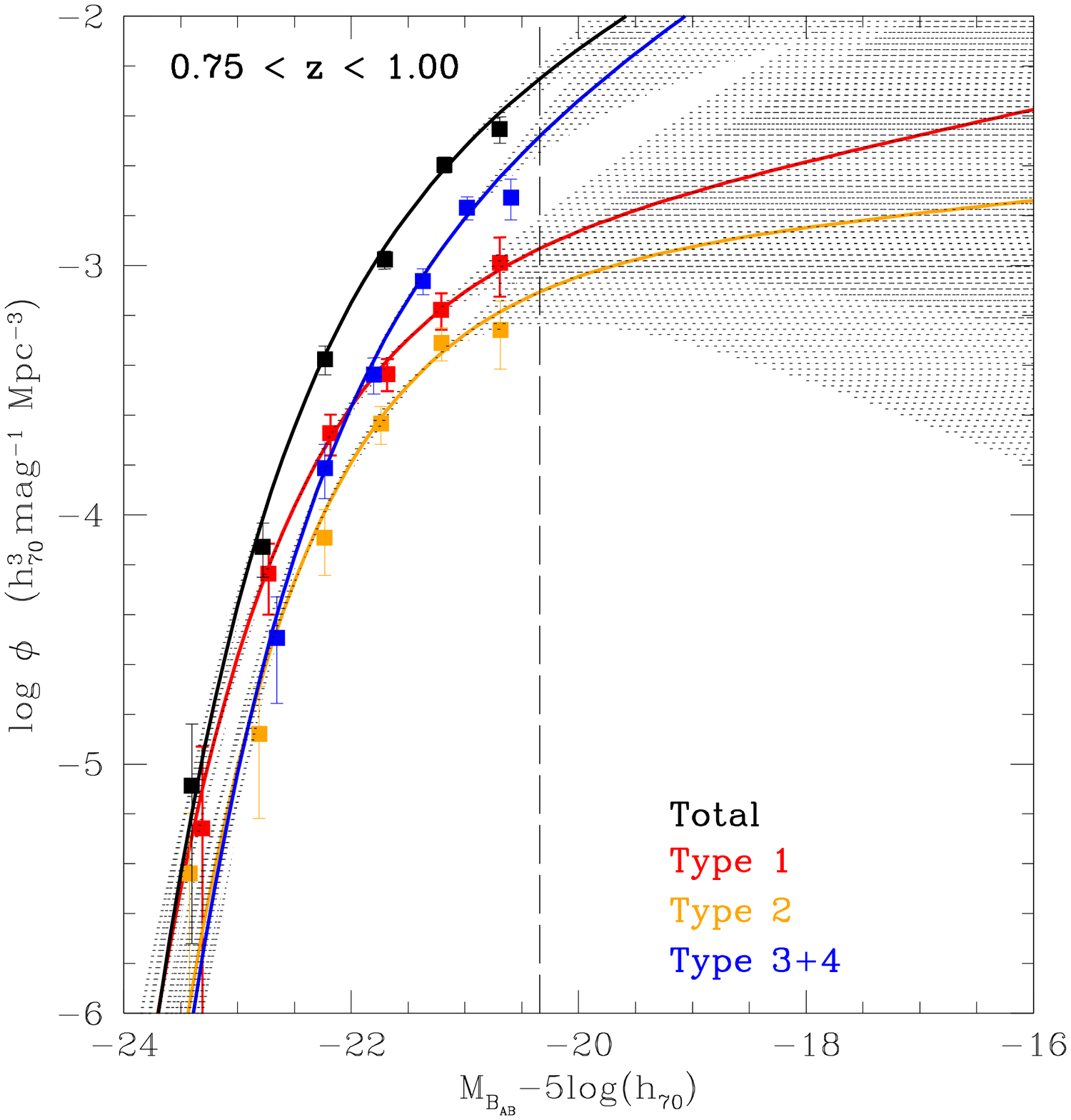}
\caption{ Luminosity functions of the different spectrophotometric types
in redshift bins (indicated in each panel): 
type 1 in red, type 2 in orange,   
type 3+4 in blue, total sample in black. The squares represent the
results from the $C^+$ and the solid lines are the results
from the  $STY$ method. 
The vertical dashed line represents the faint absolute limit considered in the
$STY$ estimate.
The shaded regions represent the $68\%$ uncertainties on the parameters
$\alpha$ and $M^*$.}
\label{LFforme_cww_new}
\end{figure*}
\begin{figure*}
\centering
\includegraphics[width=0.3\hsize]{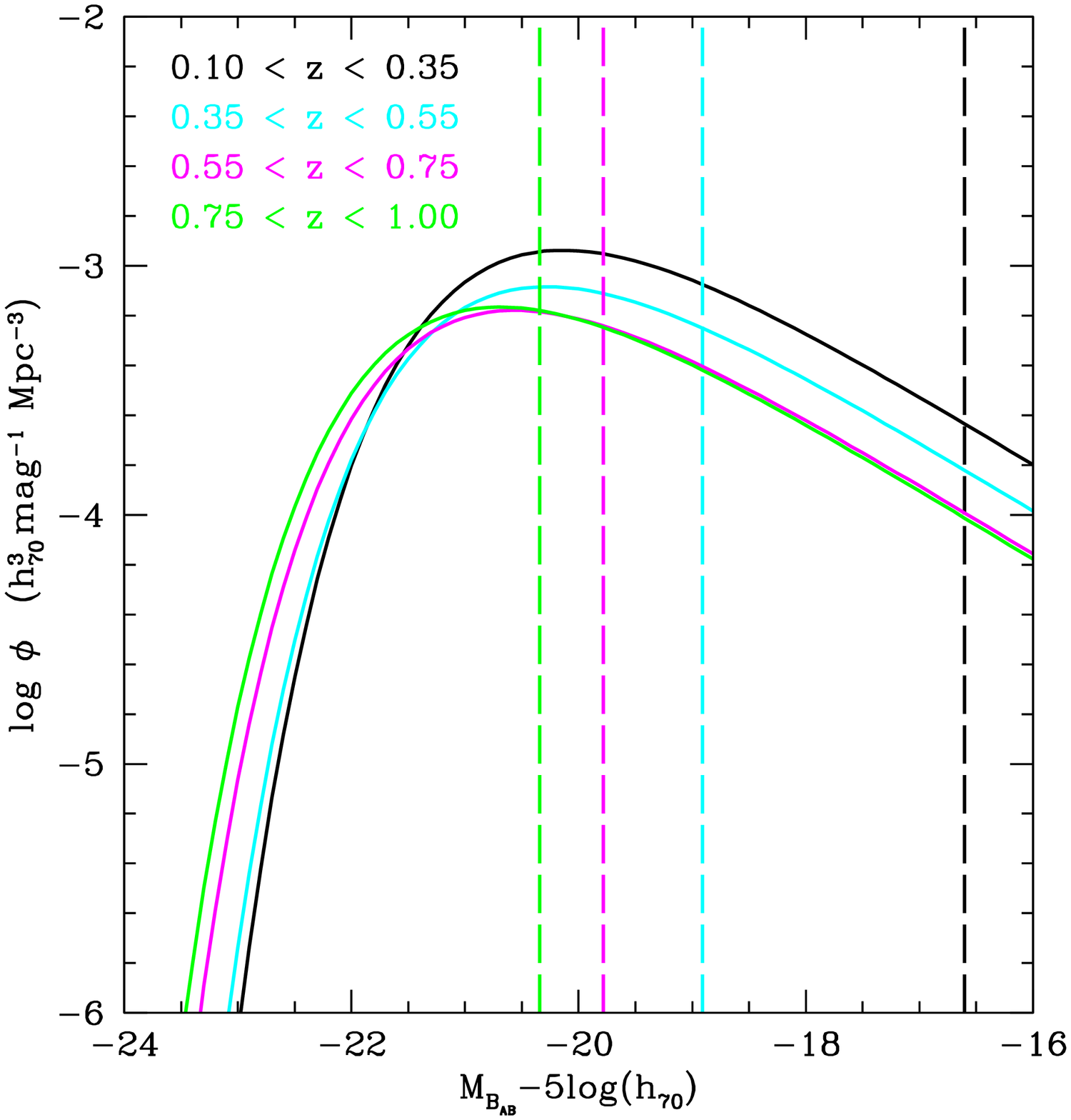}
\includegraphics[width=0.3\hsize]{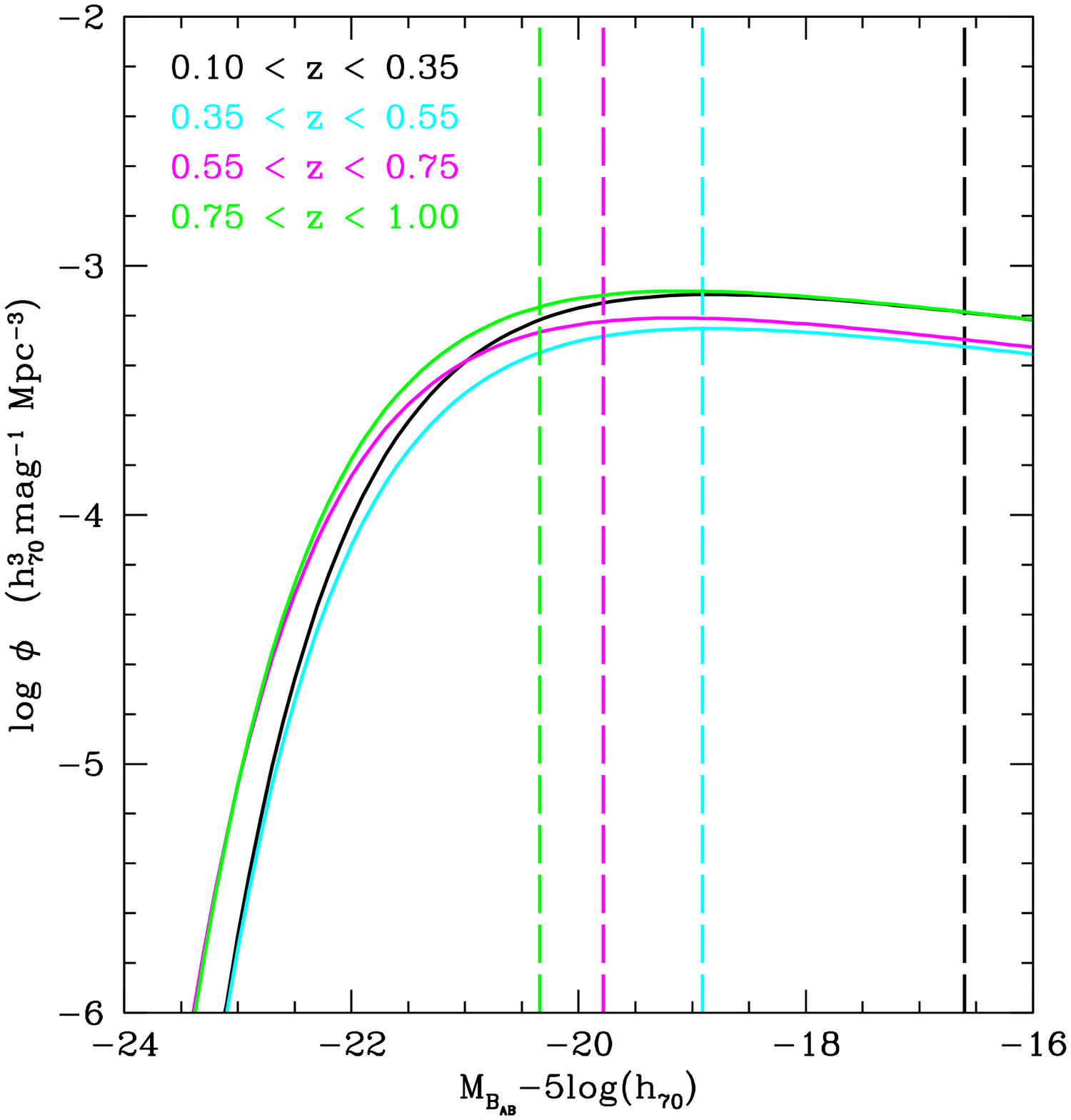}
\includegraphics[width=0.3\hsize]{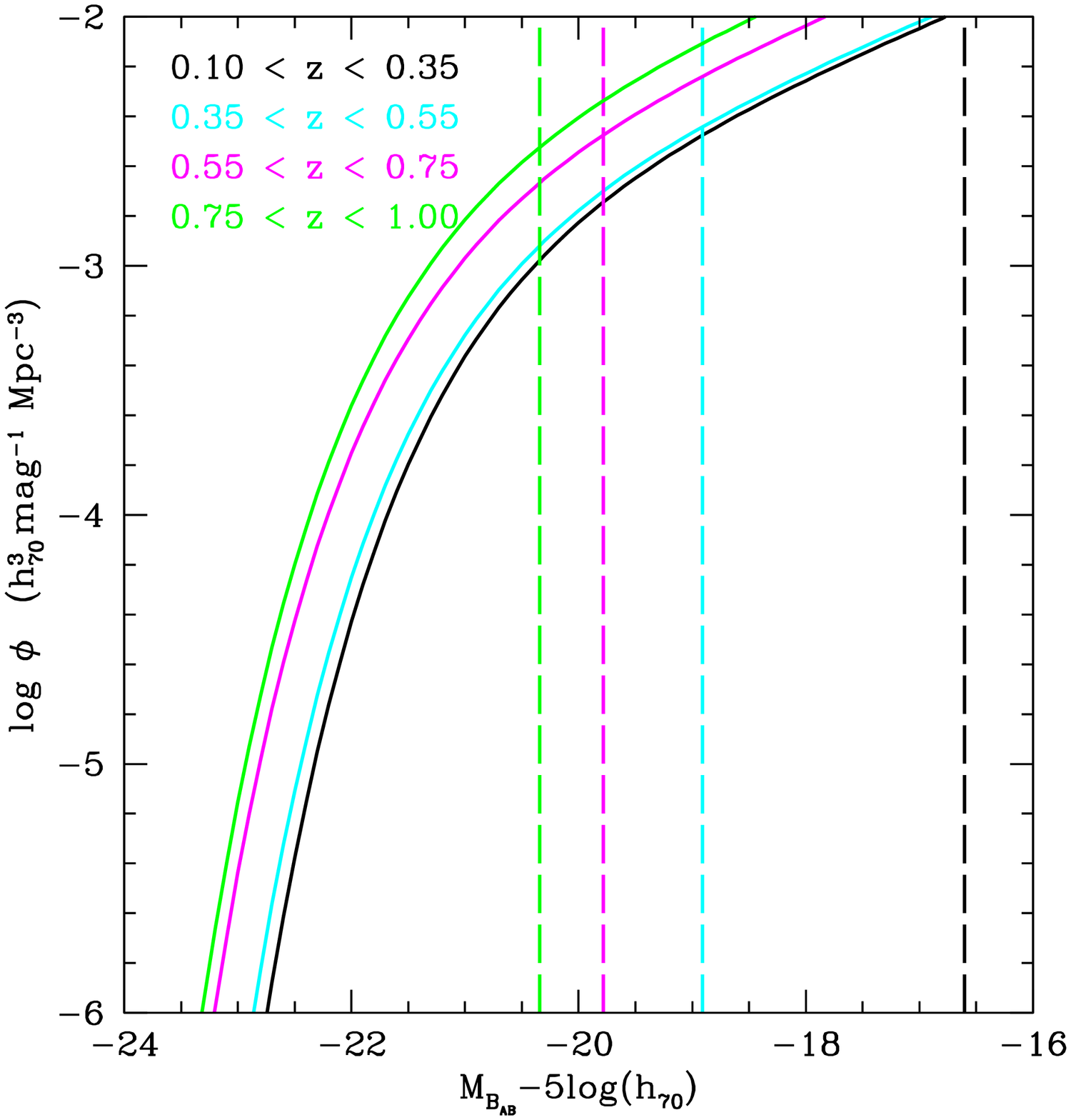}
\caption{ Evolution of the luminosity functions for different
galaxy types: 
type 1 (left panel), type 2 (middle panel) and type $3+4$
(right panel) galaxies. 
The colors refer to different redshift ranges: $[0.10-0.35]$
in black, $[0.35-0.55]$ in cyan, $[0.55-0.75]$ in magenta and
$[0.75-1.00]$ in green. The $STY$ estimates are derived with
$\alpha$ fixed.
The meaning of lines is the same as in Fig. \ref{LFforme_cww_new}: 
points from the $C^+$ estimates are not reported for clarity.
 }
\label{LFforme_z}
\end{figure*}

\begin{figure*}
\centering
\includegraphics[width=0.45\hsize]{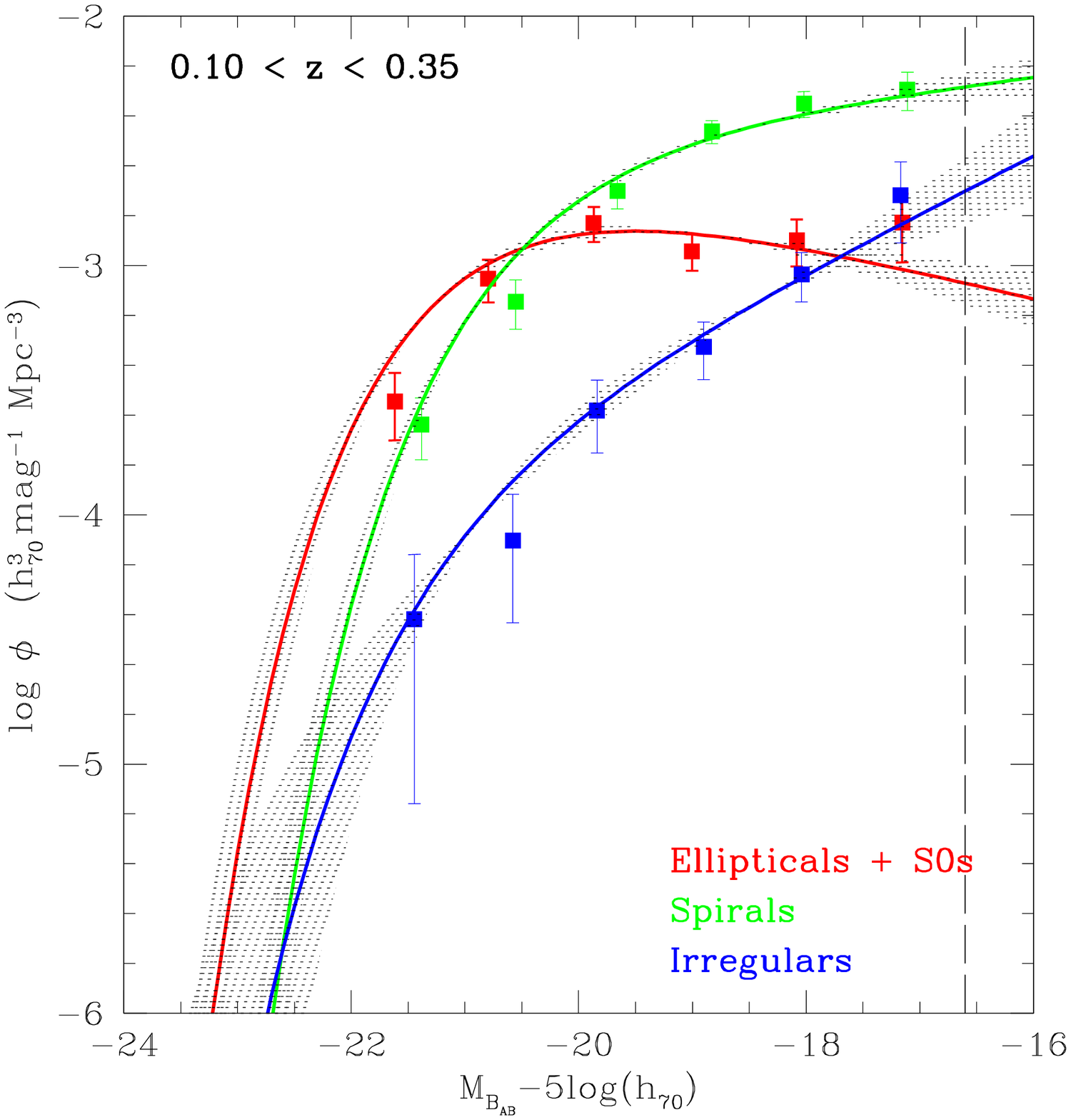}
\includegraphics[width=0.45\hsize]{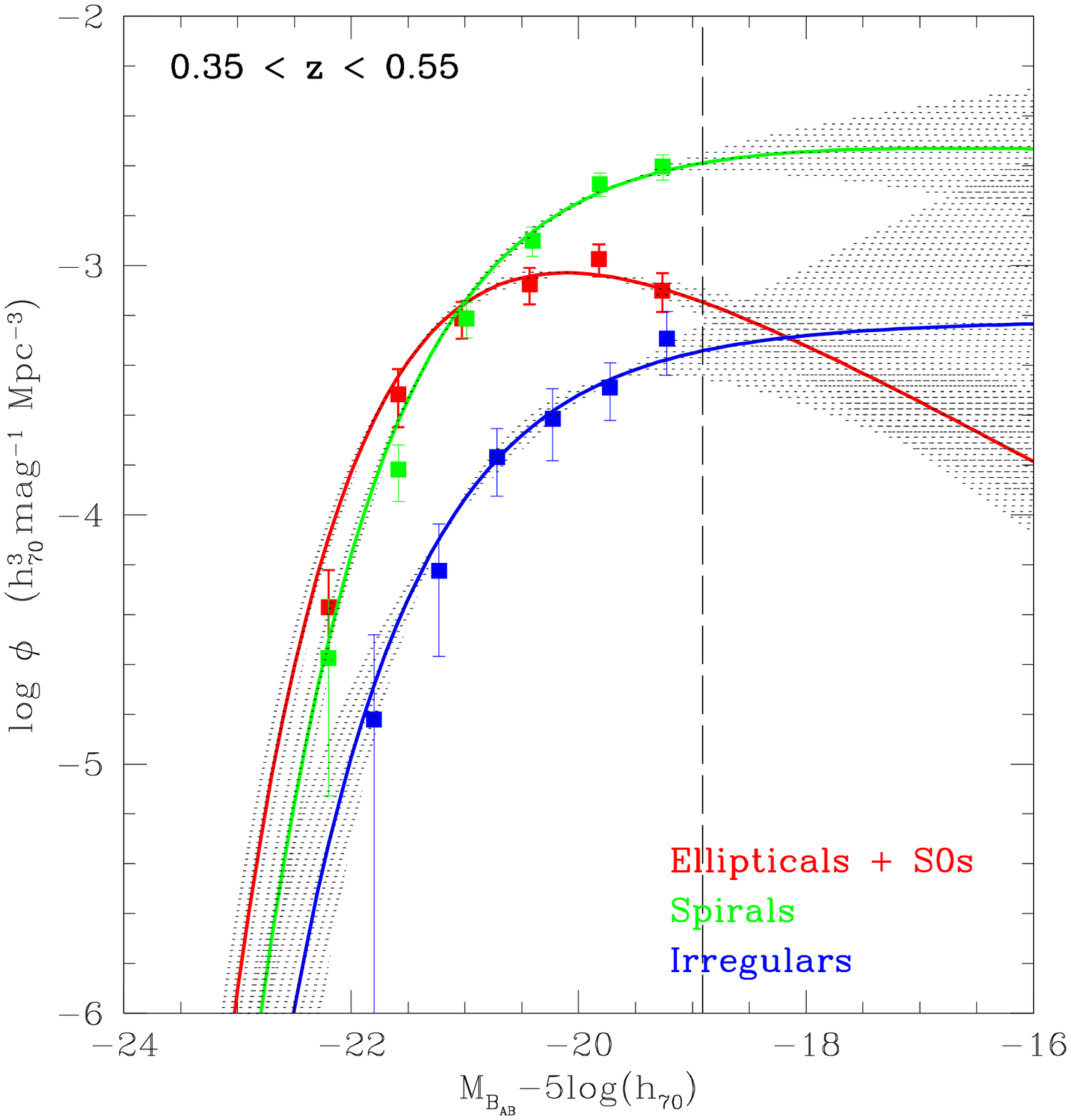}\\
\includegraphics[width=0.45\hsize]{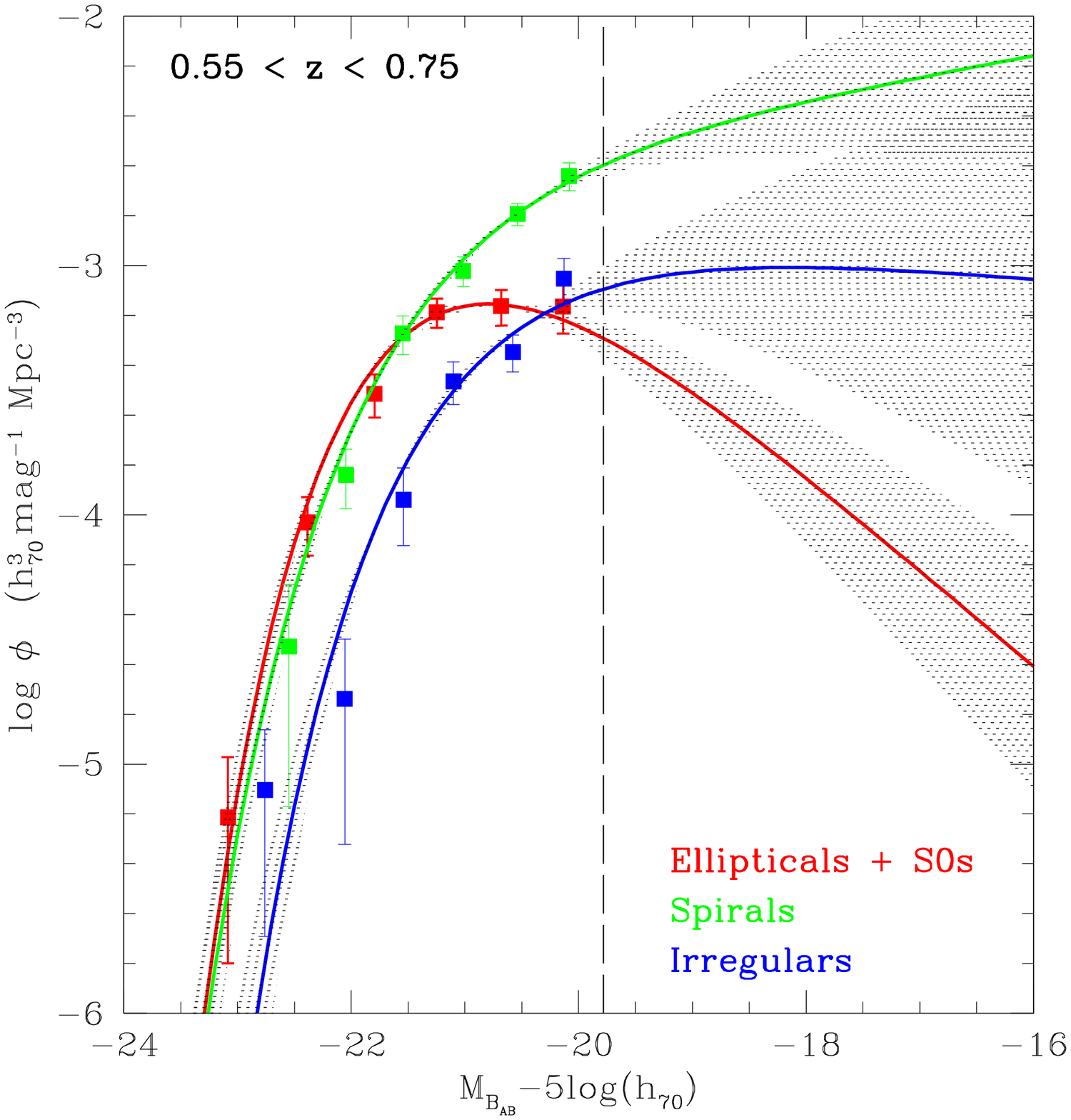}
\includegraphics[width=0.45\hsize]{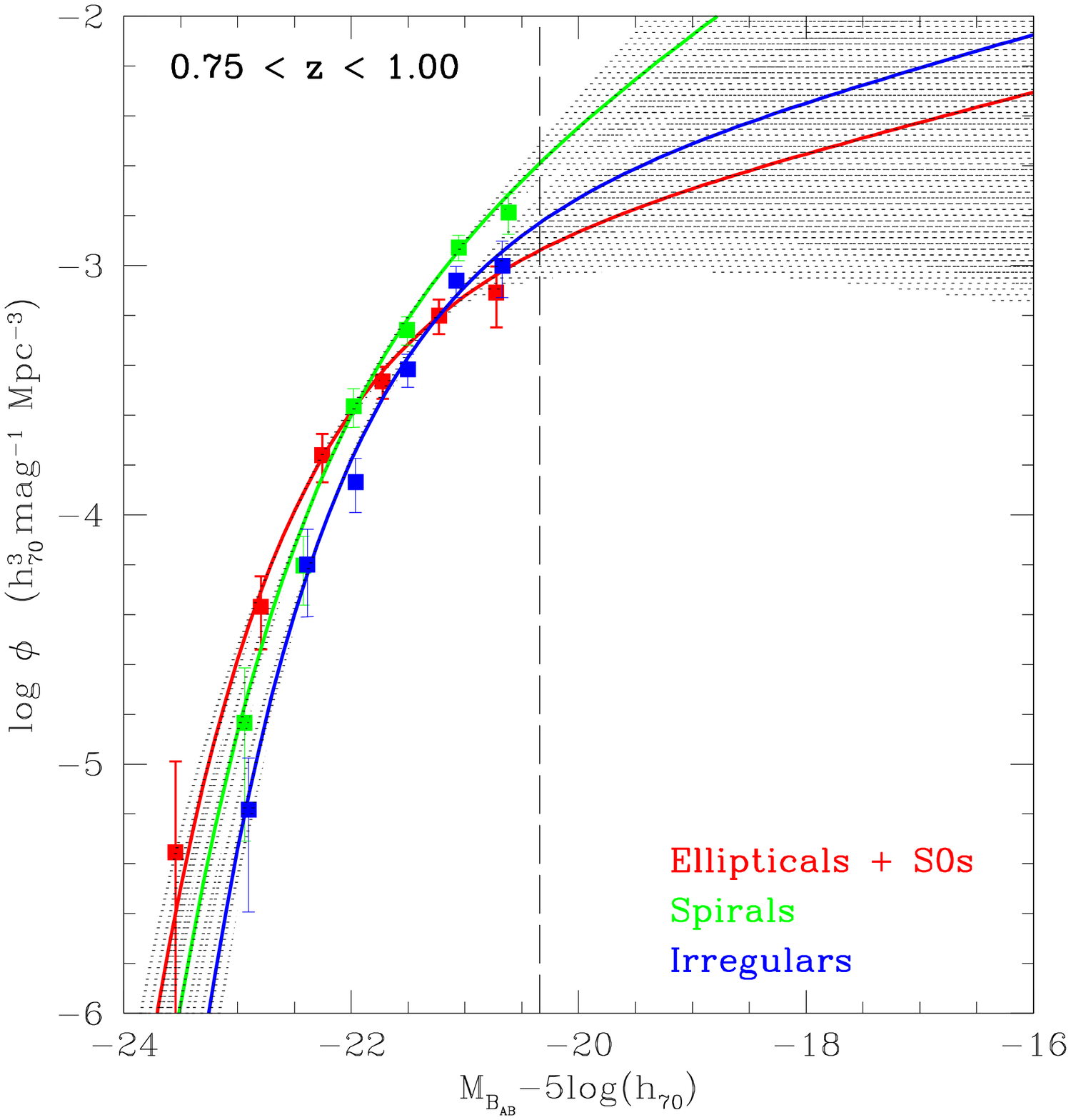}
\caption{ Luminosity functions of the different morphological types
in redshift bins: early types (ellipticals and lenticulars) 
in red, spirals in green, irregulars in
blue. The meaning of symbols and lines is the same as in 
Fig. \ref{LFforme_cww_new}.}
\label{LFforme_morpho}
\end{figure*}

\begin{figure*}
\centering
\includegraphics[width=0.49\hsize]{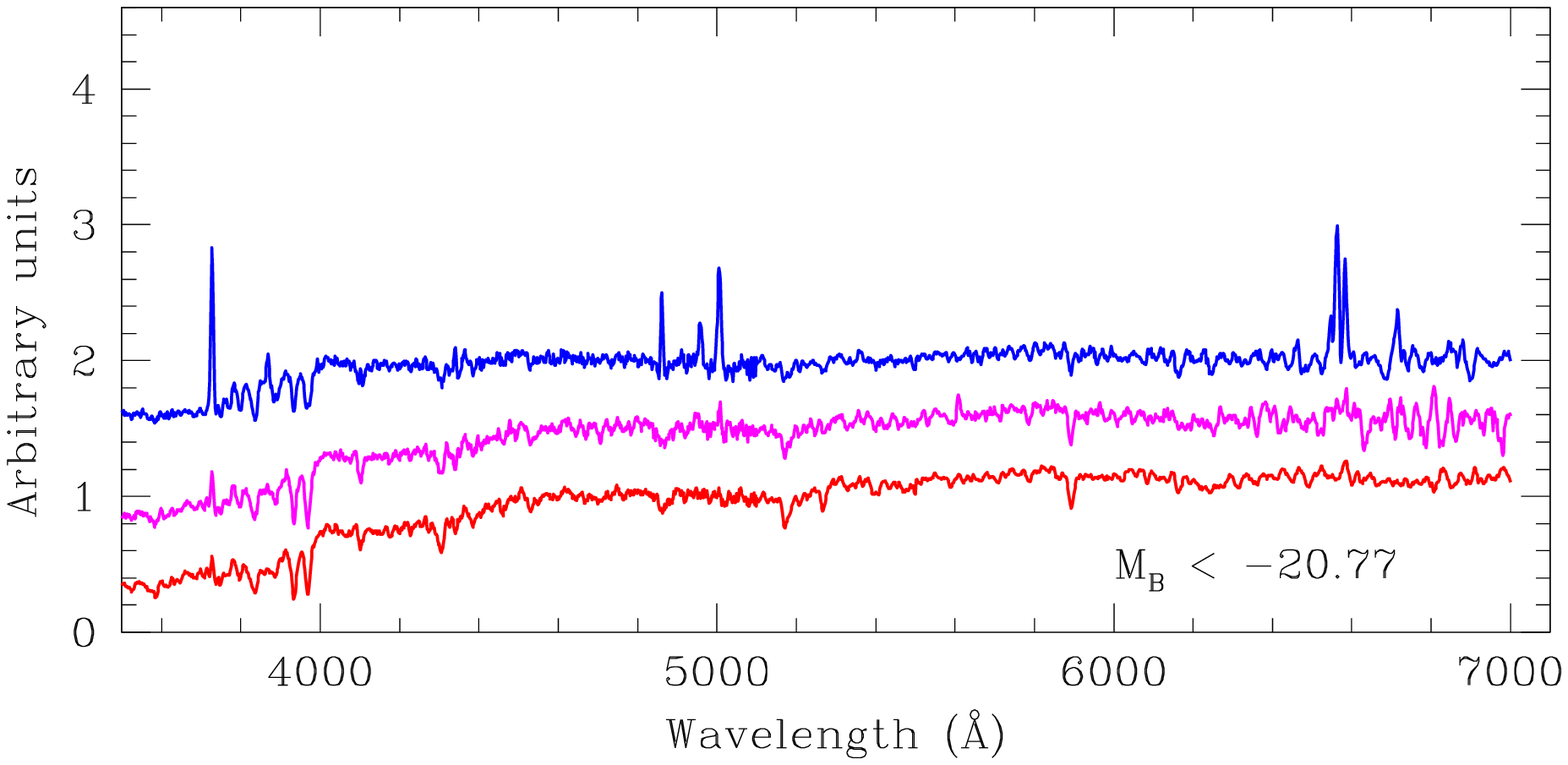}
\includegraphics[width=0.49\hsize]{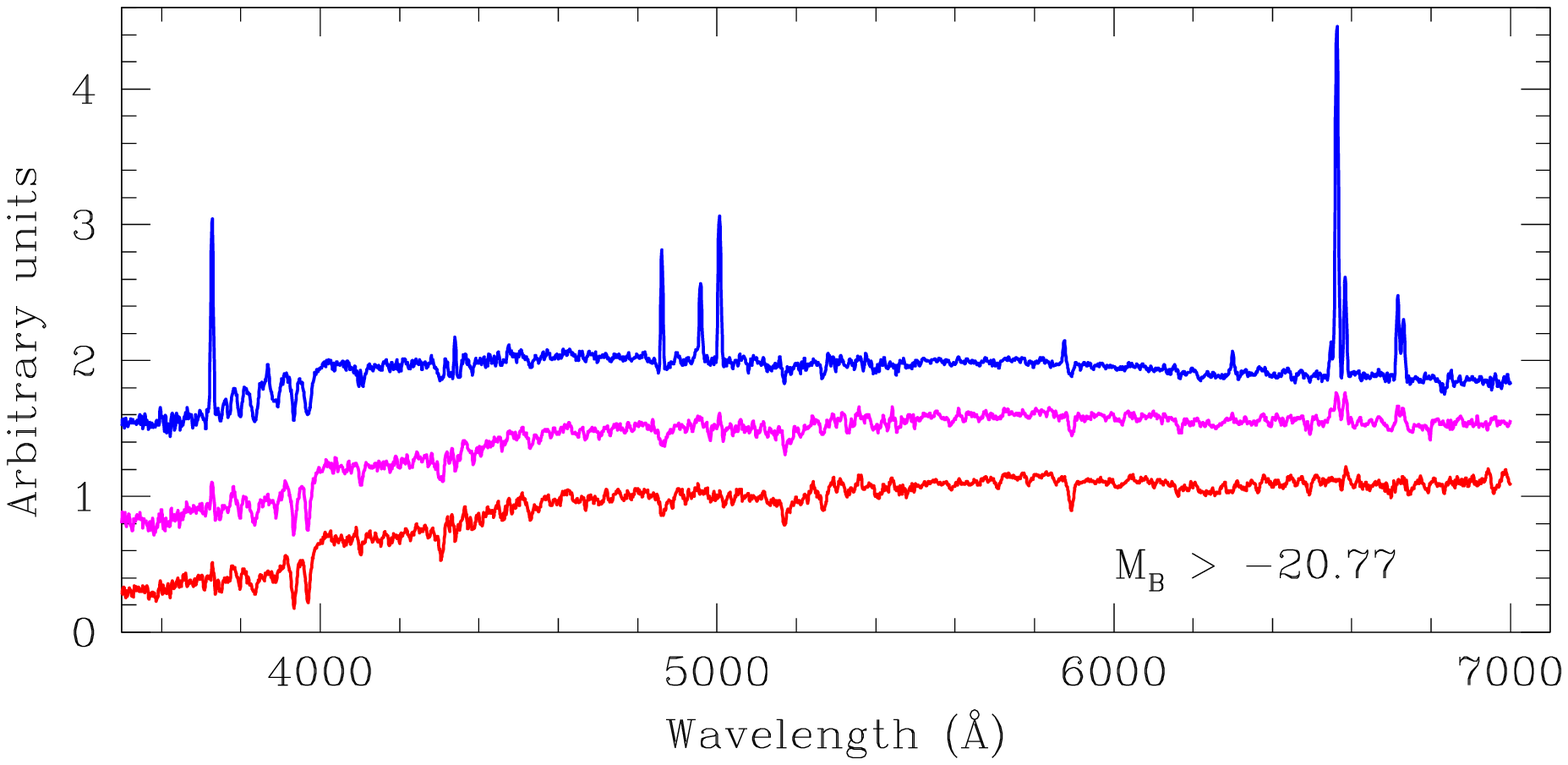}
\caption{Composite spectra of morphologically classified early types brighter
(left panel) and fainter (right panel) than $M_{B} = -20.77$.
Morphological early types are divided on the basis of their spectrophotometric
type (see the text for details): ``blue" early types in blue,
``red" early types in magenta, ``very red" early types in red.
The flux on the Y-axis is in arbitrary units and
the spectra have been vertically shifted for clarity. }
\label{composite}
\end{figure*}

\begin{figure*}
\centering
\includegraphics[width=0.45\hsize]{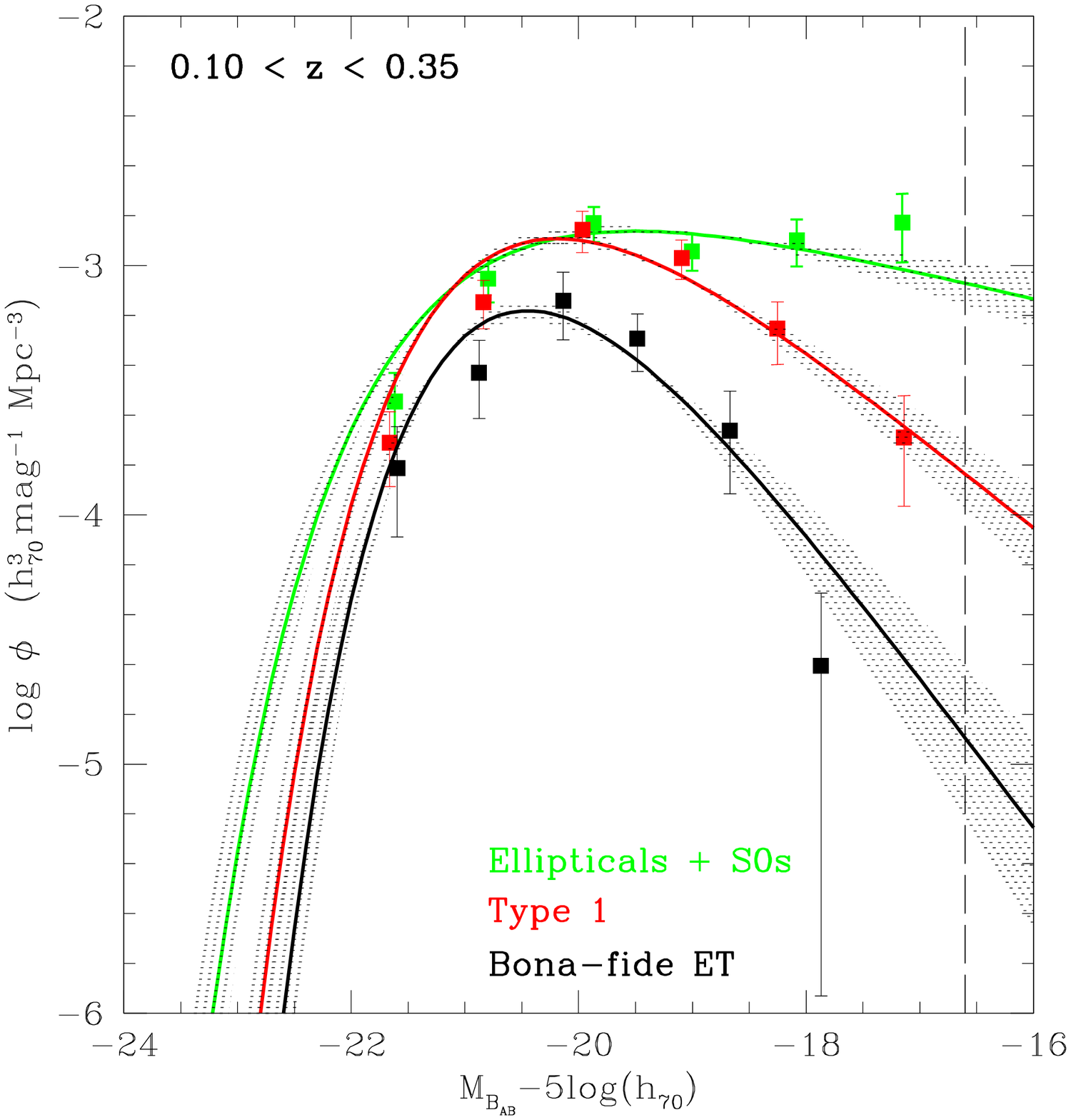}
\includegraphics[width=0.45\hsize]{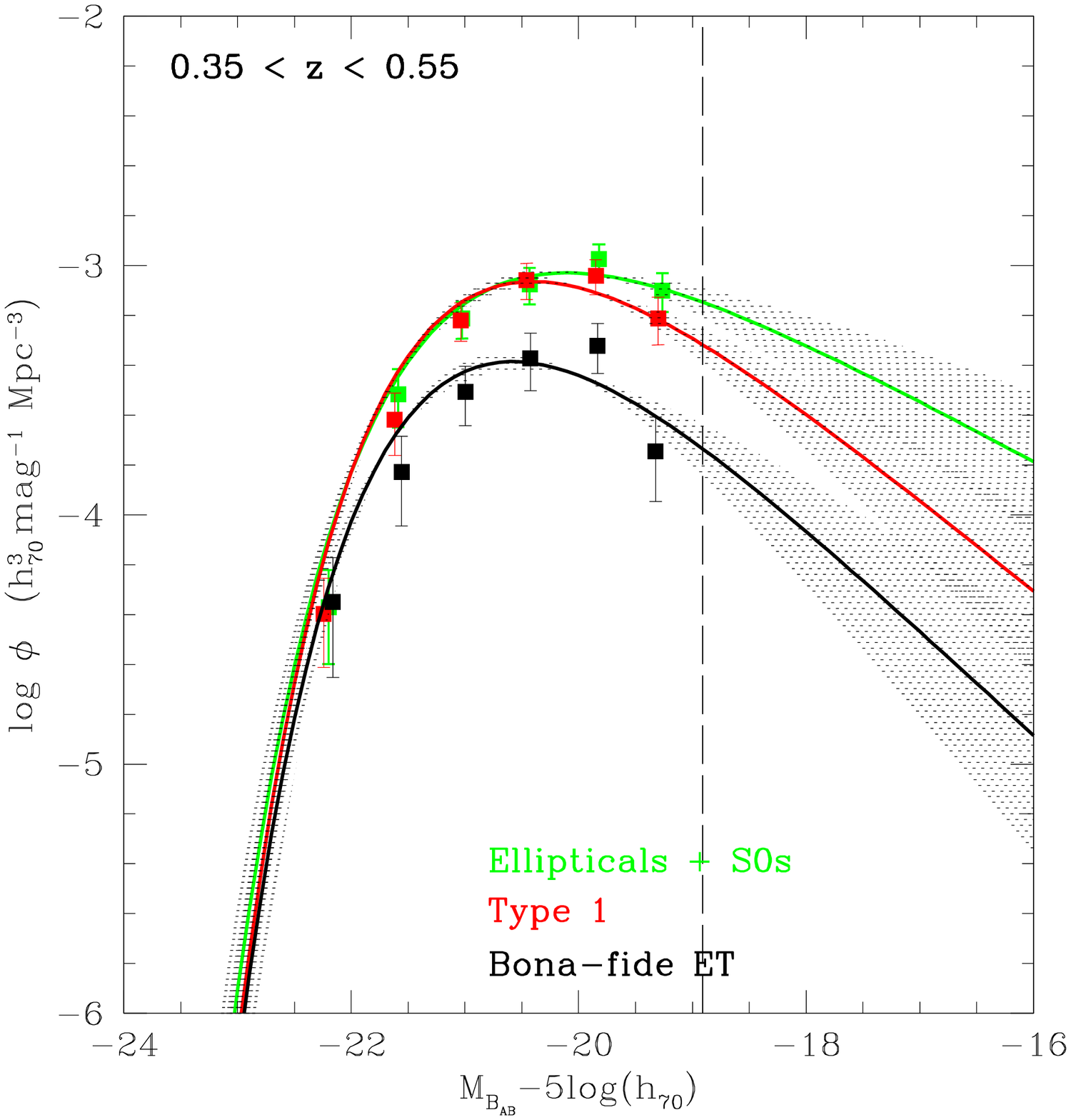}\\
\includegraphics[width=0.45\hsize]{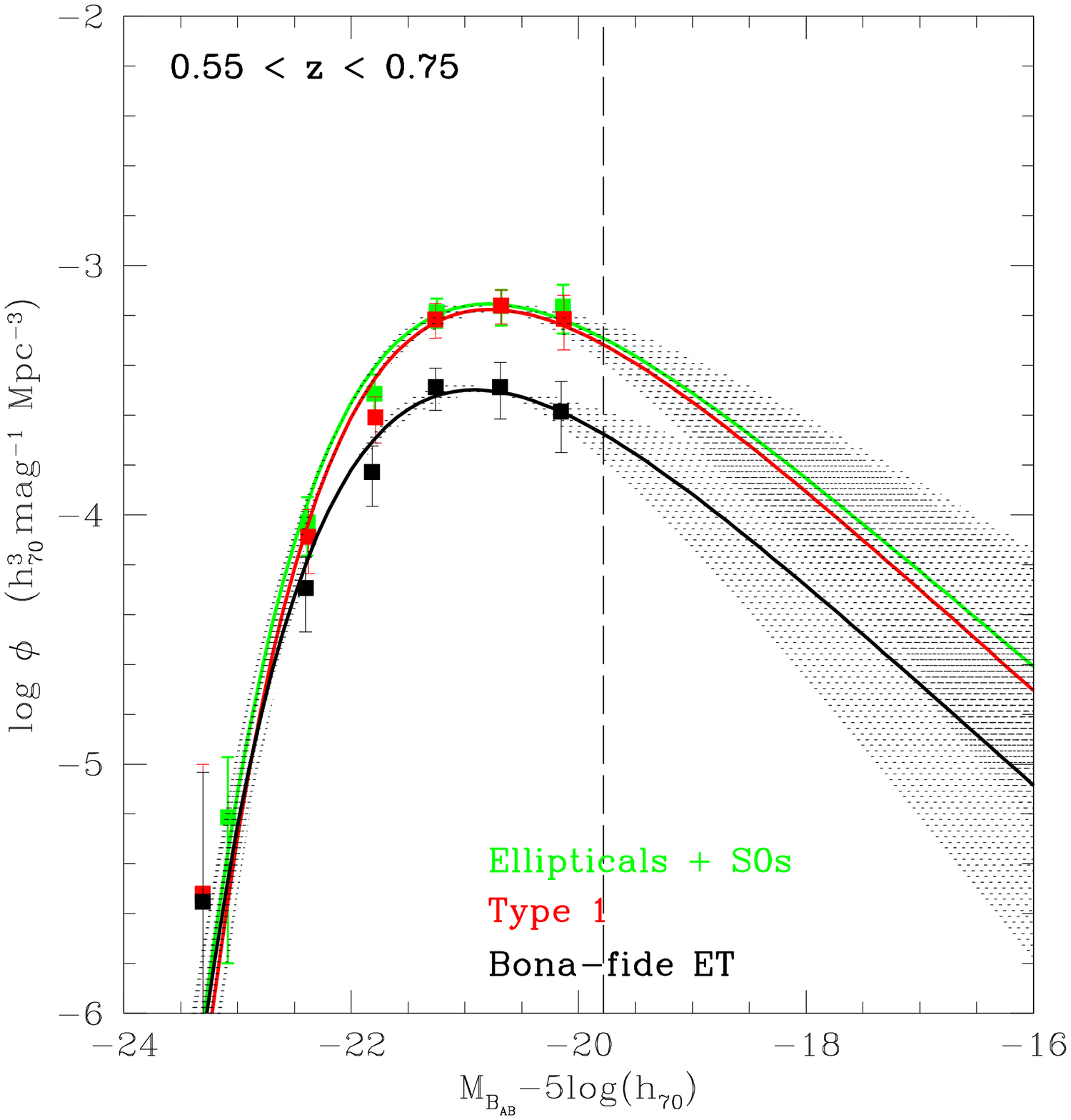}
\includegraphics[width=0.45\hsize]{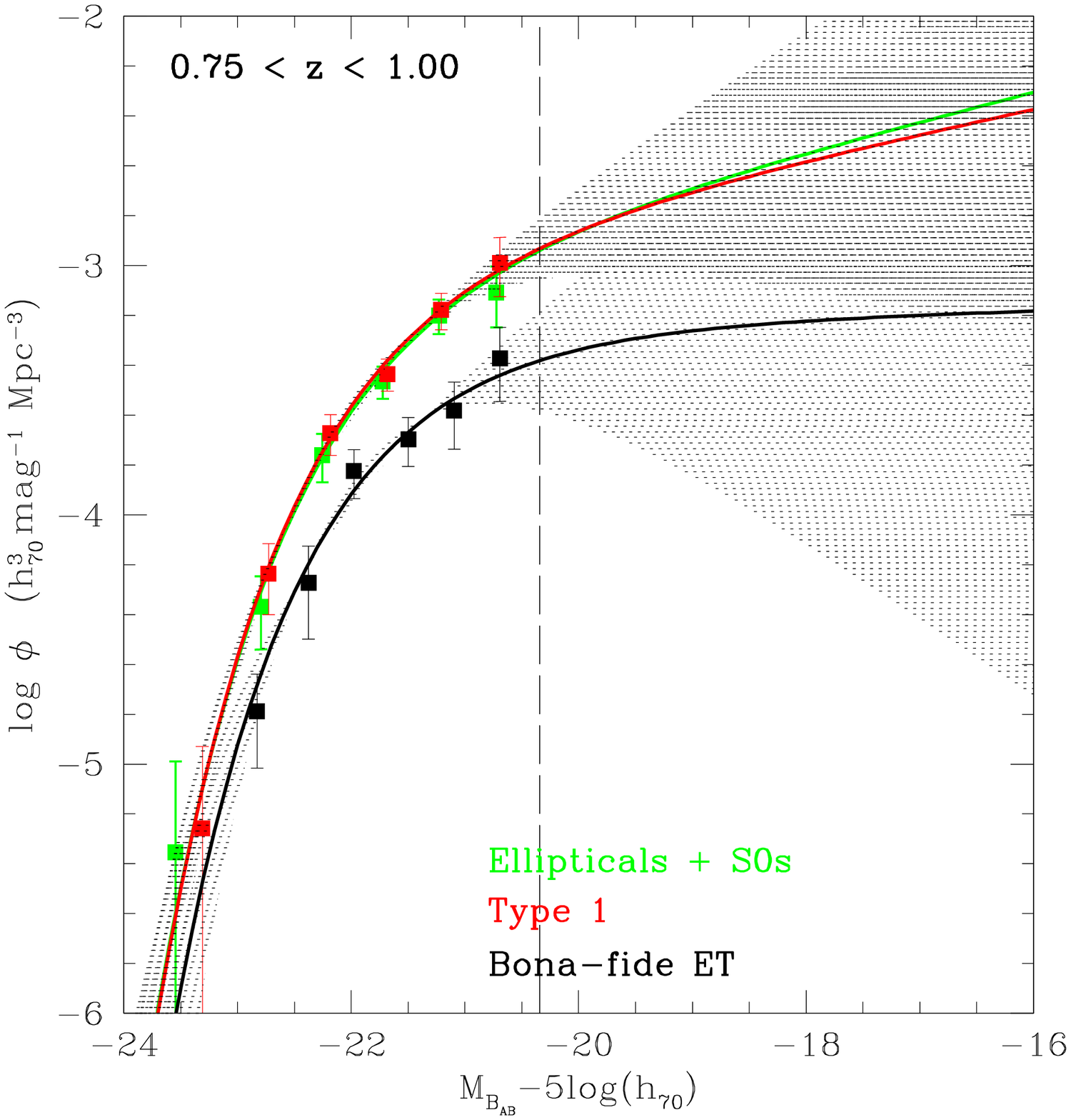}
\caption{ Luminosity functions of early-type galaxies: {\it bona-fide} ET (black),
type 1 (red) and morphological early-type (in green) galaxies.
The meaning of symbols and lines is the same as in Fig. \ref{LFforme_cww_new}. }
\label{LFforme_early}
\end{figure*}


\subsection{The weighting scheme}\label{sectwei}

To take account of unknown redshifts (for unobserved objects and poor 
quality spectra), it is necessary to apply a weight to each galaxy
(Zucca et al. \cite{zucca94}, Ilbert et al. \cite{vvdsLF}).
This weight is a combination of two different contributions: the target
sampling rate (TSR) and the spectroscopic success rate (SSR).  
\\
For each galaxy $i$, we computed the weight $w_i$, defined to be
the product of two factors: 
\\
a) $w_i^{\rm TSR}=1/{\rm TSR}=N_{\rm phot}/N_{\rm spec}$, where 
$N_{\rm spec}$ is the number of observed sources in the spectroscopic
survey and $N_{\rm phot}$ is the number of sources contained in the
parent catalogue used to select the targets; the TSR thus describes
the fraction of sources observed in the spectroscopic survey.
\\
b) $w_i^{\rm SSR}=1/{\rm SSR}=N_{\rm spec}^{\rm gal}/(N_{\rm
spec}^{\rm gal}-N_{\rm spec}^{\rm fail})$, where $N_{\rm spec}^{\rm gal}$ 
is the number of galaxies observed spectroscopically (i.e., excluding 
spectroscopically confirmed stars and Broad Line AGNs)
and $N_{\rm spec}^{\rm fail}$ is the number of objects without a
reliable measure of redshift, i.e., ``failures''. 
\\
Since the galaxies observed spectroscopically were randomly
selected from the parent sample, the TSR is independent of both the
apparent magnitude and other observed quantities, with an
approximately  constant value of $\sim 33\%$.
\\
In contrast, the SSR is a function of the selection magnitude, 
which is linked to the signal-to-noise ratio of the spectrum and  
ranges between $97.5$\% at bright magnitudes and $82$\% at the faintest ones 
for our subsample of galaxies. 
Moreover, from the photometric redshift distribution is evident that low-confidence 
redshift flags or complete failures correspond to objects at high redshift or in
redshift ranges where most of the prominent lines fall outside the
observed wavelength range (Lilly et al. \cite{lilly08}). 
For this reason,  we used the Ilbert et al. (\cite{ilbert09}) release 
of $z_{\rm phot}$ and computed the SSR in $\Delta z=0.2$ redshift bins.
Finally, the characteristic emission or absorption lines are
different for different galaxy types, as shown in Lilly et al. (\cite{lilly08}).
We further split the computation of SSR in each redshift bin
separating red and blue galaxies, selected on the basis of their
rest-frame $U-V$ color.  We computed the final weights $w_i=w_i^{\rm
TSR}\times w_i^{\rm SSR}$ considering all the described dependencies.
\\
This scheme was applied to each observed galaxy, a part from 
two groups of special objects. 
The spectroscopic catalogue contains not 
only randomly targeted objects, because a small fraction of sources 
(mainly X-ray sources) were flagged as ``compulsory'' when 
preparing the masks; the TSR of these sources is much higher than 
the global one and was computed separately.  
In some cases, a slit also contained objects in addition to the
primary target. For these ``secondary targets'', the SSR was found 
to be far lower than the global relation, because most of these sources 
had not been centered well in the slit, resulting in faint
spectra. For these objects, the SSR was computed separately.
The contribution of these two classes of objects to the total sample
was only $\sim 2\%$ and $\sim 3\%$, respectively.


\section{The global luminosity function}\label{sectLFtot}

The global luminosity function was computed as a function of redshift 
to $z=1.3$, adopting the same redshift bins used in the analysis
of the VVDS deep field (Ilbert et al. \cite{vvdsLF}), 
to allow a direct comparison to be made.
In the following, we show the results for the Johnson $B$ rest-frame band.
\\
Figure \ref{LFtot} shows the global luminosity function in redshift bins, 
obtained with the $C^+$ and $STY$ methods. The luminosity
functions derived with the other two methods ($1/V_{max}$  and $SWML$)
are consistent with those shown in the figures, but are not drawn for
clarity.  The dotted line represents the luminosity function estimated 
in the redshift range $[0.2-0.4]$ and is shown in each panel for reference.
The $STY$ estimates from the VVDS sample are plotted with dashed lines.
Vertical lines represent the bias limit described in Sect. \ref{sectalf}.
\\
The global luminosity function agrees well with the VVDS 
estimates: there are small differences in the normalization
in some redshift bins, due to the presence of underdense and overdense 
regions in both samples, which is particularly strong in the zCOSMOS sample  
(see Lilly et al. \cite{lilly08}). In particular, the high zCOSMOS LF 
normalization in the redshift bin $[0.8-1.0]$ is due to a prominent
structure, discussed in detail in Kova\v{c} et al. (\cite{kovac08}).
We note also that the VVDS luminosity functions were derived with
slightly different cosmological parameters ($\Omega_m = 0.3$ and
$\Omega_\Lambda = 0.7$).
\\
Because of the larger area, 
the zCOSMOS sample is more suitable for constraining the
bright part of the luminosity function with respect to the
VVDS deep sample. With the zCOSMOS data, it is also possible 
to derive some constraints on the $M^*$ value in the first redshift bin, where
for the VVDS sample it was necessary to fix $M^*$ to the local 
SDSS value. 
On the other hand, the VVDS fainter magnitude limit 
allowed us to estimate the slope $\alpha$ more reliably at high redshift,
where the estimate from the zCOSMOS sample is almost unconstrained.
For this reason, we fixed $\alpha$ to the VVDS value in
the redshift bins $[0.8-1.0]$ and $[1.0-1.3]$.
\\
Having shown full consistency between the zCOSMOS and VVDS
luminosity functions, 
we then derived the luminosity function in the redshift bins
$[0.10-0.35]$, $[0.35-0.55]$, $[0.55-0.75]$ and $[0.75-1.00]$, 
following the choice of Pozzetti et al. (\cite{pozzetti08}) for
their study of the zCOSMOS mass function, and 
in Table \ref{param_tot_types} we report the $STY$ parameters for 
each bin.
\\
The errors reported in this table represent a $1\sigma$ confidence
level for a 1-parameter estimate ($\Delta\chi^2 = 1.0$).
These errors underestimate the true errors.
In particular, they are smaller than 
the projection onto the parameter axes of the 2-parameter $68\%$ 
confidence ellipses, which, on the other hand, are 
always an overestimate of the true errors, especially when the 
errors on the two parameters are highly correlated. 
\\
The luminosity function evolves by $\sim 0.7$ mag in $M^*$
from the first ($[0.10-0.35]$) to the last ($[0.75-1.00]$) 
redshift bins. This result agrees with the VVDS results for the same redshift range. 
For the evolution in slope, we find that $\alpha$ is 
almost constant in the first three redshift bins, and then steepens 
in the last bin.
\\
To understand how the different galaxy populations contribute
to this evolution in the global luminosity function, we now quantify 
the contribution of the different galaxy types, 
by deriving  their luminosity functions separately. 


\section{The contribution of the different galaxy types}\label{sectLFtype}


\subsection{Spectrophotometric types}\label{sectLFcww}

Galaxies were divided into four spectrophotometric classes
(see Sect. \ref{sectmag} above).
In the left panel of Fig. \ref{fracTOT}, we plot the observed fraction of bright 
galaxies for each type as a function of redshift. We identified and selected objects with
$M_{B} < -20.77$ to be galaxies visible for the entire 
redshift range $[0-1.0]$. The same cut was adopted for the VVDS,
but in that case it was possible to sample the range $[0-1.5]$ because of
the fainter magnitude limit of the survey.  
From this figure, it is clear that the bright late-type
population becomes increasingly dominant at higher redshifts,  
while, correspondingly, the fraction of bright early-type galaxies decreases.
These trends in the fraction of the bright galaxy population with redshift are 
similar to those found in the VVDS (in the common redshift range),
although with some differences in the normalizations.
The decrease of early-type galaxies in the zCOSMOS sample 
appears less significant than in the VVDS sample: starting from similar
values at $z\sim 0.3$, the zCOSMOS curve remains significantly (at
$\sim 2\sigma$ level) higher by $\sim 20\%$
than the VVDS one. This fraction decreases by a
factor $\sim 2$ at $z\sim 0.9$ in zCOSMOS, while it decreases by a factor of 
$\sim 2.3$ at the same redshift in the VVDS.
In the same redshift range, the fraction of bright type 3 galaxies increases by
a factor $\sim 6$; the strong increase in type 4 galaxies detected in the
VVDS is not visible here, due to the lower redshift range.
\\
Luminosity functions were derived for each type in different
redshift bins.
For each type, we derived the luminosity function also by fixing $\alpha$ to
the value obtained in the redshift range $[0.30-0.80]$. This choice
allowed us to better constrain the evolution of $M^*$ with redshift 
and is acceptable because most of the $\alpha$ values estimated in the
various redshift bins are consistent with the $[0.30-0.80]$ value. 
This consistency is marginal for the highest redshift bin of type 1 galaxies,
but in this bin the $STY$ estimate is poorly constrained because the faint
end of the luminosity function is inadequately sampled, due to the
magnitude limit of the survey.
\\
For type 1, 2 and 3 galaxies, this value of $\alpha$ agrees with 
the VVDS estimate.
For type 4 galaxies, the parameter $\alpha$ is unconstrained for $z>0.3$:
for this reason, we fixed $\alpha$ to be the value derived from the first 
redshift bin, which is consistent with the VVDS value for type 4 galaxies.
\\
The evolution of $M^*$ and $\phi^*$ with redshift for the different types
are consistent with the VVDS results in the common redshift range.
\\
Given the uncertainties in the luminosity function estimate of type 4 galaxies,
we repeated the analysis by grouping together type 3 and 4 galaxies:
the results are shown in Fig. \ref{LFforme_cww_new} and the $STY$ parameters 
for each bin are reported in Table \ref{param_tot_types}, for both $\alpha$
free and $\alpha$ fixed. Parameters for the reference bin $[0.30-0.80]$ are also
reported in the table.
In Fig. \ref{LFforme_cww_new} the squares represent the
results from the $C^+$ and the solid lines are the results
from the  $STY$ method: type 1 galaxies are shown in red, type 2 galaxies in
orange, type 3+4 galaxies in blue. The total sample is shown in black.
The shaded regions represent the $68\%$ uncertainties in the parameters
$\alpha$ and $M^*$.
From this figure, it is clear that at low redshifts ($z<0.35$) late-type
galaxies dominate for faint magnitudes ($M_{B} > -20$), while
the bright end is populated mainly by type 1 galaxies.
At higher redshift, late-type galaxies evolve strongly 
and at redshift $z > 0.55$ the contributions of the various types
to the bright end of the luminosity function are comparable. 
The faint end remains dominated by late-type galaxies
over the entire redshift range.
\\
To visualize the evolution with redshift, in Fig. \ref{LFforme_z} 
we plot the luminosity functions in different redshift bins for
each galaxy type: type 1 galaxies in the left panel, type 2
galaxies in the middle panel, and type 3+4 galaxies in the right panel.
The different colors represent different redshift bins:
$[0.10-0.35]$ in black, $[0.35-0.55]$ in cyan, $[0.55-0.75]$ in magenta,
and $[0.75-1.00]$ in green. To follow the evolution
in $M^*$ and $\phi^*$, we show the $STY$ estimates
obtained with $\alpha$ fixed.
For type 1 galaxies, evolution occurs in both luminosity
and normalization: $M^*$ brightens by $\sim 0.6$ mag
and $\phi^*$ decreases by a factor $\sim 1.7$ between the first and the last 
redshift bin.
Type 3+4 galaxies also evolve both in luminosity and normalization, 
but with an opposite trend for the normalization: a brightening with
redshift of $\sim 0.5$ mag is evident in $M^*$, while $\phi^*$ increases 
by a factor $\sim 1.8$.
Type 2 galaxies exhibit a milder evolution, involving a brightening of $\sim 0.25$
mag in $M^*$ and no significant evolution in $\phi^*$.
\\
The galaxy stellar mass functions of the various types (Pozzetti et al.
\cite{pozzetti08}) show differences in the massive part stronger that
the differences we find in the bright part of the luminosity functions.
This is due to the fact that the mass-to-light ratio of early-type
galaxies is on average higher than that of late-type galaxies:
in fact, the contribution to the bright end of our $B$ band luminosity
functions comes from both massive red galaxies and blue galaxies
with strong star formation. 
This fact also implies that the bimodality observed in the global galaxy
stellar mass function by Pozzetti et al. (\cite{pozzetti08}) is not
detected in the global luminosity function.


\subsection{Morphological types}

A major advantage of the zCOSMOS survey is the availability of
galaxy morphologies obtained from the HST ACS images (Koekemoer et al.
\cite{koe07}).
Galaxies were divided into early types (including ellipticals and
lenticulars), spirals, and
irregulars following the classification described in Sect. \ref{sectacs}
above. In the right panel of Fig. \ref{fracTOT}, we plot the observed 
fraction of bright galaxies of each morphological type as a function 
of redshift. Unclassified galaxies are a very small fraction
of the sample ($\sim 3\%$) and are uniformly distributed with redshift.
\\
As observed for spectrophotometric
types, late-type galaxies steadily increase their fraction with increasing
redshift, while the fraction of early types decreases. 
However, at low redshift the fraction of morphologically classified early types
is higher than that of type 1 galaxies, as is evident by comparing
the left and right panels in Fig. \ref{fracTOT}.
\\
The luminosity functions of the different types are
shown in Fig. \ref{LFforme_morpho} (early types in red,
spirals in green, irregulars in blue) and the $STY$ parameters for 
each bin are reported in Table \ref{param_tot_types}, where we show
results for 
both $\alpha$ free and $\alpha$ fixed to the value determined in 
the redshift range $[0.30-0.80]$.
\\
At low redshift ($z<0.35$), early-type galaxies dominate the
bright end of the luminosity function, while spiral galaxies 
dominate the faint end. Irregular galaxies increase their
contribution at the lowest luminosities.
At intermediate redshift ($[0.35-0.75]$), spiral galaxies increase
their luminosities and their contribution to the bright end of
the luminosity function is similar to that of the early types.
At high redshift ($z>0.75$), irregular galaxies evolve strongly and
the three morphological types contribute almost equally 
to the total luminosity function. 
Irregular galaxies show an evolution of a factor $\sim 3.3$ in $\phi^*$
from low to high redshift.
This evolution occurs mainly in the last redshift bin, while
for $z<0.75$, the contribution of these galaxies to
the global luminosity function is significantly lower than
that of spirals and early types.
\\
Scarlata et al. (\cite{scarlata07}) derived luminosity functions for 
COSMOS galaxies, using photometric redshifts and the ZEST morphological
classification. 
The shape parameters we find for early types and spirals are consistent
with those found by Scarlata et al. (\cite{scarlata07}) for  
their early-type and disk galaxies, but are significantly different
for irregular galaxies, which have a much flatter slope in 
Scarlata et al. (\cite{scarlata07}).
These differences are likely due to the different morphological
classification applied. 
\\
Although the general trend in the luminosity functions of 
the different morphological types 
is similar to the results obtained in the previous section
for spectrophotometric types, some differences are present.
In particular, there are more morphological early-type than type 1 galaxies
at the faint end of the luminosity function ($M_{B} \simgt -19.5$
in the first redshift bin).   
In the following, we discuss the relationship between 
spectrophotometric and morphological types, paying particular
attention to early-type galaxies.


\section{Spectrophotometric versus morphological types}\label{sect2types}

Although there is a broad agreement between our results for spectrophotometric 
and morphological types, it is reasonable that there should also be 
some differences.  
Spectrophotometric types are based on the galaxy SEDs and
therefore depend on the star formation history, while morphological
types reflect mainly the dynamical history of the galaxy.
These classifications can also be affected by different
observational and methodological biases.
\\
We compared the spectrophotometric and morphological results for
each galaxy.
Considering early-type galaxies, we have 2387 morphological early types,
1504 of which ($63\%$) are classified as type 1 galaxies.
In contrast, $71\%$ of the 2105 type 1 galaxies are classified
as morphological early types. 
If we consider only galaxies whose SED is most accurately reproduced 
by the most extreme (i.e. reddest) type 1 template, the fraction 
increases to $80\%$.
The remaining fraction of type 1 galaxies that are not classified
as morphological early types is in part due to objects being without morphological
classification ($\sim 3\%$) and in part due to a population of 
``red" spirals, many of which on visual inspection appear to be 
edge-on spiral galaxies, often dominated by a strong dust lane
(see Tasca et al. \cite{tasca08} for a detailed discussion).
The red SED for these galaxies is probably caused by a significant
amount of dust extinction. 
\\
We then considered in more detail the $\sim 37\%$ of morphological
early-type galaxies that were not classified as spectrophotometric
type 1. We termed ``blue" early types those with spectrophotometric 
type 2, 3 or 4; among the morphological early types with spectrophotometric 
type 1, we termed ``very red" those best fitted by the reddest type 1 template
and ``red" those best fitted by the other type 1 templates.
These ``blue" early types were visually inspected (see details in
Tasca et al. \cite{tasca08}) revealing a class of face-on late-type 
galaxies with morphological parameters typical of an early-type population.
\\
To explore the properties of these classes of morphological 
early types, composite spectra were generated for these objects
(following Mignoli et al. \cite{mignoli08}) by averaging their
spectra and dividing them into bins brighter and fainter than 
$M_{B} = -20.77$.  
\\
These spectra are shown in Fig. \ref{composite}, where 
the Y-axis is rescaled arbitrary 
for clarity, for the bright sample on
the left and for the faint sample on the right.
Blue, magenta, and
red spectra correspond, respectively, to ``blue", ``red", and
``very red" early types. 
From this figure, clear differences
are visible in the spectra, in particular prominent emission
lines are present in ``blue" early types.
By comparing the bright and the faint ``blue" early types,
we find that the $H\alpha$ line equivalent width increases for
fainter galaxies and that the ratio between $[NII]$ and $H\alpha$
lines is stronger for brighter galaxies. 
The $[NII]$ / $H\alpha$ and $[OIII]$ / $H\beta$ ratios of the
bright ``blue" early types are consistent with those of liners.
Bright ``blue" early types are $\sim 30\%$ of the total number of
bright early types; by considering the faint early types,
the fraction of ``blue" early types increases to $\sim 44\%$.
\\ 
Similar spectra for ``blue" early types at low (\logM$<9$) and
high (\logM$>10.5$) masses are shown by Pozzetti et al. (\cite{pozzetti08}), 
who find a more evident separation between the properties
of objects in these two subclasses.
\\ 
As a final check we also considered a sample of 
{\it bona-fide} early-type (ET) galaxies (Moresco et al. \cite{moresco09}) 
selected in a more conservative way by combining information on morphologies, 
spectrophotometric types, colors, and emission line equivalent widths 
(see Pozzetti et al. \cite{pozzetti08} for the exact criteria adopted).
This conservative selection reduces the number of objects in the
early-type sample to 981: therefore, we use {\it bona-fide} ET galaxies
mainly as a comparison sample.
\\
We now explore the luminosity function of early-type galaxies.

\begin{figure*}
\centering
\includegraphics[width=0.3\hsize]{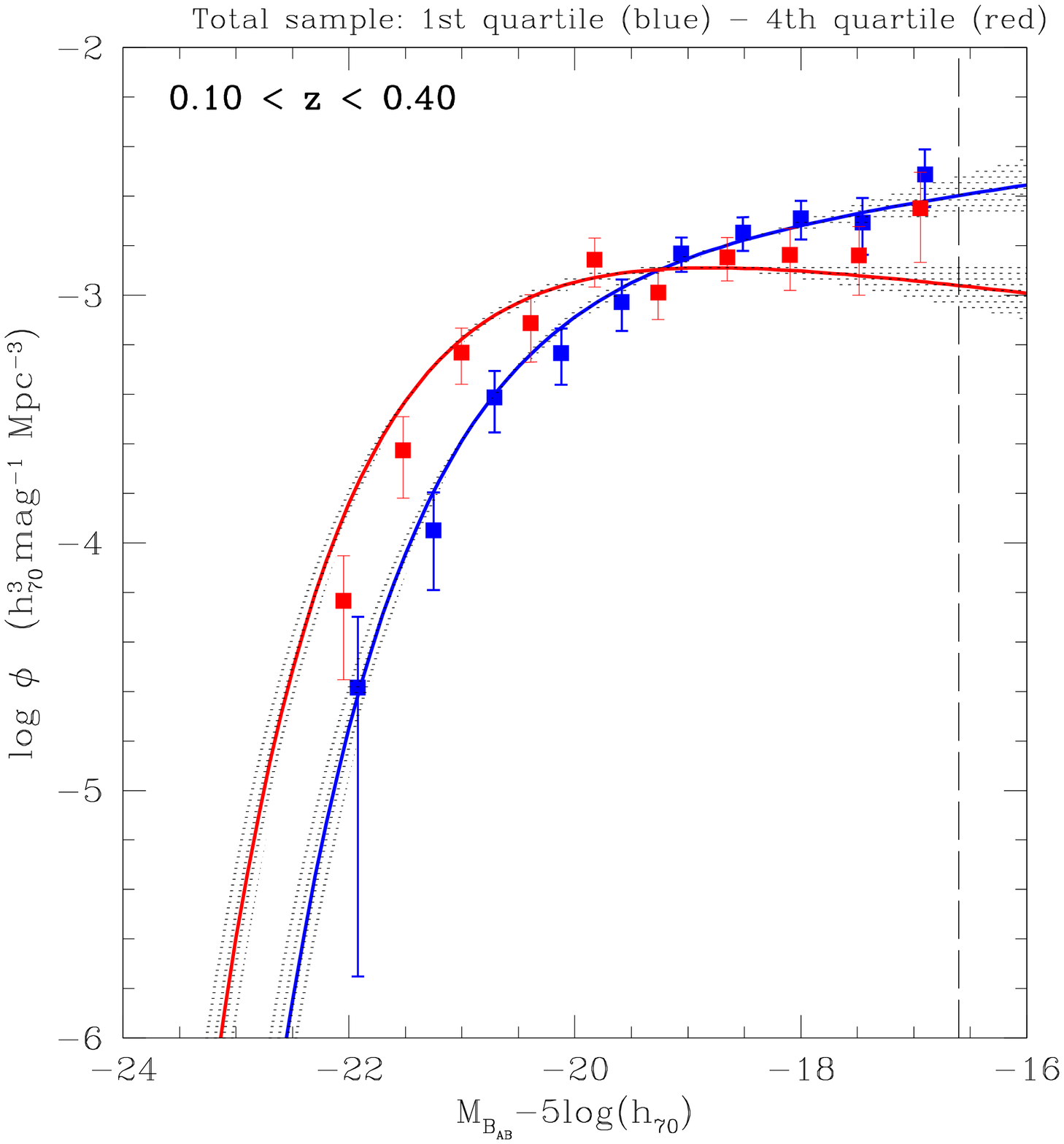}
\includegraphics[width=0.3\hsize]{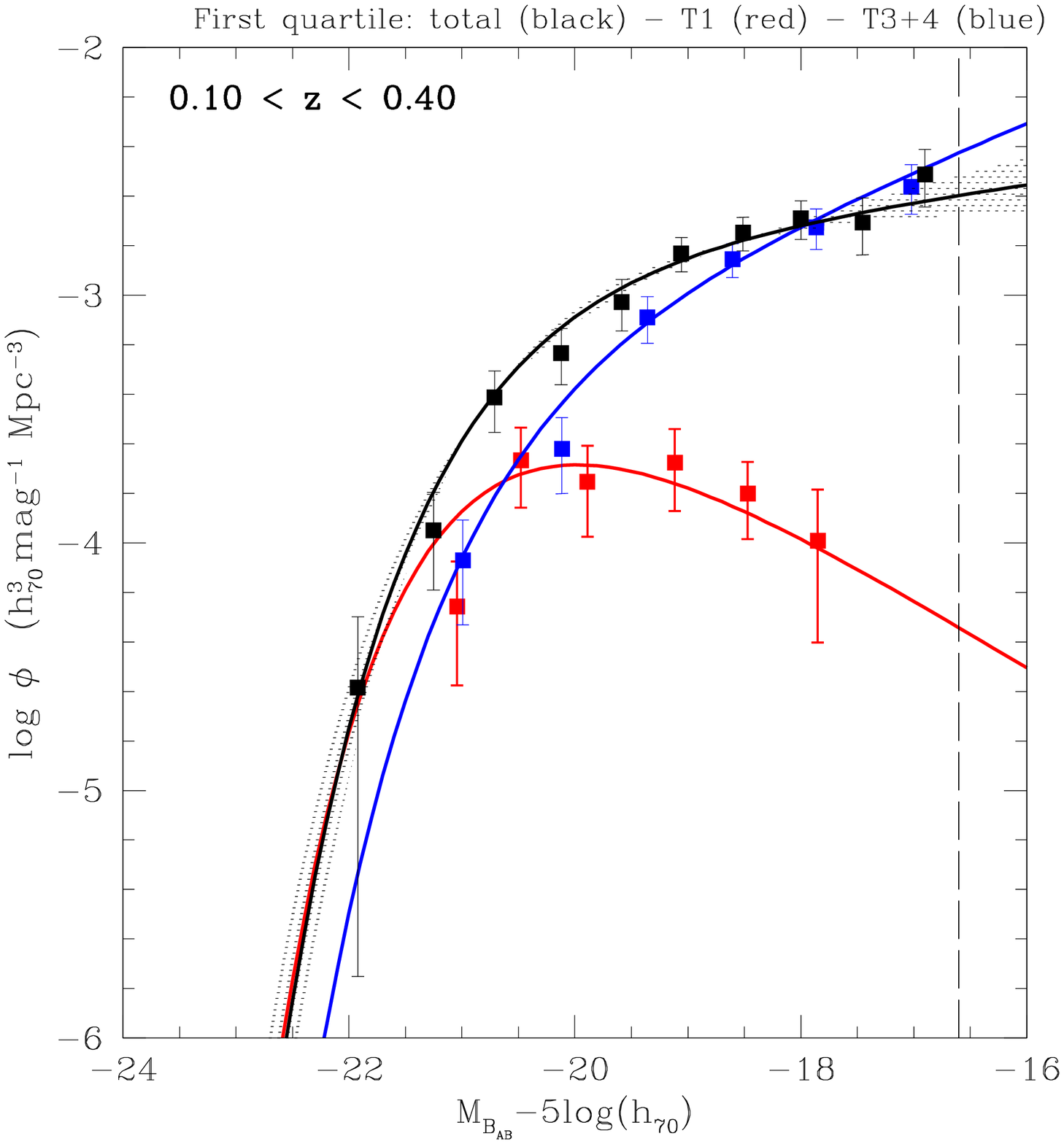}
\includegraphics[width=0.3\hsize]{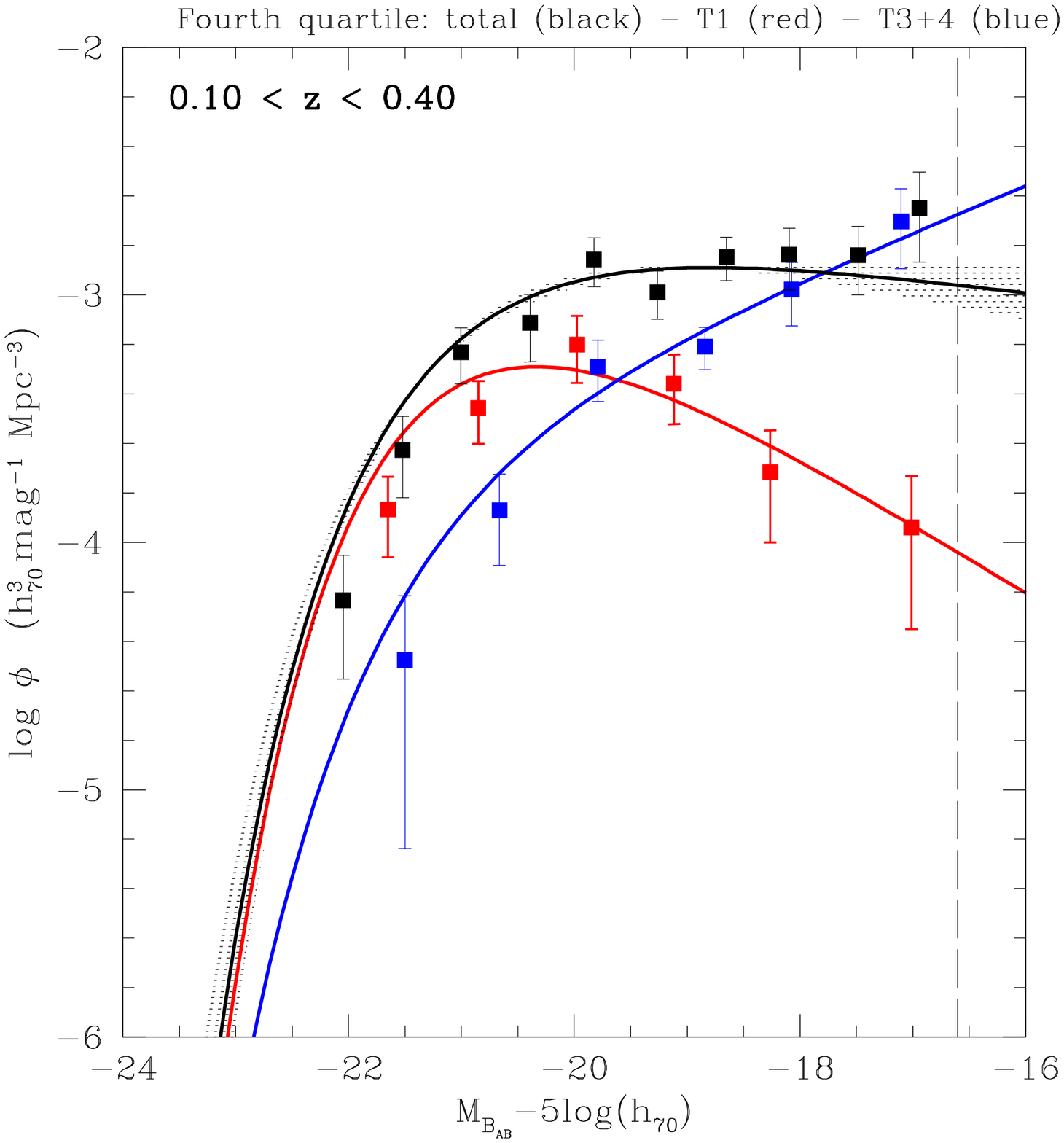}\\
\includegraphics[width=0.3\hsize]{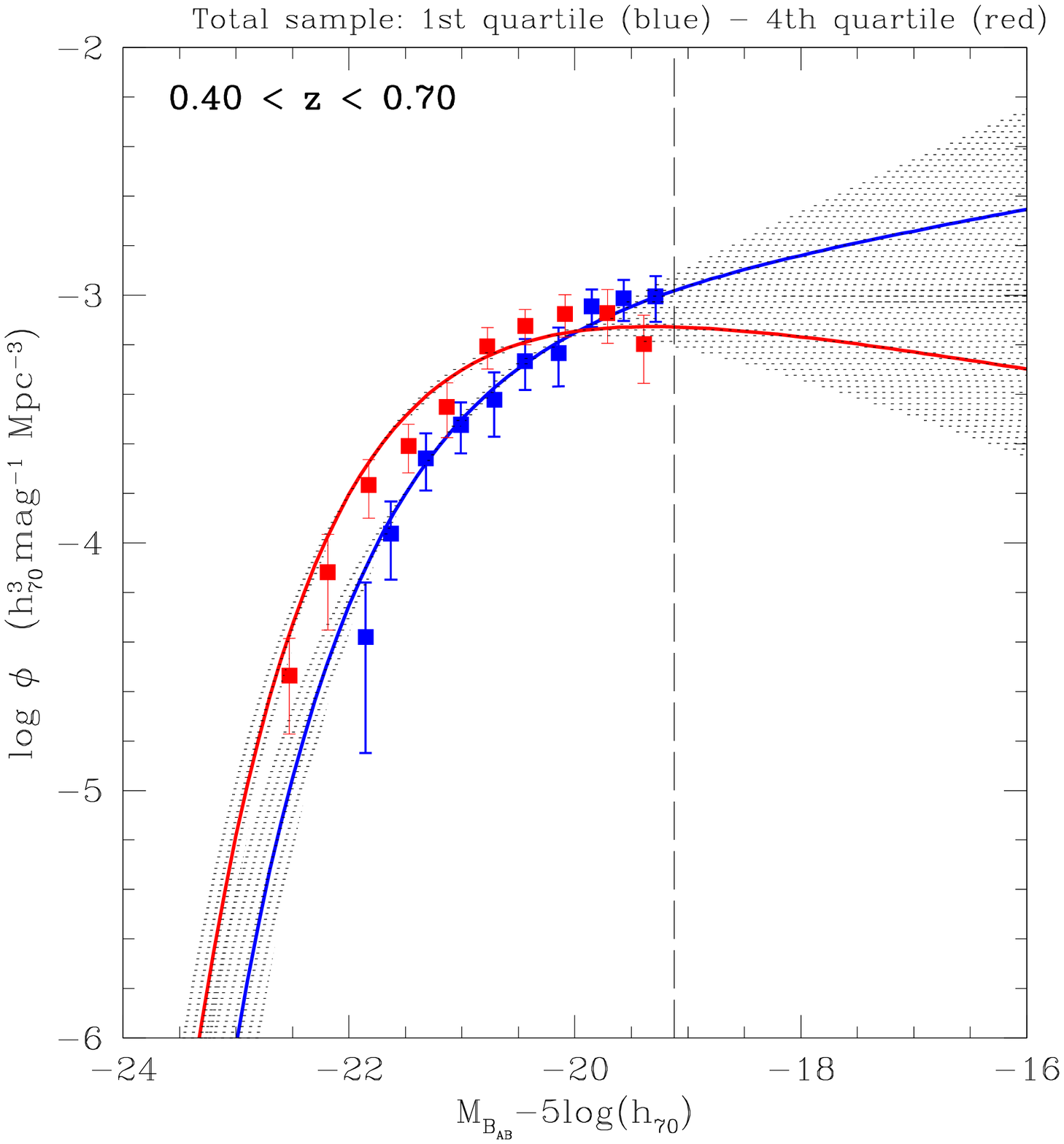}
\includegraphics[width=0.3\hsize]{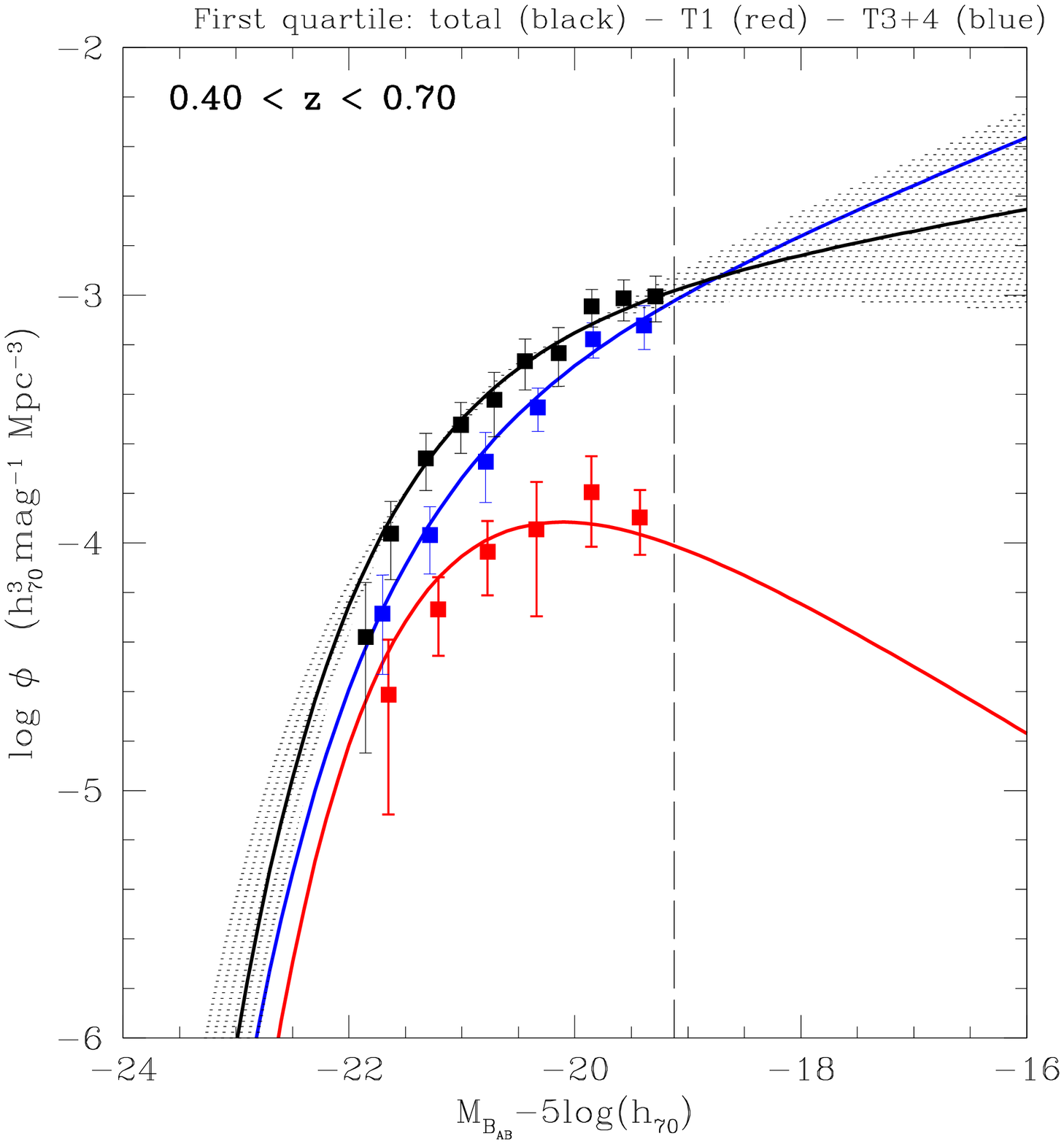}
\includegraphics[width=0.3\hsize]{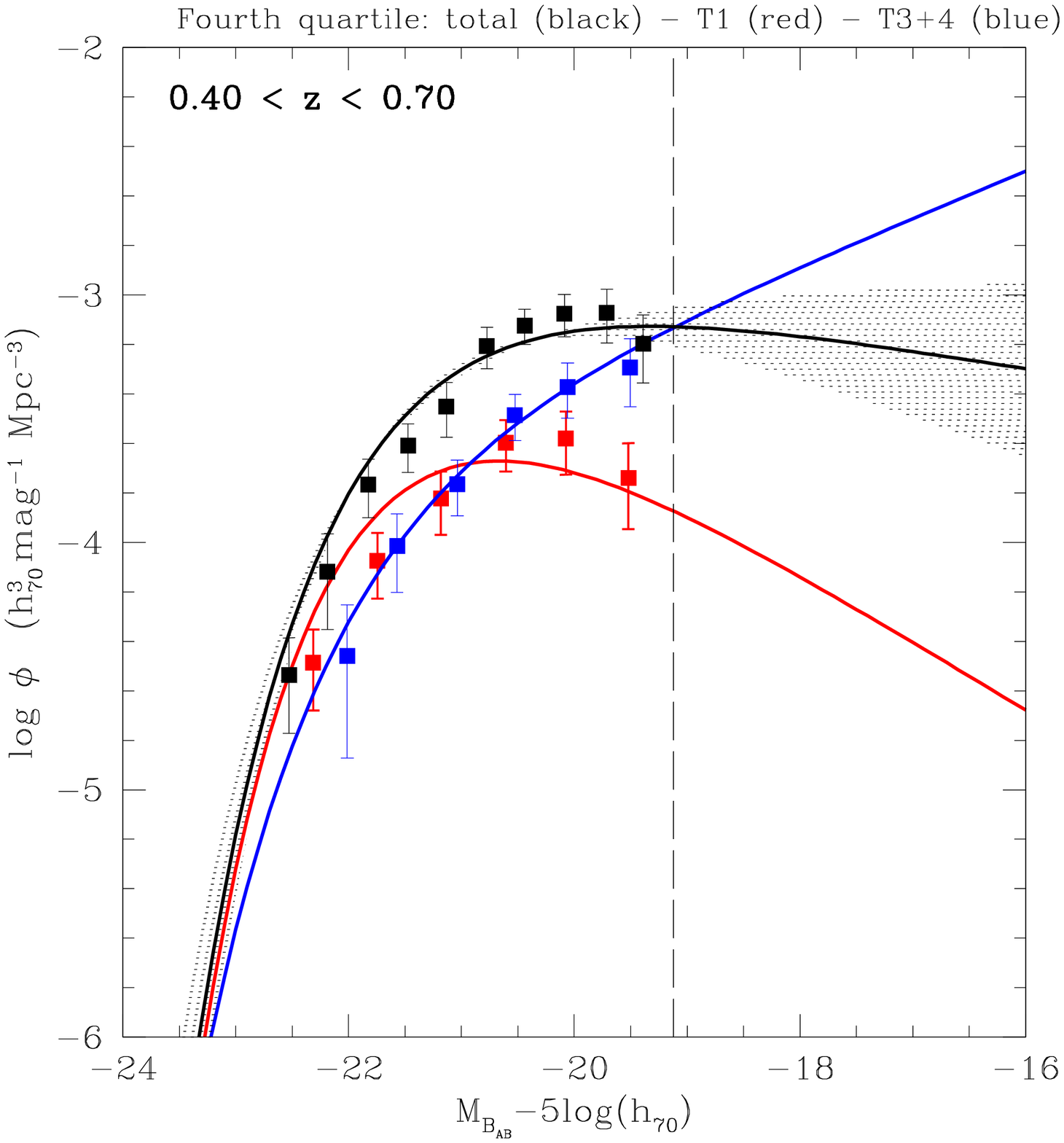}\\
\includegraphics[width=0.3\hsize]{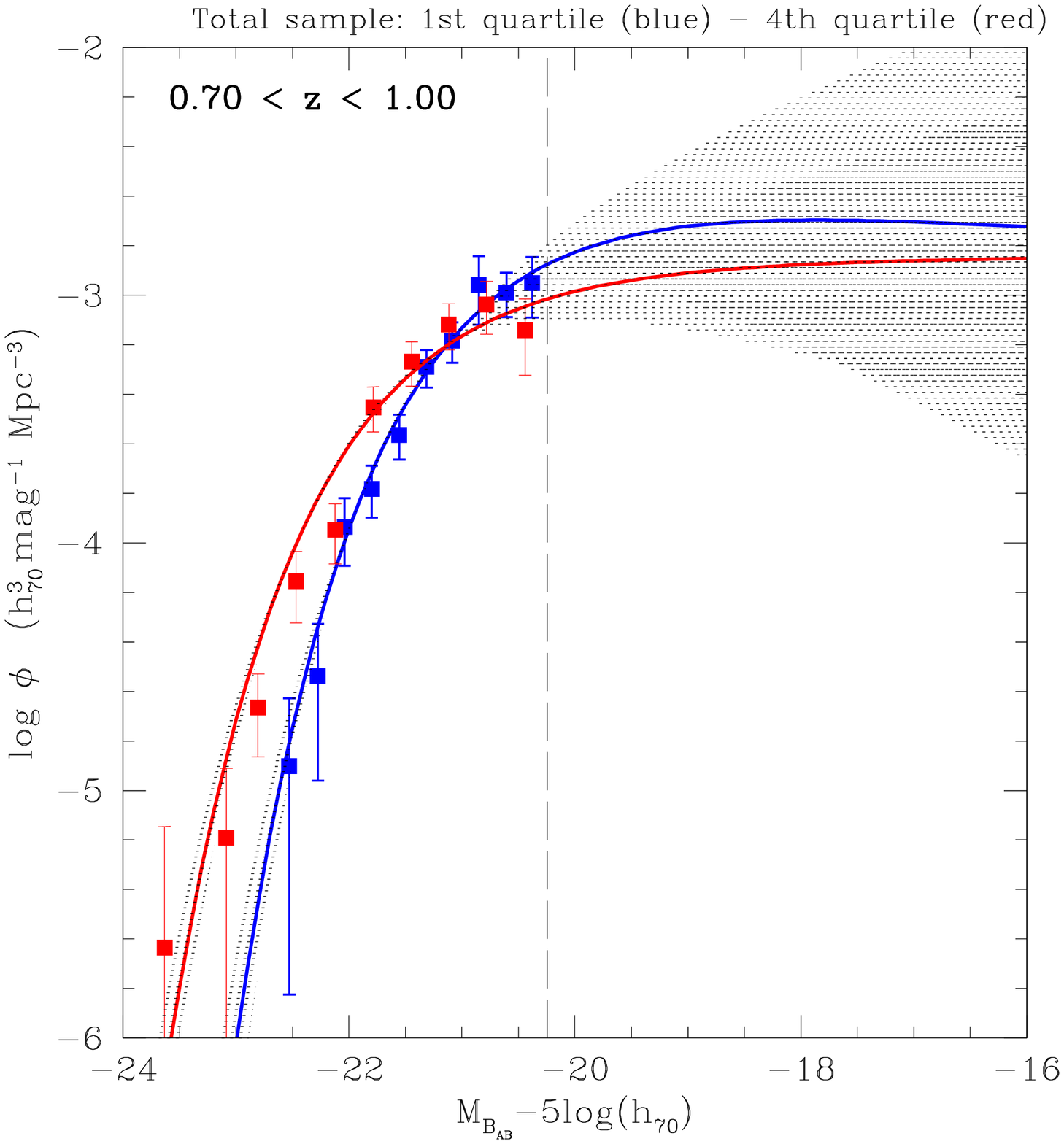}
\includegraphics[width=0.3\hsize]{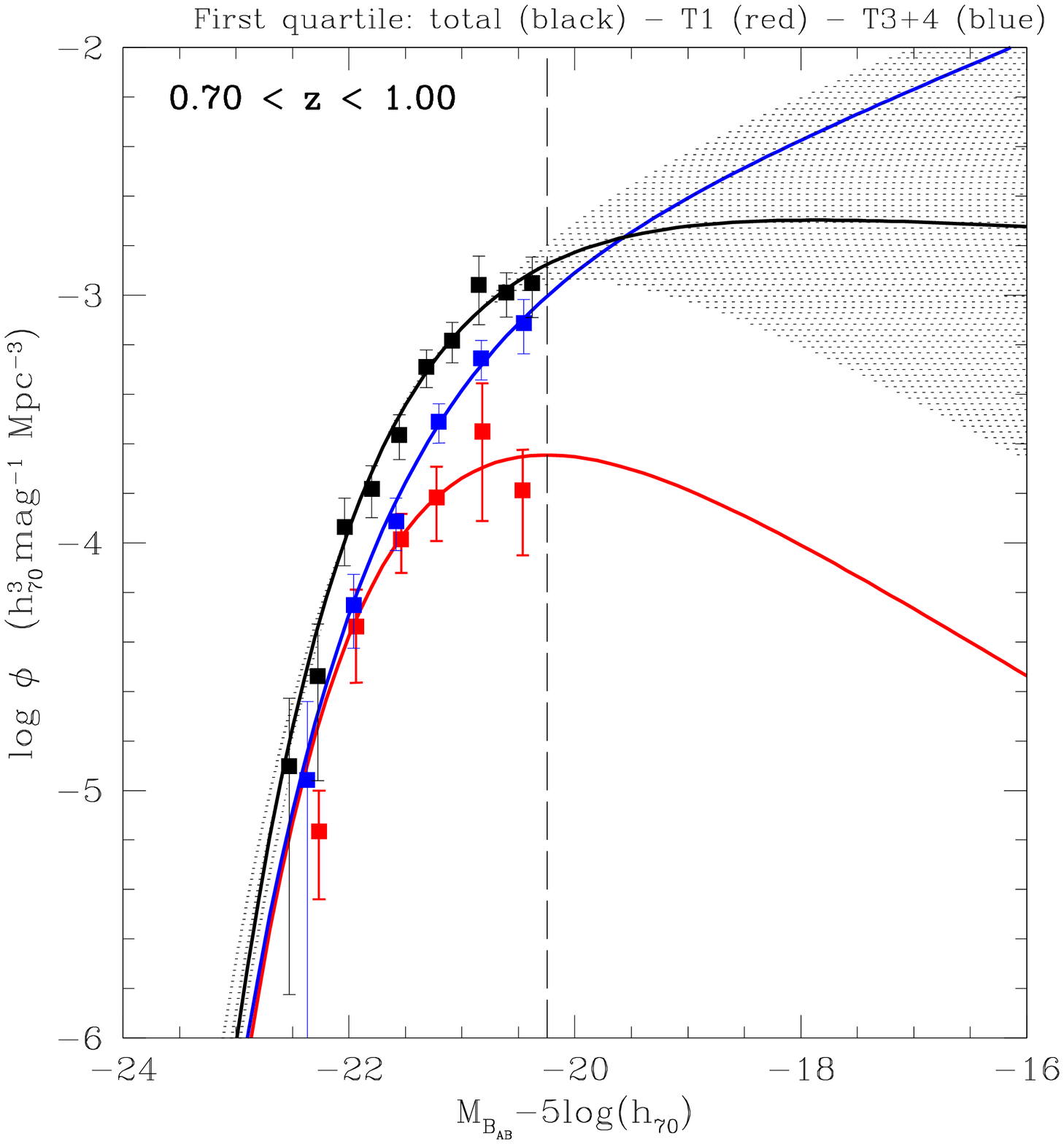}
\includegraphics[width=0.3\hsize]{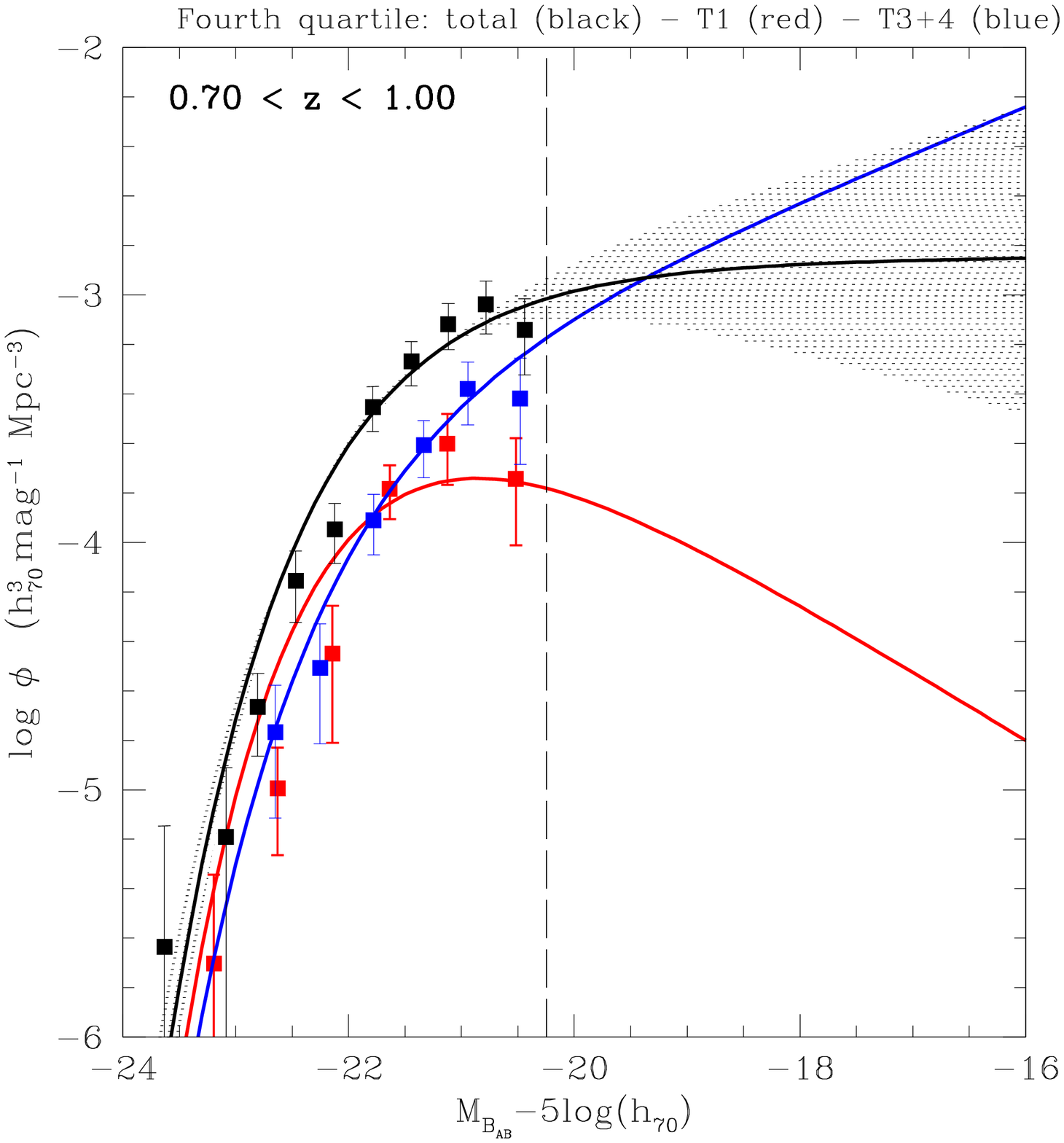}\\
\caption{Left column: luminosity functions in the lowest (blue) and highest (red) quartile
of the density distribution in redshift bins for the total sample. 
Overdensities are derived from the 5 nearest neighbours.
Middle and right columns: luminosity functions in the lowest (middle column) and highest 
(right column) quartile of the density distribution in redshift bins. 
In each panel the global luminosity function is shown in black, while
the luminosity function of type 1 and type $3+4$ galaxies are drawn
in red and blue, respectively. 
The meaning of symbols and lines is the same as in Fig. \ref{LFforme_cww_new}.
The shaded regions represent the $68\%$ uncertainties on the parameters
$\alpha$ and $M^*$. For type 1 and type 3+4 samples the $STY$ estimates with
$\alpha$ fixed are shown and therefore the shaded area is not drawn.
}
\label{LFenv_type}
\end{figure*}

\begin{figure*}
\centering
\includegraphics[width=0.3\hsize]{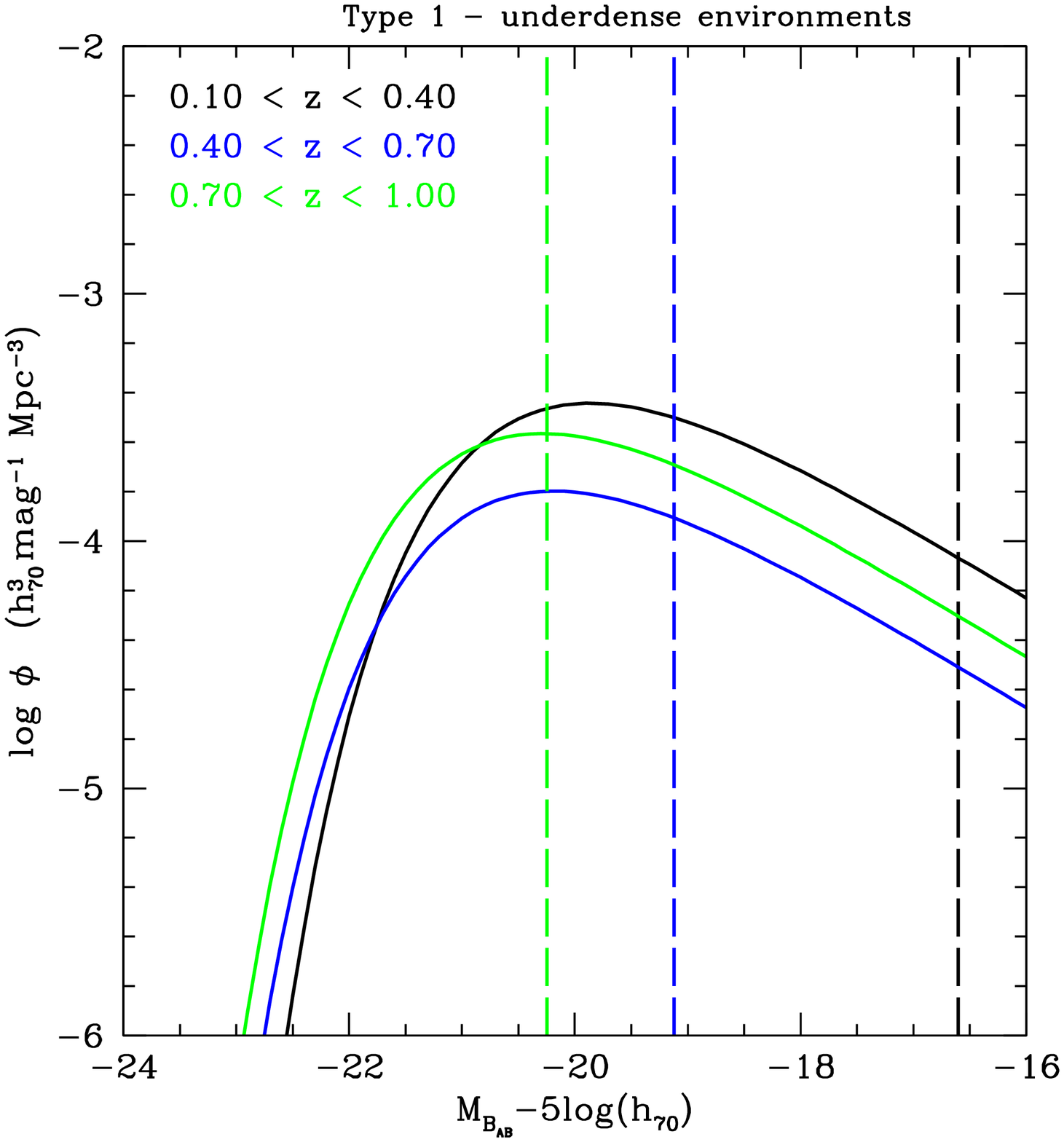}
\includegraphics[width=0.3\hsize]{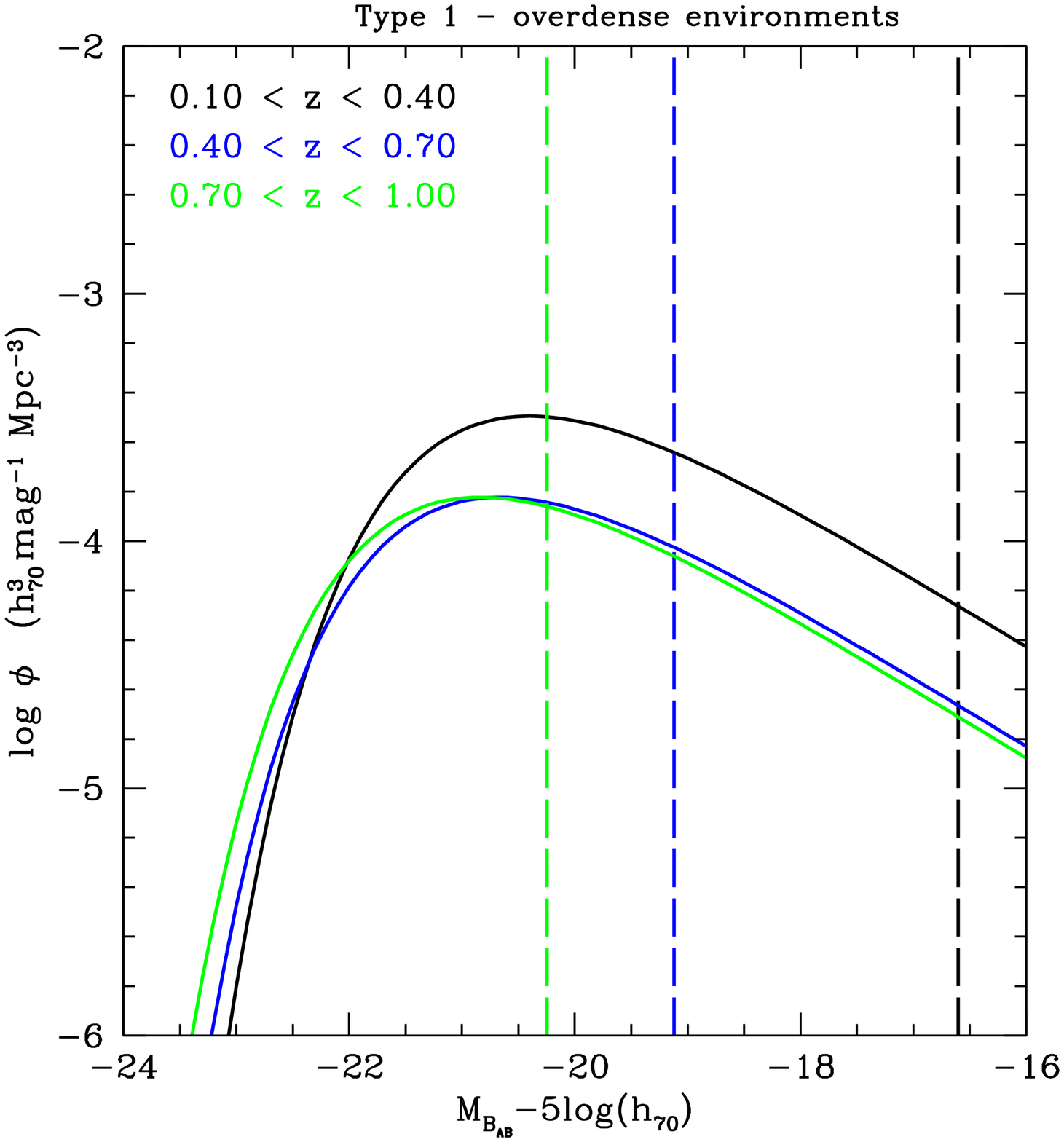}\\
\includegraphics[width=0.3\hsize]{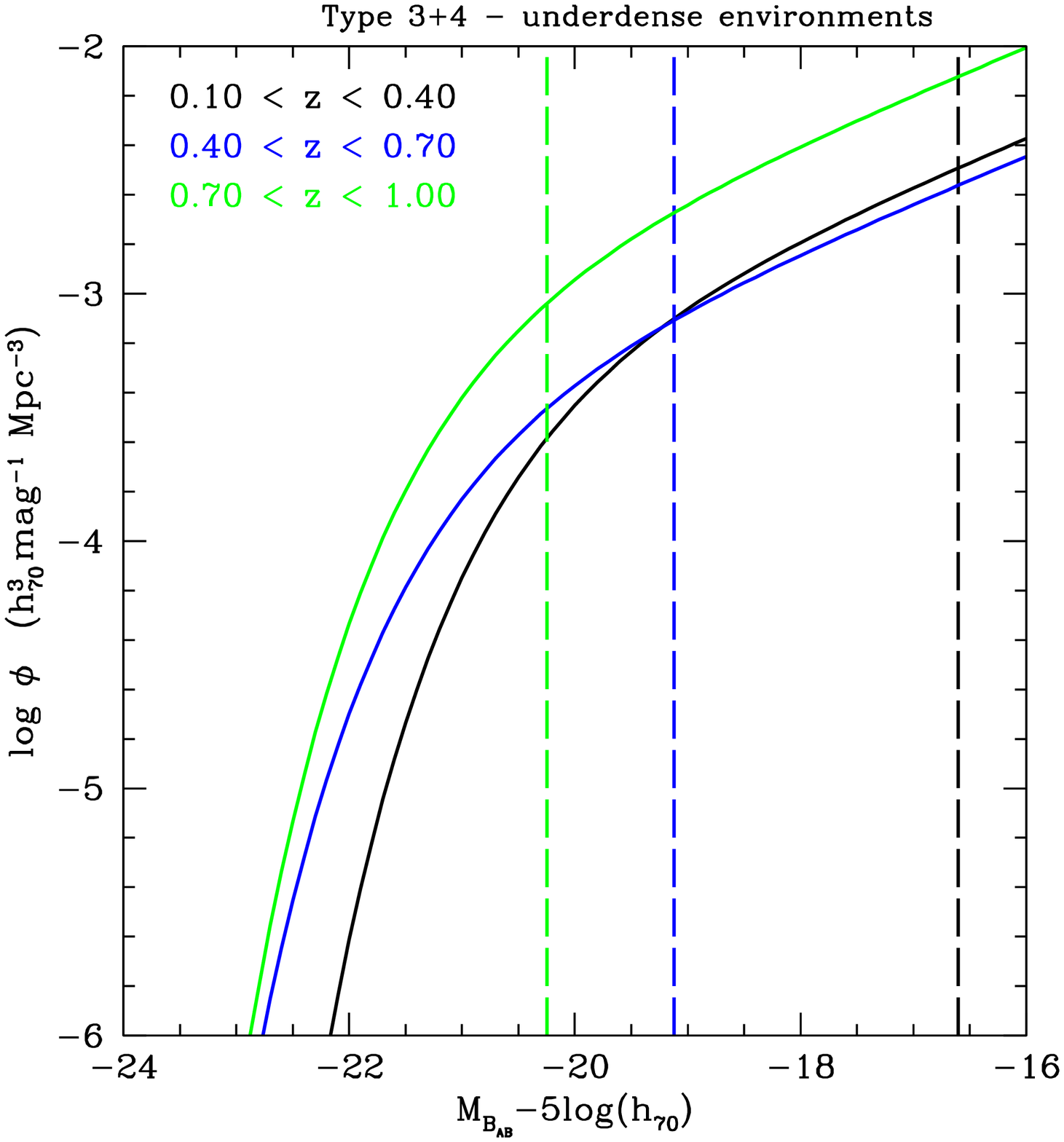}
\includegraphics[width=0.3\hsize]{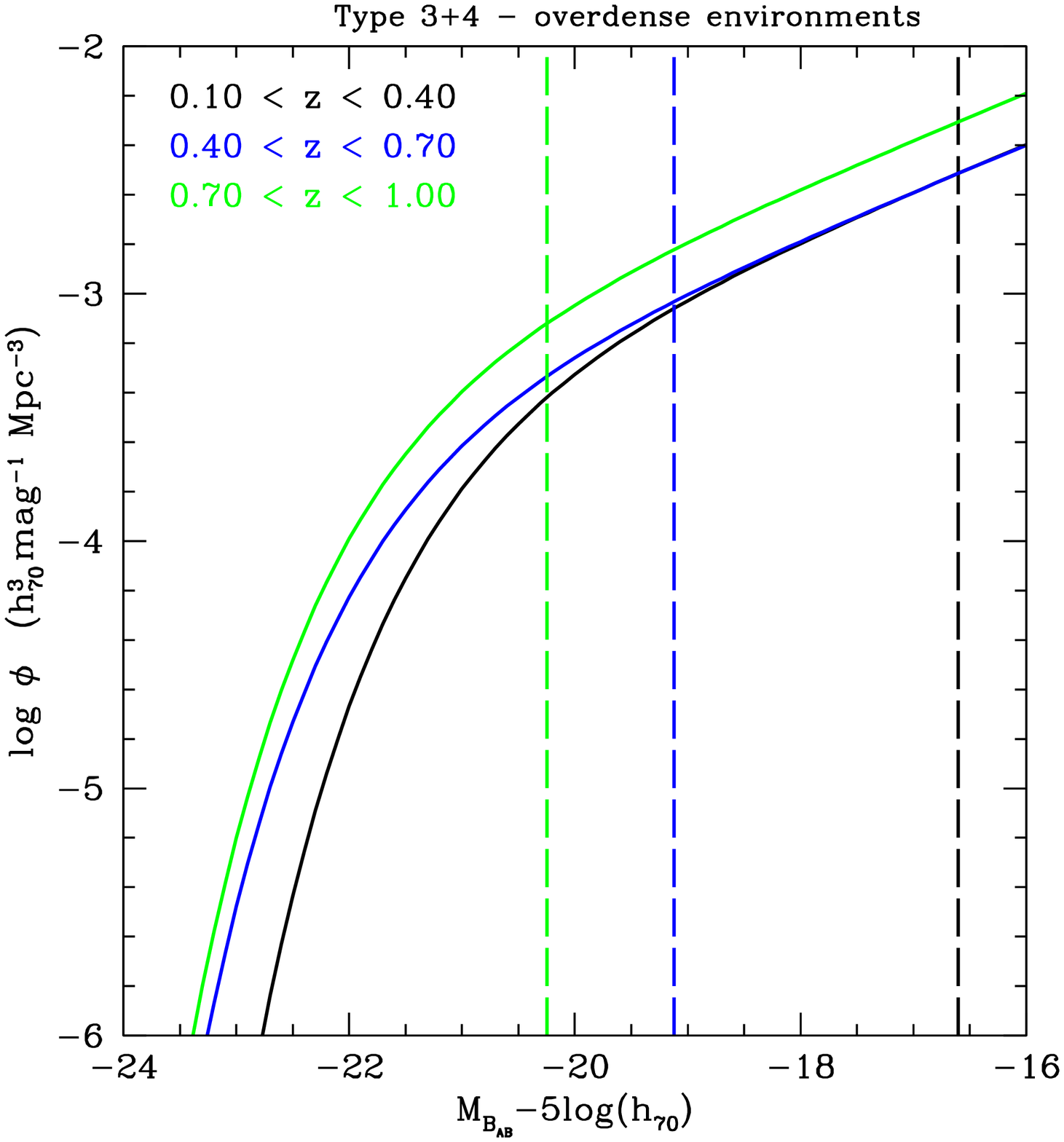}
\caption{ Evolution of the luminosity functions for different
galaxy types in different environments (low density on the left,
high density on the right): 
type 1 (upper panels) and type $3+4$ (lower panels) galaxy samples. 
The color refer to different redshift ranges: $[0.10-0.40]$
in black, $[0.40-0.70]$ in blue and $[0.70-1.00]$ in green. 
The $STY$ estimates are derived with $\alpha$ fixed.
The meaning of lines is the same as in Fig. \ref{LFforme_cww_new}: 
points from the $C^+$ estimates are not reported for clarity.
 }
\label{LFforme_env_z}
\end{figure*}


\subsection{The luminosity function of early-type galaxies}\label{sectLFet}

In Fig. \ref{LFforme_early} we show the luminosity function of 
early-type galaxies from different samples: {\it bona-fide} ET (in black),
type 1 (in red), and morphological early-type (in green) galaxies.
\\
The most apparent feature in this figure is the excess of faint
($M_{B} \simgt -19.5$) morphological early-type galaxies at $z<0.55$.
This effect is not evident at higher redshift because this faint 
population is cut by the magnitude limit of the survey.
This excess is due to the population of ``blue" early types
discussed in the previous section.
A similar effect was already noticed by Ilbert et al. (\cite{ilbertlauger}),
who found that the main contribution to the faint end of the luminosity function 
of bulge-dominated galaxies comes from blue objects. 
\\
Moreover, it is interesting that the bright ends of the
luminosity functions of type 1 and morphological early-type galaxies are almost 
indistinguishable at all redshifts.
\\
We also note that type 1 and {\it bona-fide} ET galaxies
luminosity functions have similar shapes, but different
normalizations.
The slopes are consistent to within the errors, as well as 
the $M^*$ values; the difference in $\phi^*$ is
about a factor of 2, due to the selection
criteria adopted in constructing the {\it bona-fide} ET sample.
The similarity between these luminosity functions implies that,
even if the type 1 galaxy sample is contaminated by
a fraction of ``red" spirals, the contribution of this
population is approximately constant with both luminosity and redshift. 


\section{The role of the environment}\label{sectenv}

To understand the effect of the environment, 
we derived the luminosity functions for galaxies in underdense and
overdense regions, using the density estimates described in Sect. \ref{sectacs}.
We repeated the analysis using various density estimators, finding that
the general trends are similar in all cases.
In the following, we show the results obtained for overdensities derived
with the 5th nearest neighbour estimator. 
\\
To emphasize the differences between the underdense and overdense
environments, we consider only the two extreme tails of
the overdensity distribution. We also ensure that we had
roughly the same number of objects in each subsample, 
to avoid spurious normalization effects. For this reason,
we divided the galaxy density distribution into quartiles, within each
redshift bin, and derived the luminosity functions for each subsample. 
In the following, we compare the results for the lowest and highest 
quartiles of the overdensity distributions.
The use of both quartiles and subsamples reduces
the number of usable objects, and we therefore decided to
use only three redshift bins, $[0.1-0.4]$, $[0.4-0.7]$, and 
$[0.7-1.0]$. 
\\
The most appropriate choice of quartile values depends on the topic being 
investigated. To measure the contribution of different galaxy
types to the global luminosity function in different environments, it is necessary
to derive the quartile values for the total sample and use the
same density cuts for all the subsamples.
However, if the aim is to compare the behaviour of the luminosity
function of specific subsamples of galaxies in 
underdense and overdense environments, we must carefully select
for each subsample the appropriate quartiles (i.e., as derived
for that particular subsample).
\\
The use of quartiles makes difficult to derive a number density from
the $\phi^*$ parameter of the luminosity function, because this kind of
selection does not conserve information about the volume occupied by
the overdensity associated to each galaxy. 
Therefore, when analysing luminosity functions in different environments,
only a shape comparison is allowed.
However, the comparison of $\phi^*$ values is correct when luminosity 
functions for galaxies in the same quartile are compared.    
\\
In the following, we show results both for the total sample and 
for early- and late-type galaxies, using the type 1 and type 3+4 subsamples.
For the early-type galaxies we decided to use the type 1 instead of the {\it bona-fide} ET sample
for three reasons: 1. the more reliable statistical analysis possible for a
higher number of objects;
2. the homogeneity in the classification compared to the late-type galaxy sample;
3. the similarity between the type 1 and {\it bona-fide} ET galaxy luminosity functions 
(see previous section).
However, we checked that the results obtained with the {\it bona-fide} ET sample
(although with larger uncertainties) are consistent with those derived
in the following for the type 1 sample.
\\
In the first column of Fig. \ref{LFenv_type}, we show the luminosity function 
for the total sample in the lowest (blue lines and points) and highest (red lines 
and points) quartiles of the density distribution in redshift bins. 
The luminosity function of galaxies in overdense
regions is consistently of brighter $M^*$: the slope in underdense regions is steeper
in the first and second redshift bins, while the $\alpha$ values in the highest 
redshift bin, in both environments, are consistent with each other, 
in the significant margins of large error.
Increasing the redshift, the contribution from the luminosity function
of galaxies in underdense environments increases and the crossing point between
the two LFs progressively moves towards brighter magnitudes.
To explore whether this behaviour is due to the different
morphological mixes in the two environments, in the second and third column of
Fig. \ref{LFenv_type} we show the luminosity function in underdense 
and overdense environments: the total luminosity
function is shown in black, while the contribution from type 1 and
type 3+4 galaxies are drawn in red and blue, respectively.
In this case, the quartile values are those derived for the total sample.
The parameters of the $STY$  estimates are reported in 
Table \ref{param_overd_rel}.
The $STY$ parameters for type 1 and type 3+4 galaxies samples, although 
formally constrained, are poorly determined. We therefore repeated the
$STY$ estimate by fixing $\alpha$ for each type to the global value
derived in Sect. \ref{sectLFcww}, after verifying that this choice is
acceptable.
In low density environments, the main contribution to the luminosity
function is from type 3+4 galaxies, while for high density
environments an important contribution is that of type 1 galaxies
at the bright end.
The differences between the global luminosity function in the two
environments are due not only to the different relative numbers
of type 1 and type 3+4 galaxies, but also to their relative 
luminosity distributions.
The value of $M^*$ in underdense regions is always
fainter than in overdense environments, by $\sim 0.50$ mag and $\sim 0.65$ mag
for type 1 and type 3+4 galaxies, respectively. 
For $\phi^*$, between underdense and overdense environments
there is a decrease of a factor $\sim 2$ for type 3+4 galaxies and an increase
of $\sim 2$ for type 1 galaxies (except in the last redshift
bin, where the $STY$ estimate is poorly determined).
These results indicate that galaxies of the same type in
different environments have different properties.
\\
We now investigate if there is also a differential evolution
for different environments within each class of objects.
To achieve this, we need to use, for each subsample, the appropriate 
quartiles (i.e., those derived for that particular subsample).
The parameters of the $STY$ estimates obtained with this choice 
are reported in Table \ref{param_overd}.
\\
To visualize the evolutionary effects more clearly, in 
Fig. \ref{LFforme_env_z} we plot in the same panel the luminosity functions in 
different redshift bins for each galaxy type in the two environments,
showing for clarity only the $STY$ estimates.
For type 3+4 galaxies, the evolution in the luminosity function
within underdense and overdense regions is similar for the
first and the second redshift bin, and occurs mainly in luminosity: 
there is a brightening of $\sim 0.5$ mag in $M^*$ for all
cases, and a slight variation in $\phi^*$.
In contrast, between the second and the third redshift bins
there is little luminosity evolution, but a significant evolution in $\phi^*$,
which differs in the two environments: in overdense regions $\phi^*$
increases by a factor $\sim 1.6$, while in underdense regions
this increase reaches a factor $\sim 2.8$.
\\
By analyzing the type 3+4 galaxy population in more detail, 
we find that the evolution in the relative ratio of type 3 to type 4
galaxies differs between the two environments: in particular, 
moving from the second to the third redshift bin, the ratio $N_{type 4}/N_{type 3}$
for bright ($M_{B} < -20.25$) galaxies 
changes from $33\%$ to $46\%$ in low density environments,
but remains almost constant ($\sim 35\%$) in high density environments.
Therefore, the strong evolution observed for type 3+4 galaxies
from $z\sim 0.5$ to $z\sim 0.9$ is mainly due to type 4 galaxies.
The adopted limit in absolute magnitude corresponds to the bias
limit (see Sect. \ref{sectalf}) in the highest redshift bin and 
allows a direct comparison
of galaxy numbers at different redshifts.
\\
For type 1 galaxies, the situation is more unclear, because
the lower number of objects introduces larger uncertainties in the
luminosity function estimates. 
Between the first and the second redshift bin, there is a
similar evolution in both environments, while passing from the
second to the third redshift bins there are indications of a more
significant evolution in underdense regions.
\\
To check if the choice of different quartiles for each redshift bin
can affect these results, we repeated the analysis using in
each redshift bin the value of the quartiles derived for
the whole redshift range $[0.1-1]$, finding that the trends do
not change.
\\
These results agree with those found studying the galaxy stellar
mass function in different environments (Bolzonella et al.
\cite{bolzonella08}). In particular, in high density environments
there is a strong contribution of early-type galaxies, which are
dominant at high masses in the galaxy stellar mass function and
at bright magnitudes in the luminosity function: however, this
dominance is clear at all redshifts in the galaxy stellar mass
function, but decreases with increasing redshifts in the
luminosity function. This is due to the fact that at high
redshift blue galaxies brighten because they increase their star formation
activity. 
Finally, Bolzonella et al. (\cite{bolzonella08}) find that the contribution
of early-type galaxies to the total galaxy stellar mass function 
increases more rapidly in high density environments: consistently,
we find that the relative contribution of type 1 and type 3+4 
galaxies to the bright end of the luminosity function is similar at high redshift 
in both environments, but at low redshift type 1 galaxies dominate 
mainly at high density.


\section{Summary and conclusions}\label{sectconc}

An unbiased and detailed characterization of the luminosity function 
is the basic requirement in many astrophysical issues and it is of particular
interest to assess the role of environment on the evolution 
in the luminosity function of different galaxy types.
With this aim, we have studied the evolution in the luminosity function to
redshift $z\sim 1$ for the zCOSMOS 10k sample, for which both
accurate galaxy classifications and a detailed description of the
density field are available.
\medskip
\\
The main results of this analysis are the following:
\medskip
\\
{\it -- Global luminosity function:} 
The global luminosity function shows a brightening of $\sim 0.7$ mag
in $M^*$ from $z\sim 0.2 $ to $z\sim 0.9$, in agreement with the
VVDS results in similar redshift ranges.
\medskip
\\
{\it -- Luminosity functions by spectrophotometric types:}
To quantify the contribution of the different spectrophotometric
types, we have considered three classes of objects: type 1 galaxies, corresponding
to early-type SEDs, type 3+4 galaxies, corresponding to late-type SEDs, and
the intermediate class of type 2 galaxies.
At low redshift ($z<0.35$), type 3+4 galaxies dominate the luminosity
function at faint magnitudes ($M_{B} > -20$), 
while the bright end is populated mainly by type 1 galaxies.
At higher redshift, type 3+4 galaxies evolve strongly 
and therefore, at redshift $z\sim 1$, the contribution
to the bright end of the luminosity function of the various types is
comparable. The faint end remains dominated by type 3+4 galaxies
over the entire redshift range.
For type 1 galaxies, the evolution occurs in both luminosity
and normalization: not only does $M^*$ brighten by $\sim 0.6$ mag
but $\phi^*$ also decreases by a factor $\sim 1.7$
between the first and the last redshift bin.
Type 3+4 galaxies also evolve in both luminosity and normalization, 
but with an opposite trend for the normalization: a brightening 
with redshift by $\sim 0.5$ mag is
present in $M^*$, while $\phi^*$ increases by a factor $\sim 1.8$.
Type 2 galaxies show a milder evolution, of a brightening of $\sim 0.25$
mag in $M^*$ and no significant evolution in $\phi^*$.
\medskip 
\\
{\it -- Luminosity functions by  morphological types:}
At low redshift ($z<0.35$), morphological early-type galaxies dominate the
bright end of the luminosity function, while spiral galaxies 
dominate the faint end. Irregular galaxies increase their
contribution at the lowest luminosities.
At intermediate redshift ($[0.35-0.75]$), spiral galaxies increase
their luminosities and their contribution to the bright end of
the luminosity function is similar to that of the morphological early-type one.
At high redshift ($z>0.75$), irregular galaxies evolve strongly and
therefore the three morphological types contribute almost equally 
to the total luminosity function. 
Irregular galaxies exhibit an evolution of a factor $\sim 3.3$ in $\phi^*$
from low to high redshift:
this evolution occurs mainly in the last redshift bin, while,
for $z<0.75$, the contribution of these galaxies to
the global luminosity function is significantly lower than
that of spirals and early types.
\medskip
\\
{\it -- The role of the environment:}
For the total sample, the luminosity function of galaxies in overdense
regions always has a brighter $M^*$ and a flatter slope.
Increasing the redshift, the contribution to the luminosity function
of galaxies in underdense environments increases, and the crossing point between
the two LFs moves progressively towards brighter magnitudes.
In low density environments, the main contribution to the luminosity
function originates in type 3+4 galaxies, while for high density
environments there is an important contribution from type 1 galaxies
to the bright end.
\\
The differences between the global luminosity functions in the two
environments are due not only to a difference in relative numbers
of type 1 and type 3+4 galaxies, but also to their relative 
luminosity distributions.
The value of $M^*$ in underdense regions is always
fainter than in overdense environments, by $\sim 0.50$ mag 
and $\sim 0.65$ mag for type 1 and type 3+4 galaxies, respectively. 
For $\phi^*$, between underdense and overdense environments,
there is a decrease of a factor
$\sim 2$ for type 3+4 galaxies and an increase
of $\sim 2$ for type 1 galaxies.
These results indicate that galaxies of the same type in
different environments have different properties.
\\
We have also detected differential evolution for type 3+4 galaxies
in different environments.
We found that the evolution in their luminosity function
in underdense and overdense regions is similar between 
$z\sim 0.25$ and $z\sim 0.55$, and is mainly in luminosity: 
there is a brightening of $\sim 0.5$ mag in $M^*$ in all
cases, but only a slight variation in $\phi^*$.
In contrast, between $z\sim 0.55$ and $z\sim 0.85$ 
there is little luminosity evolution but a strong evolution in $\phi^*$,
which differs in the two environments: in overdense regions, $\phi^*$
increases by a factor $\sim 1.6$, while in underdense regions
there is an increase of a factor $\sim 2.8$.
\\
Analyzing the type 3+4 galaxy population in more detail, 
we find that the evolution in the relative ratio of type 3 to type 4
galaxies differs between the two environments: in particular, 
between $z\sim 0.55$ and $z\sim 0.85$,  the ratio $N_{type 4}/N_{type 3}$
for bright ($M_{B} < -20.25$) galaxies 
changes from $33\%$ to $46\%$ in low density environments,
but remains almost constant ($\sim 35\%$) in high density environments.
Therefore, the strong evolution observed for type 3+4 galaxies
between $z\sim 0.55$ and $z\sim 0.85$ is mainly due to type 4 galaxies.
\\
For type 1 galaxies, the situation is unclear, because
the lower number of objects introduce larger uncertainties in the
luminosity function estimates. 
Between the first and the second redshift bins, there is a
similar evolution in both environments, while passing from the
second to the third redshift bins, there are indications of
a more significant evolution in underdense regions.
\\
\begin{figure}
\centering
\includegraphics[width=0.8\hsize]{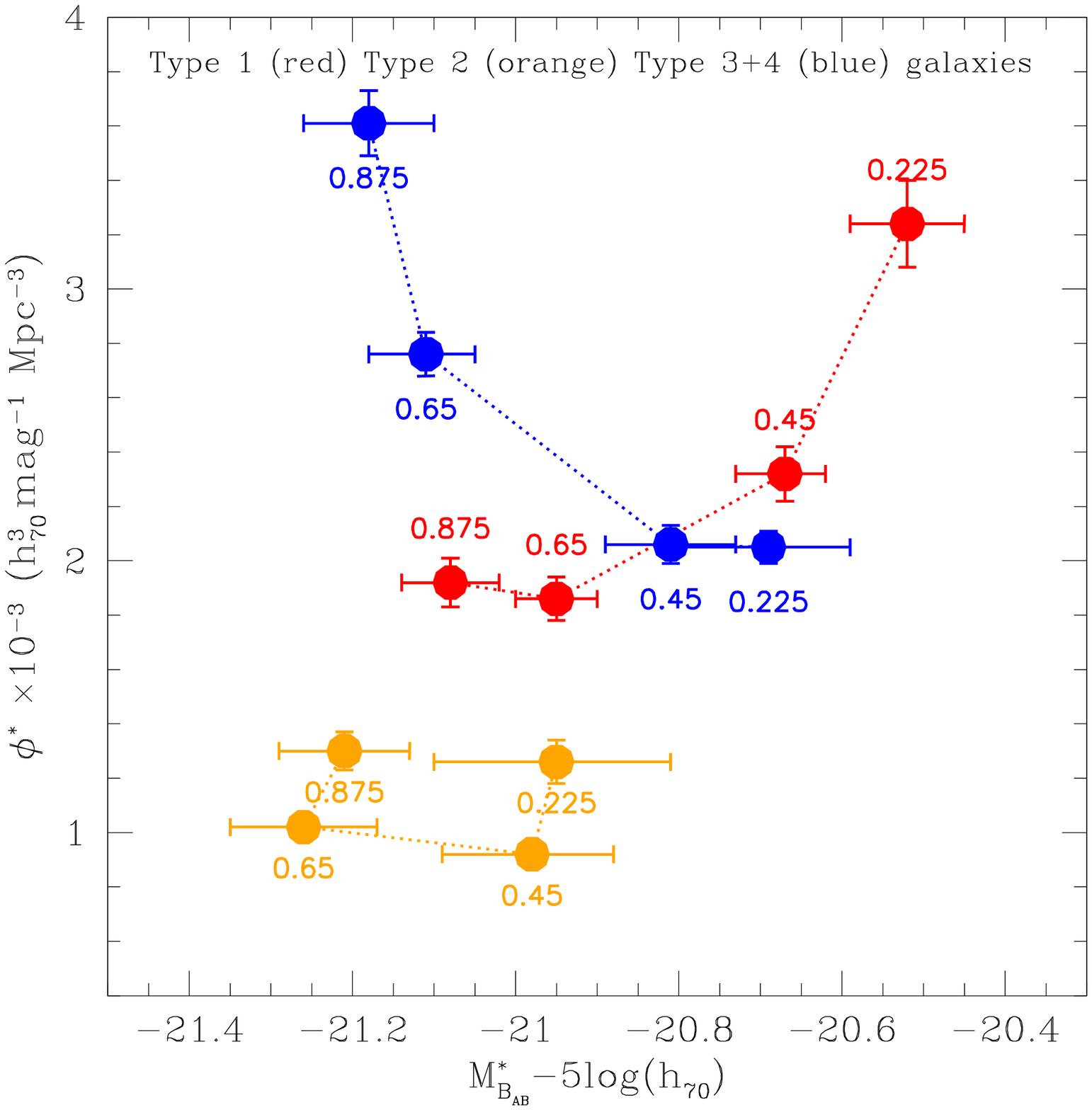}\\
\includegraphics[width=0.8\hsize]{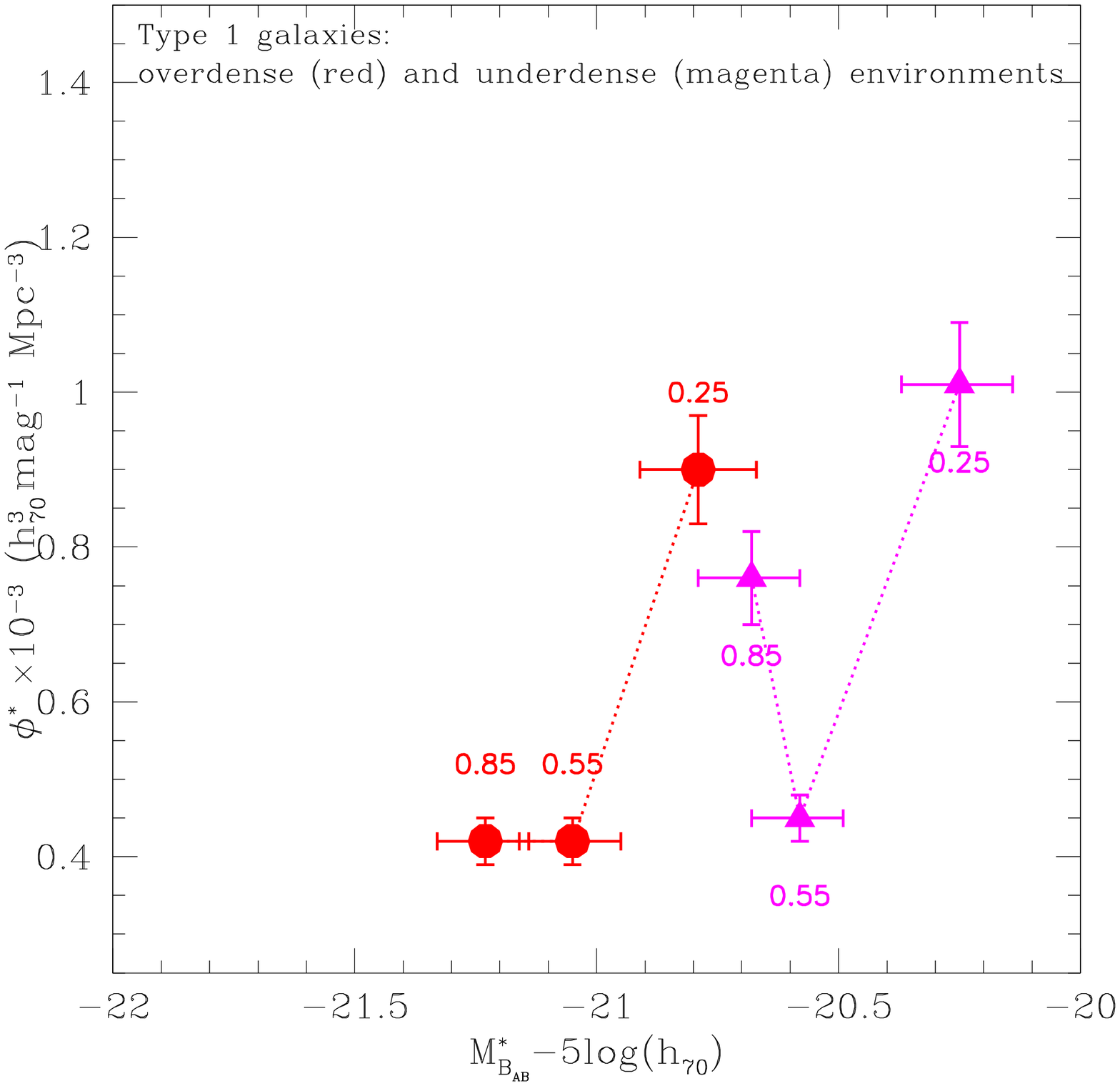}\\
\includegraphics[width=0.8\hsize]{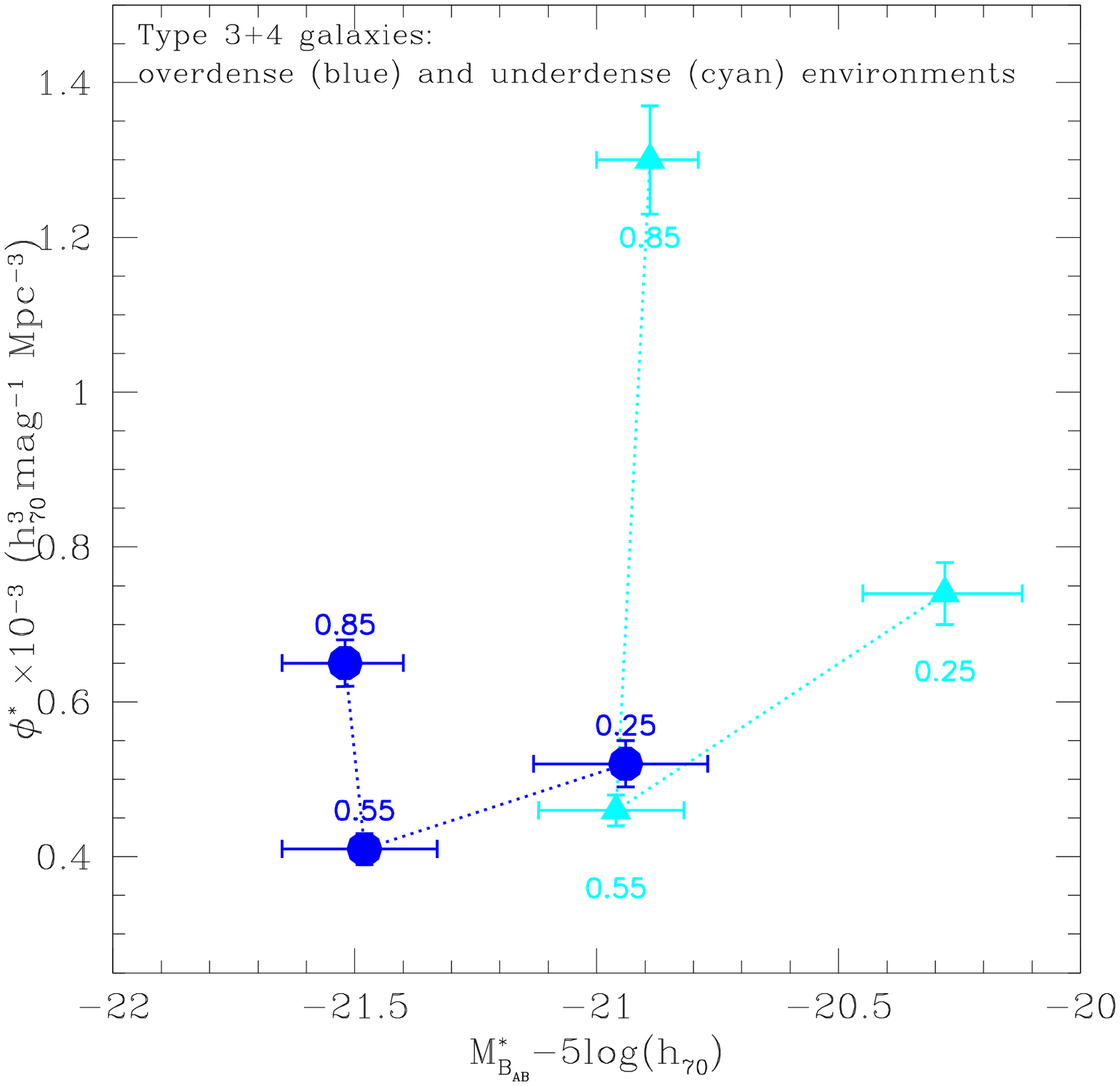}
\caption{Evolution of the $M^* - \phi^*$ parameters (derived with
$\alpha$ fixed) for different samples: the label near each point indicates
the mean redshift of the considered bin. Upper panel: type 1 (red),
type 2 (orange) and type 3+4 (blue) galaxies in the total sample. 
Middle panel: type 1 galaxies in overdense (red circles) and 
underdense (magenta triangles) environments.
Lower panel: type 3+4 galaxies in overdense (blue circles) and 
underdense (cyan triangles) environments.
 }
\label{evol_param}
\end{figure}
%
\bigskip
\\
All these results are summarized in Fig. \ref{evol_param}, where
the evolution of the parameters $M^*$ and $\phi^*$, obtained with
$\alpha$ fixed, is shown for different samples; the label near each
point indicates the mean redshift of the considered bin.
The upper panel of Fig. \ref{evol_param} refers to galaxies of different
spectrophotometric types in the total sample: type 1 galaxies in red,
type 2 galaxies in orange and type 3+4 galaxies in blue.
From this figure it is immediately evident the symmetric behaviour of type
1 and type 3+4 galaxies, as well as the almost negligible evolution
of type 2 galaxies.
While we find that the $M^*$ evolution for type 1 and type 3+4 galaxies
is consistent with passive evolution, it is more interesting the 
density evolution.
Given the fact that $\phi^*$ is proportional to the density
of $M^*$ galaxies, this figure is consistent with a scenario
where a part of type 3+4 galaxies is transformed in
type 1 galaxies with increasing cosmic time, without significant
changes in the fraction of intermediate-type galaxies.
\\
The middle and lower panels of Fig. \ref{evol_param} show the
evolution of the parameters $M^*$ and $\phi^*$ for galaxies of different
types in different environments: type 1 galaxies in underdense (magenta
triangles) and overdense (red circles) environments in the middle 
panel, type 3+4 galaxies in underdense (cyan triangles) and overdense
(blue circles) environments in the lower panel.
The behaviour of type 1 and type 3+4 galaxies in overdense environments is
consistent with the transformation scenario described above for the 
total sample.
In underdense environments, a similar scenario is acceptable, with the
exception of the highest redshift bin, where type 1 galaxies show a
strong increase of $\phi^*$ with respect to the previous bin.
\\
The lower panel of Fig. \ref{evol_param} clearly shows the different
evolution which type 3+4 galaxies are undergoing in the different
environments, with a much stronger density evolution in 
underdense regions.
\\
This indicates that the bulk of the tranformation described above 
from blue to red galaxies in overdense regions probably
happened before $z\sim 1$, while it is still ongoing at lower
redshifts in underdense environments.


\begin{acknowledgements}
We acknowledge support from an INAF contract PRIN-2007/1.06.10.08 and 
an ASI grant ASI/COFIS/WP3110 I/026/07/0.
\end{acknowledgements}


\begin{table*}
\caption[]{$STY$ parameters (with $1\sigma$ errors) in different redshift bins for total sample and different galaxy type subsamples.}
\begin{flushleft}
\begin{tabular}{r r r r r r} 
\hline
z-bin &  Number$^{(a)}$ &  Number$^{(b)}$ &   $\alpha$ &  $M^*_{B_{AB}}-5{\rm log}(h_{70})$ & $\phi^*$($10^{-3}\ h_{70}^3\ {\rm Mpc}^{-3}$)   
\\ \hline
\multicolumn{6}{l}{ Total sample} \\
\hline
0.10-0.35 &  1968 &  1876 &  -1.09$^{{\rm + 0.04}}_{{\rm -0.04}}$ & -20.85$^{{\rm + 0.10}}_{{\rm -0.11}}$ &   5.62$^{{\rm + 0.58}}_{{\rm - 0.56}}$ \\ 
          &       &       &  -1.03      fixed                     & -20.73$^{{\rm + 0.05}}_{{\rm -0.06}}$ &   6.45$^{{\rm + 0.15}}_{{\rm - 0.15}}$ \\ 
0.35-0.55 &  2059 &  1841 &  -0.82$^{{\rm + 0.08}}_{{\rm -0.08}}$ & -20.67$^{{\rm + 0.09}}_{{\rm -0.10}}$ &   6.40$^{{\rm + 0.58}}_{{\rm - 0.59}}$ \\ 
          &       &       &  -1.03      fixed                     & -20.91$^{{\rm + 0.05}}_{{\rm -0.05}}$ &   4.90$^{{\rm + 0.11}}_{{\rm - 0.11}}$ \\	  
0.55-0.75 &  2163 &  2086 &  -0.85$^{{\rm + 0.11}}_{{\rm -0.11}}$ & -20.98$^{{\rm + 0.09}}_{{\rm -0.10}}$ &   6.59$^{{\rm + 0.57}}_{{\rm - 0.61}}$ \\ 
          &       &       &  -1.03      fixed                     & -21.14$^{{\rm + 0.04}}_{{\rm -0.04}}$ &   5.57$^{{\rm + 0.12}}_{{\rm - 0.12}}$ \\	   
0.75-1.00 &  1769 &  1750 &  -1.59$^{{\rm + 0.16}}_{{\rm -0.16}}$ & -21.57$^{{\rm + 0.13}}_{{\rm -0.15}}$ &   4.32$^{{\rm + 0.96}}_{{\rm - 0.93}}$ \\ 
          &       &       &  -1.03      fixed                     & -21.17$^{{\rm + 0.04}}_{{\rm -0.04}}$ &   7.15$^{{\rm + 0.17}}_{{\rm - 0.17}}$ \\ 	  
0.30-0.80 &  5249 &  4972 &  -1.03$^{{\rm + 0.04}}_{{\rm -0.04}}$ & -21.02$^{{\rm + 0.05}}_{{\rm -0.05}}$ &   5.42$^{{\rm + 0.32}}_{{\rm - 0.32}}$ \\ 
\hline
\multicolumn{6}{l}{ Spectrophotometric type 1 galaxies} \\
\hline
0.10-0.35 &   418 &   416 &  -0.07$^{{\rm + 0.10}}_{{\rm -0.10}}$ & -20.26$^{{\rm + 0.11}}_{{\rm -0.12}}$ &   3.78$^{{\rm + 0.17}}_{{\rm - 0.20}}$ \\  
          &       &       &  -0.30     fixed                      & -20.52$^{{\rm + 0.07}}_{{\rm -0.07}}$ &   3.24$^{{\rm + 0.16}}_{{\rm - 0.16}}$ \\ 
0.35-0.55 &   580 &   552 &  -0.07$^{{\rm + 0.15}}_{{\rm -0.15}}$ & -20.48$^{{\rm + 0.12}}_{{\rm -0.13}}$ &   2.54$^{{\rm + 0.11}}_{{\rm - 0.11}}$ \\ 
          &       &       &  -0.30     fixed                      & -20.67$^{{\rm + 0.05}}_{{\rm -0.06}}$ &   2.32$^{{\rm + 0.10}}_{{\rm - 0.10}}$ \\ 
0.55-0.75 &   577 &   562 &   0.03$^{{\rm + 0.21}}_{{\rm -0.20}}$ & -20.74$^{{\rm + 0.12}}_{{\rm -0.13}}$ &   1.97$^{{\rm + 0.08}}_{{\rm - 0.08}}$ \\ 
          &       &       &  -0.30     fixed                      & -20.95$^{{\rm + 0.05}}_{{\rm -0.05}}$ &   1.86$^{{\rm + 0.08}}_{{\rm - 0.08}}$ \\ 
0.75-1.00 &   454 &   451 &  -1.25$^{{\rm + 0.30}}_{{\rm -0.29}}$ & -21.65$^{{\rm + 0.22}}_{{\rm -0.27}}$ &   1.27$^{{\rm + 0.37}}_{{\rm - 0.40}}$ \\ 
          &       &       &  -0.30     fixed                      & -21.08$^{{\rm + 0.06}}_{{\rm -0.06}}$ &   1.92$^{{\rm + 0.09}}_{{\rm - 0.09}}$ \\  
0.30-0.80 &  1404 &  1374 &  -0.30$^{{\rm + 0.08}}_{{\rm -0.08}}$ & -20.79$^{{\rm + 0.07}}_{{\rm -0.07}}$ &   2.02$^{{\rm + 0.08}}_{{\rm - 0.09}}$ \\  
\hline
\multicolumn{6}{l}{ Spectrophotometric type 2 galaxies} \\
\hline
0.10-0.35 &   288 &   279 &  -0.97$^{{\rm + 0.10}}_{{\rm -0.10}}$ & -21.23$^{{\rm + 0.29}}_{{\rm -0.38}}$ &   0.97$^{{\rm + 0.25}}_{{\rm - 0.24}}$ \\  
          &       &       &  -0.86     fixed                      & -20.95$^{{\rm + 0.14}}_{{\rm -0.15}}$ &   1.26$^{{\rm + 0.08}}_{{\rm - 0.08}}$ \\
0.35-0.55 &   346 &   311 &  -0.52$^{{\rm + 0.20}}_{{\rm -0.20}}$ & -20.61$^{{\rm + 0.19}}_{{\rm -0.22}}$ &   1.30$^{{\rm + 0.17}}_{{\rm - 0.21}}$ \\ 
          &       &       &  -0.86     fixed                      & -20.98$^{{\rm + 0.10}}_{{\rm -0.11}}$ &   0.92$^{{\rm + 0.05}}_{{\rm - 0.05}}$ \\ 
0.55-0.75 &   400 &   386 &  -0.59$^{{\rm + 0.26}}_{{\rm -0.25}}$ & -21.03$^{{\rm + 0.20}}_{{\rm -0.23}}$ &   1.25$^{{\rm + 0.17}}_{{\rm - 0.23}}$ \\ 
          &       &       &  -0.86     fixed                      & -21.26$^{{\rm + 0.09}}_{{\rm -0.09}}$ &   1.02$^{{\rm + 0.05}}_{{\rm - 0.05}}$ \\  
0.75-1.00 &   337 &   336 &  -1.12$^{{\rm + 0.37}}_{{\rm -0.36}}$ & -21.37$^{{\rm + 0.24}}_{{\rm -0.28}}$ &   1.13$^{{\rm + 0.28}}_{{\rm - 0.35}}$ \\ 
          &       &       &  -0.86     fixed                      & -21.21$^{{\rm + 0.08}}_{{\rm -0.08}}$ &   1.30$^{{\rm + 0.07}}_{{\rm - 0.07}}$ \\
0.30-0.80 &   929 &   893 &  -0.86$^{{\rm + 0.10}}_{{\rm -0.10}}$ & -21.12$^{{\rm + 0.12}}_{{\rm -0.12}}$ &   1.00$^{{\rm + 0.12}}_{{\rm - 0.12}}$ \\ 
\hline
\multicolumn{6}{l}{ Spectrophotometric type $3+4$ galaxies} \\
\hline
0.10-0.35 &  1262 &  1181 &  -1.34$^{{\rm + 0.06}}_{{\rm -0.06}}$ & -20.38$^{{\rm + 0.15}}_{{\rm -0.17}}$ &   3.08$^{{\rm + 0.57}}_{{\rm - 0.53}}$ \\   
          &       &       &  -1.47     fixed                      & -20.69$^{{\rm + 0.10}}_{{\rm -0.11}}$ &   2.05$^{{\rm + 0.06}}_{{\rm - 0.06}}$ \\
0.35-0.55 &  1133 &   978 &  -1.23$^{{\rm + 0.14}}_{{\rm -0.14}}$ & -20.51$^{{\rm + 0.16}}_{{\rm -0.18}}$ &   3.18$^{{\rm + 0.70}}_{{\rm - 0.67}}$ \\  
          &       &       &  -1.47     fixed                      & -20.81$^{{\rm + 0.08}}_{{\rm -0.08}}$ &   2.06$^{{\rm + 0.07}}_{{\rm - 0.07}}$ \\ 
0.55-0.75 &  1186 &  1138 &  -1.34$^{{\rm + 0.18}}_{{\rm -0.17}}$ & -20.98$^{{\rm + 0.16}}_{{\rm -0.18}}$ &   3.31$^{{\rm + 0.77}}_{{\rm - 0.76}}$ \\ 
          &       &       &  -1.47     fixed                      & -21.11$^{{\rm + 0.06}}_{{\rm -0.07}}$ &   2.76$^{{\rm + 0.08}}_{{\rm - 0.08}}$ \\ 
0.75-1.00 &   978 &   963 &  -1.71$^{{\rm + 0.28}}_{{\rm -0.27}}$ & -21.36$^{{\rm + 0.22}}_{{\rm -0.26}}$ &   2.72$^{{\rm + 1.09}}_{{\rm - 1.03}}$ \\ 
          &       &       &  -1.47     fixed                      & -21.18$^{{\rm + 0.06}}_{{\rm -0.06}}$ &   3.61$^{{\rm + 0.12}}_{{\rm - 0.12}}$ \\ 
0.30-0.80 &  2916 &  2705 &  -1.47$^{{\rm + 0.07}}_{{\rm -0.07}}$ & -21.00$^{{\rm + 0.09}}_{{\rm -0.09}}$ &   2.50$^{{\rm + 0.34}}_{{\rm - 0.32}}$ \\ 
\hline
\multicolumn{6}{l}{ Morphological early-type galaxies} \\
\hline
0.10-0.35 &   576 &   566 &  -0.72$^{{\rm + 0.07}}_{{\rm -0.07}}$ & -20.90$^{{\rm + 0.15}}_{{\rm -0.17}}$ &   2.81$^{{\rm + 0.34}}_{{\rm - 0.33}}$ \\
          &       &       &  -0.56     fixed                      & -20.62$^{{\rm + 0.07}}_{{\rm -0.07}}$ &   3.57$^{{\rm + 0.15}}_{{\rm - 0.15}}$ \\ 	  
0.35-0.55 &   644 &   605 &  -0.38$^{{\rm + 0.14}}_{{\rm -0.13}}$ & -20.63$^{{\rm + 0.13}}_{{\rm -0.14}}$ &   2.54$^{{\rm + 0.19}}_{{\rm - 0.22}}$ \\  
          &       &       &  -0.56     fixed                      & -20.80$^{{\rm + 0.06}}_{{\rm -0.06}}$ &   2.23$^{{\rm + 0.09}}_{{\rm - 0.09}}$ \\  
0.55-0.75 &   619 &   607 &  -0.02$^{{\rm + 0.19}}_{{\rm -0.19}}$ & -20.81$^{{\rm + 0.12}}_{{\rm -0.13}}$ &   2.07$^{{\rm + 0.08}}_{{\rm - 0.08}}$ \\  
          &       &       &  -0.56     fixed                      & -21.19$^{{\rm + 0.06}}_{{\rm -0.06}}$ &   1.73$^{{\rm + 0.07}}_{{\rm - 0.07}}$ \\ 	  
0.75-1.00 &   455 &   453 &  -1.29$^{{\rm + 0.29}}_{{\rm -0.29}}$ & -21.69$^{{\rm + 0.23}}_{{\rm -0.27}}$ &   1.16$^{{\rm + 0.35}}_{{\rm - 0.37}}$ \\ 
          &       &       &  -0.56     fixed                      & -21.21$^{{\rm + 0.06}}_{{\rm -0.06}}$ &   1.78$^{{\rm + 0.08}}_{{\rm - 0.08}}$ \\ 
0.30-0.80 &  1553 &  1504 &  -0.56$^{{\rm + 0.07}}_{{\rm -0.07}}$ & -20.99$^{{\rm + 0.08}}_{{\rm -0.08}}$ &   1.93$^{{\rm + 0.11}}_{{\rm - 0.12}}$ \\  
\hline
\multicolumn{6}{l}{ Morphological spiral galaxies} \\
\hline
0.10-0.35 &  1123 &  1064 &  -1.14$^{{\rm + 0.06}}_{{\rm -0.06}}$ & -20.46$^{{\rm + 0.14}}_{{\rm -0.15}}$ &   3.59$^{{\rm + 0.53}}_{{\rm - 0.50}}$ \\
          &       &       &  -1.27     fixed                      & -20.75$^{{\rm + 0.09}}_{{\rm -0.10}}$ &   2.50$^{{\rm + 0.08}}_{{\rm - 0.08}}$ \\
0.35-0.55 &  1178 &  1020 &  -0.97$^{{\rm + 0.13}}_{{\rm -0.12}}$ & -20.51$^{{\rm + 0.13}}_{{\rm -0.14}}$ &   3.68$^{{\rm + 0.55}}_{{\rm - 0.55}}$ \\  
          &       &       &  -1.27     fixed                      & -20.86$^{{\rm + 0.07}}_{{\rm -0.07}}$ &   2.36$^{{\rm + 0.07}}_{{\rm - 0.07}}$ \\ 
0.55-0.75 &  1131 &  1078 &  -1.21$^{{\rm + 0.17}}_{{\rm -0.17}}$ & -21.08$^{{\rm + 0.16}}_{{\rm -0.18}}$ &   2.92$^{{\rm + 0.61}}_{{\rm - 0.62}}$ \\ 
          &       &       &  -1.27     fixed                      & -21.14$^{{\rm + 0.06}}_{{\rm -0.06}}$ &   2.68$^{{\rm + 0.08}}_{{\rm - 0.08}}$ \\ 
0.75-1.00 &   754 &   743 &  -1.78$^{{\rm + 0.27}}_{{\rm -0.27}}$ & -21.59$^{{\rm + 0.24}}_{{\rm -0.29}}$ &   1.58$^{{\rm + 0.74}}_{{\rm - 0.65}}$ \\  
          &       &       &  -1.27     fixed                      & -21.20$^{{\rm + 0.06}}_{{\rm -0.06}}$ &   2.84$^{{\rm + 0.10}}_{{\rm - 0.10}}$ \\  
0.30-0.80 &  2876 &  2682 &  -1.27$^{{\rm + 0.06}}_{{\rm -0.06}}$ & -21.00$^{{\rm + 0.08}}_{{\rm -0.08}}$ &   2.64$^{{\rm + 0.28}}_{{\rm - 0.27}}$ \\   
\hline
\multicolumn{6}{l}{ Morphological irregular galaxies} \\
\hline
0.10-0.35 &   217 &   195 &  -1.57$^{{\rm + 0.14}}_{{\rm -0.14}}$ & -21.14$^{{\rm + 0.49}}_{{\rm -0.80}}$ &   0.20$^{{\rm + 0.15}}_{{\rm - 0.12}}$ \\  
          &       &       &  -1.20     fixed                      & -20.23$^{{\rm + 0.17}}_{{\rm -0.18}}$ &   0.67$^{{\rm + 0.05}}_{{\rm - 0.05}}$ \\ 
0.35-0.55 &   183 &   165 &  -1.01$^{{\rm + 0.36}}_{{\rm -0.35}}$ & -20.51$^{{\rm + 0.36}}_{{\rm -0.50}}$ &   0.61$^{{\rm + 0.26}}_{{\rm - 0.27}}$ \\ 
          &       &       &  -1.20     fixed                      & -20.73$^{{\rm + 0.17}}_{{\rm -0.19}}$ &   0.47$^{{\rm + 0.04}}_{{\rm - 0.04}}$ \\ 
0.55-0.75 &   359 &   351 &  -0.90$^{{\rm + 0.35}}_{{\rm -0.33}}$ & -20.66$^{{\rm + 0.23}}_{{\rm -0.27}}$ &   1.48$^{{\rm + 0.29}}_{{\rm - 0.39}}$ \\ 
          &       &       &  -1.20     fixed                      & -20.88$^{{\rm + 0.09}}_{{\rm -0.10}}$ &   1.15$^{{\rm + 0.06}}_{{\rm - 0.06}}$ \\
0.75-1.00 &   509 &   505 &  -1.32$^{{\rm + 0.37}}_{{\rm -0.36}}$ & -21.15$^{{\rm + 0.23}}_{{\rm -0.27}}$ &   2.04$^{{\rm + 0.55}}_{{\rm - 0.67}}$ \\  
          &       &       &  -1.20     fixed                      & -21.08$^{{\rm + 0.07}}_{{\rm -0.07}}$ &   2.22$^{{\rm + 0.10}}_{{\rm - 0.10}}$ \\ 
0.30-0.80 &   687 &   661 &  -1.20$^{{\rm + 0.15}}_{{\rm -0.15}}$ & -20.86$^{{\rm + 0.15}}_{{\rm -0.17}}$ &   0.93$^{{\rm + 0.19}}_{{\rm - 0.18}}$ \\ 
\hline
\multicolumn{6}{l}{(a) Number of galaxies in the redshift bin }\\
\multicolumn{6}{l}{(b) Number of galaxies brighter than the bias limit 
(sample used for $STY$ estimate; see the text for details)}\\
\end{tabular}
\end{flushleft}
\label{param_tot_types}
\end{table*}

\begin{table*}
\caption[]{$STY$ parameters (with $1\sigma$ errors) for different galaxy samples in different environments: total sample quartiles. }
\begin{flushleft}
\begin{tabular}{l r r r r r r} 
\hline
sample & z-bin &  Number$^{(a)}$ &  Number$^{(b)}$ &   $\alpha$ &  $M^*_{B_{AB}}-5{\rm log}(h_{70})$ & $\phi^*$($10^{-3}\ h_{70}^3\ {\rm Mpc}^{-3}$)   
\\ \hline
\multicolumn{7}{l}{ Under-dense environments } \\
\hline
Total sample        & 0.10-0.40  &    680 &   655 &  -1.16$^{{\rm + 0.08}}_{{\rm -0.08}}$ & -20.44$^{{\rm + 0.17}}_{{\rm -0.19}}$ &   1.62$^{{\rm + 0.31}}_{{\rm - 0.29}}$ \\  
Type 1 galaxies     & 0.10-0.40  &     92 &    92 &  -0.28$^{{\rm + 0.29}}_{{\rm -0.28}}$ & -20.34$^{{\rm + 0.39}}_{{\rm -0.58}}$ &   0.59$^{{\rm + 0.09}}_{{\rm - 0.14}}$ \\ 
                    &            &        &       &  -0.30     fixed                      & -20.37$^{{\rm + 0.18}}_{{\rm -0.21}}$ &   0.58$^{{\rm + 0.06}}_{{\rm - 0.06}}$ \\  
Type $3+4$ galaxies & 0.10-0.40  &    499 &   477 &  -1.33$^{{\rm + 0.12}}_{{\rm -0.11}}$ & -20.02$^{{\rm + 0.22}}_{{\rm -0.26}}$ &   1.25$^{{\rm + 0.37}}_{{\rm - 0.34}}$ \\  
                    &            &        &       &  -1.47     fixed                      & -20.31$^{{\rm + 0.15}}_{{\rm -0.16}}$ &   0.85$^{{\rm + 0.04}}_{{\rm - 0.04}}$ \\                  
\hline
Total sample        & 0.40-0.70  &    675 &   628 &  -1.20$^{{\rm + 0.20}}_{{\rm -0.20}}$ & -20.97$^{{\rm + 0.26}}_{{\rm -0.33}}$ &   0.96$^{{\rm + 0.33}}_{{\rm - 0.32}}$ \\ 
Type 1 galaxies     & 0.40-0.70  &    131 &   127 &  -0.88$^{{\rm + 0.47}}_{{\rm -0.46}}$ & -21.19$^{{\rm + 0.62}}_{{\rm -1.46}}$ &   0.21$^{{\rm + 0.12}}_{{\rm - 0.14}}$ \\ 
                    &            &        &       &  -0.30     fixed                      & -20.50$^{{\rm + 0.13}}_{{\rm -0.15}}$ &   0.34$^{{\rm + 0.03}}_{{\rm - 0.03}}$ \\  
Type $3+4$ galaxies & 0.40-0.70  &    449 &   413 &  -1.64$^{{\rm + 0.26}}_{{\rm -0.25}}$ & -21.26$^{{\rm + 0.41}}_{{\rm -0.63}}$ &   0.37$^{{\rm + 0.29}}_{{\rm - 0.22}}$ \\ 
                    &            &        &       &  -1.47     fixed                      & -20.98$^{{\rm + 0.13}}_{{\rm -0.14}}$ &   0.55$^{{\rm + 0.03}}_{{\rm - 0.03}}$ \\                 
\hline
Total sample        & 0.70-1.00  &    606 &   594 &  -0.93$^{{\rm + 0.36}}_{{\rm -0.35}}$ & -20.73$^{{\rm + 0.19}}_{{\rm -0.23}}$ &   2.82$^{{\rm + 0.41}}_{{\rm - 0.59}}$ \\ 
Type 1 galaxies     & 0.70-1.00  &    129 &   129 &   0.38$^{{\rm + 0.84}}_{{\rm -0.82}}$ & -20.34$^{{\rm + 0.27}}_{{\rm -0.38}}$ &   0.54$^{{\rm + 0.17}}_{{\rm - 0.18}}$ \\  
                    &            &        &       &  -0.30     fixed                      & -20.64$^{{\rm + 0.11}}_{{\rm -0.12}}$ &   0.64$^{{\rm + 0.06}}_{{\rm - 0.06}}$ \\                  
Type $3+4$ galaxies & 0.70-1.00  &    365 &   357 &  -1.51$^{{\rm + 0.49}}_{{\rm -0.48}}$ & -20.93$^{{\rm + 0.31}}_{{\rm -0.42}}$ &   1.34$^{{\rm + 0.60}}_{{\rm - 0.68}}$ \\ 
                    &            &        &       &  -1.47     fixed                      & -20.90$^{{\rm + 0.10}}_{{\rm -0.11}}$ &   1.40$^{{\rm + 0.07}}_{{\rm - 0.07}}$ \\                  
\hline
\multicolumn{7}{l}{ Over-dense environments } \\
\hline
Total sample        & 0.10-0.40  &    713 &   678 &  -0.86$^{{\rm + 0.07}}_{{\rm -0.07}}$ & -20.90$^{{\rm + 0.14}}_{{\rm -0.15}}$ &   2.14$^{{\rm + 0.27}}_{{\rm - 0.26}}$ \\ 
Type 1 galaxies     & 0.10-0.40  &    264 &   262 &  -0.12$^{{\rm + 0.13}}_{{\rm -0.13}}$ & -20.50$^{{\rm + 0.15}}_{{\rm -0.17}}$ &   1.63$^{{\rm + 0.10}}_{{\rm - 0.13}}$ \\  
                    &            &        &       &  -0.30     fixed                      & -20.72$^{{\rm + 0.09}}_{{\rm -0.09}}$ &   1.44$^{{\rm + 0.09}}_{{\rm - 0.09}}$ \\ 
Type $3+4$ galaxies & 0.10-0.40  &    327 &   298 &  -1.34$^{{\rm + 0.12}}_{{\rm -0.12}}$ & -20.78$^{{\rm + 0.29}}_{{\rm -0.36}}$ &   0.51$^{{\rm + 0.19}}_{{\rm - 0.17}}$ \\  
                    &            &        &       &  -1.47     fixed                      & -21.10$^{{\rm + 0.21}}_{{\rm -0.25}}$ &   0.33$^{{\rm + 0.02}}_{{\rm - 0.02}}$ \\                   
\hline
Total sample        & 0.40-0.70  &    703 &   684 &  -0.81$^{{\rm + 0.15}}_{{\rm -0.15}}$ & -21.14$^{{\rm + 0.16}}_{{\rm -0.18}}$ &   1.34$^{{\rm + 0.21}}_{{\rm - 0.22}}$ \\ 
Type 1 galaxies     & 0.40-0.70  &    263 &   256 &  -0.59$^{{\rm + 0.24}}_{{\rm -0.24}}$ & -21.36$^{{\rm + 0.28}}_{{\rm -0.36}}$ &   0.49$^{{\rm + 0.09}}_{{\rm - 0.12}}$ \\ 
                    &            &        &       &  -0.30     fixed                      & -21.05$^{{\rm + 0.09}}_{{\rm -0.09}}$ &   0.60$^{{\rm + 0.04}}_{{\rm - 0.04}}$ \\  
Type $3+4$ galaxies & 0.40-0.70  &    306 &   298 &  -1.04$^{{\rm + 0.29}}_{{\rm -0.28}}$ & -20.93$^{{\rm + 0.28}}_{{\rm -0.36}}$ &   0.67$^{{\rm + 0.22}}_{{\rm - 0.24}}$ \\
                    &            &        &       &  -1.47     fixed                      & -21.49$^{{\rm + 0.17}}_{{\rm -0.19}}$ &   0.32$^{{\rm + 0.02}}_{{\rm - 0.02}}$ \\                   
\hline
Total sample        & 0.70-1.00  &    656 &   636 &  -1.01$^{{\rm + 0.22}}_{{\rm -0.22}}$ & -21.44$^{{\rm + 0.16}}_{{\rm -0.18}}$ &   1.44$^{{\rm + 0.24}}_{{\rm - 0.28}}$ \\
Type 1 galaxies     & 0.70-1.00  &    207 &   202 &  -0.48$^{{\rm + 0.38}}_{{\rm -0.37}}$ & -21.35$^{{\rm + 0.23}}_{{\rm -0.27}}$ &   0.49$^{{\rm + 0.05}}_{{\rm - 0.09}}$ \\ 
                    &            &        &       &  -0.30     fixed                      & -21.24$^{{\rm + 0.08}}_{{\rm -0.09}}$ &   0.51$^{{\rm + 0.04}}_{{\rm - 0.04}}$ \\ 
Type $3+4$ galaxies & 0.70-1.00  &    300 &   288 &  -1.34$^{{\rm + 0.43}}_{{\rm -0.42}}$ & -21.38$^{{\rm + 0.32}}_{{\rm -0.44}}$ &   0.68$^{{\rm + 0.28}}_{{\rm - 0.33}}$ \\ 
                    &            &        &       &  -1.47     fixed                      & -21.49$^{{\rm + 0.13}}_{{\rm -0.14}}$ &   0.58$^{{\rm + 0.03}}_{{\rm - 0.03}}$ \\                   
\hline
\multicolumn{7}{l}{(a) Number of galaxies in the redshift bin }\\
\multicolumn{7}{l}{(b) Number of galaxies brighter than the bias limit 
(sample used for $STY$ estimate; see the text for details)}\\

\end{tabular}
\end{flushleft}
\label{param_overd_rel}
\end{table*}

\begin{table*}
\caption[]{$STY$ parameters (with $1\sigma$ errors) as a function of the environment (from the 5 nearest neighbours)
in different redshift bins, for different galaxy types. In each sample the appropriate quartiles are used.}
\begin{flushleft}
\begin{tabular}{r r r r r r r} 
\hline
z-bin & quartile &  Number$^{(a)}$ &  Number$^{(b)}$ &   $\alpha$ &  $M^*_{B_{AB}}-5{\rm log}(h_{70})$ & $\phi^*$($10^{-3}\ h_{70}^3\ {\rm Mpc}^{-3}$)   
\\ \hline
\multicolumn{7}{l}{ Total sample} \\
\hline
0.10-0.40    &  first &    680 &   655 &  -1.16$^{{\rm + 0.08}}_{{\rm -0.08}}$ & -20.44$^{{\rm + 0.17}}_{{\rm -0.19}}$ &   1.62$^{{\rm + 0.31}}_{{\rm - 0.29}}$ \\ 
0.10-0.40    &  fourth &   713 &   678 &  -0.86$^{{\rm + 0.07}}_{{\rm -0.07}}$ & -20.90$^{{\rm + 0.14}}_{{\rm -0.15}}$ &   2.14$^{{\rm + 0.27}}_{{\rm - 0.26}}$ \\ 
\hline
0.40-0.70    &  first &    675 &   628 &  -1.20$^{{\rm + 0.20}}_{{\rm -0.20}}$ & -20.97$^{{\rm + 0.26}}_{{\rm -0.33}}$ &   0.96$^{{\rm + 0.33}}_{{\rm - 0.32}}$ \\   
0.40-0.70    &  fourth &   703 &   684 &  -0.81$^{{\rm + 0.15}}_{{\rm -0.15}}$ & -21.14$^{{\rm + 0.16}}_{{\rm -0.18}}$ &   1.34$^{{\rm + 0.21}}_{{\rm - 0.22}}$ \\  
\hline
0.70-1.00    &  first &    606 &   594 &  -0.93$^{{\rm + 0.36}}_{{\rm -0.35}}$ & -20.73$^{{\rm + 0.19}}_{{\rm -0.23}}$ &   2.82$^{{\rm + 0.41}}_{{\rm - 0.59}}$ \\ 
0.70-1.00    &  fourth &   656 &   636 &  -1.01$^{{\rm + 0.22}}_{{\rm -0.22}}$ & -21.44$^{{\rm + 0.16}}_{{\rm -0.18}}$ &   1.44$^{{\rm + 0.24}}_{{\rm - 0.28}}$ \\ 
\hline
\multicolumn{7}{l}{ Type 1 galaxies } \\
\hline
0.10-0.40    &  first &    160 &   160 &  -0.37$^{{\rm + 0.20}}_{{\rm -0.19}}$ & -20.34$^{{\rm + 0.26}}_{{\rm -0.32}}$ &   0.96$^{{\rm + 0.14}}_{{\rm - 0.17}}$ \\ 
             &         &       &       &  -0.30     fixed                      & -20.25$^{{\rm + 0.11}}_{{\rm -0.12}}$ &   1.01$^{{\rm + 0.08}}_{{\rm - 0.08}}$ \\                    
0.10-0.40    &  fourth &   166 &   164 &  -0.20$^{{\rm + 0.16}}_{{\rm -0.15}}$ & -20.65$^{{\rm + 0.21}}_{{\rm -0.24}}$ &   0.97$^{{\rm + 0.09}}_{{\rm - 0.11}}$ \\    
             &         &       &       &  -0.30     fixed                      & -20.79$^{{\rm + 0.12}}_{{\rm -0.12}}$ &   0.90$^{{\rm + 0.07}}_{{\rm - 0.07}}$ \\                 
\hline
0.40-0.70    &  first &    183 &   179 &  -0.39$^{{\rm + 0.30}}_{{\rm -0.29}}$ & -20.65$^{{\rm + 0.24}}_{{\rm -0.29}}$ &   0.43$^{{\rm + 0.05}}_{{\rm - 0.08}}$ \\   
             &         &       &       &  -0.30     fixed                      & -20.58$^{{\rm + 0.09}}_{{\rm -0.10}}$ &   0.45$^{{\rm + 0.03}}_{{\rm - 0.03}}$ \\                    
0.40-0.70    &  fourth &   187 &   180 &  -0.99$^{{\rm + 0.28}}_{{\rm -0.27}}$ & -21.98$^{{\rm + 0.50}}_{{\rm -0.89}}$ &   0.20$^{{\rm + 0.10}}_{{\rm - 0.10}}$ \\   
             &         &       &       &  -0.30     fixed                      & -21.05$^{{\rm + 0.10}}_{{\rm -0.11}}$ &   0.42$^{{\rm + 0.03}}_{{\rm - 0.03}}$ \\                   
\hline
0.70-1.00    &  first &    160 &   160 &  -0.04$^{{\rm + 0.75}}_{{\rm -0.73}}$ & -20.55$^{{\rm + 0.30}}_{{\rm -0.40}}$ &   0.74$^{{\rm + 0.14}}_{{\rm - 0.20}}$ \\ 
             &         &       &       &  -0.30     fixed                      & -20.68$^{{\rm + 0.10}}_{{\rm -0.11}}$ &   0.76$^{{\rm + 0.06}}_{{\rm - 0.06}}$ \\                   
0.70-1.00    &  fourth &   166 &   163 &  -0.61$^{{\rm + 0.42}}_{{\rm -0.41}}$ & -21.41$^{{\rm + 0.27}}_{{\rm -0.32}}$ &   0.39$^{{\rm + 0.06}}_{{\rm - 0.10}}$ \\    
             &         &       &       &  -0.30     fixed                      & -21.23$^{{\rm + 0.09}}_{{\rm -0.10}}$ &   0.42$^{{\rm + 0.03}}_{{\rm - 0.03}}$ \\                   
\hline
\multicolumn{7}{l}{ Type $3+4$ galaxies } \\
\hline
0.10-0.40    &  first &    422 &   406 &  -1.32$^{{\rm + 0.13}}_{{\rm -0.13}}$ & -19.98$^{{\rm + 0.23}}_{{\rm -0.28}}$ &   1.11$^{{\rm + 0.36}}_{{\rm - 0.32}}$ \\  
             &         &       &       &  -1.47     fixed                      & -20.28$^{{\rm + 0.16}}_{{\rm -0.17}}$ &   0.74$^{{\rm + 0.04}}_{{\rm - 0.04}}$ \\                   
0.10-0.40    &  fourth &   439 &   407 &  -1.34$^{{\rm + 0.11}}_{{\rm -0.11}}$ & -20.64$^{{\rm + 0.24}}_{{\rm -0.29}}$ &   0.78$^{{\rm + 0.24}}_{{\rm - 0.22}}$ \\
             &         &       &       &  -1.47     fixed                      & -20.94$^{{\rm + 0.17}}_{{\rm -0.19}}$ &   0.52$^{{\rm + 0.03}}_{{\rm - 0.03}}$ \\                  
\hline
0.40-0.70    &  first &    375 &   343 &  -1.73$^{{\rm + 0.28}}_{{\rm -0.27}}$ & -21.40$^{{\rm + 0.49}}_{{\rm -0.86}}$ &   0.23$^{{\rm + 0.25}}_{{\rm - 0.17}}$ \\   
             &         &       &       &  -1.47     fixed                      & -20.96$^{{\rm + 0.14}}_{{\rm -0.16}}$ &   0.46$^{{\rm + 0.02}}_{{\rm - 0.02}}$ \\                 
0.40-0.70    &  fourth &   391 &   382 &  -1.04$^{{\rm + 0.25}}_{{\rm -0.25}}$ & -20.92$^{{\rm + 0.25}}_{{\rm -0.31}}$ &   0.84$^{{\rm + 0.25}}_{{\rm - 0.26}}$ \\ 
             &         &       &       &  -1.47     fixed                      & -21.48$^{{\rm + 0.15}}_{{\rm -0.17}}$ &   0.41$^{{\rm + 0.02}}_{{\rm - 0.02}}$ \\                    
\hline
0.70-1.00    &  first &    328 &   324 &  -1.63$^{{\rm + 0.52}}_{{\rm -0.50}}$ & -21.00$^{{\rm + 0.34}}_{{\rm -0.48}}$ &   1.11$^{{\rm + 0.62}}_{{\rm - 0.65}}$ \\ 
             &         &       &       &  -1.47     fixed                      & -20.89$^{{\rm + 0.10}}_{{\rm -0.11}}$ &   1.30$^{{\rm + 0.07}}_{{\rm - 0.07}}$ \\                  
0.70-1.00    &  fourth &   357 &   342 &  -1.44$^{{\rm + 0.38}}_{{\rm -0.38}}$ & -21.49$^{{\rm + 0.32}}_{{\rm -0.43}}$ &   0.67$^{{\rm + 0.31}}_{{\rm - 0.34}}$ \\     
             &         &       &       &  -1.47     fixed                      & -21.52$^{{\rm + 0.12}}_{{\rm -0.13}}$ &   0.65$^{{\rm + 0.03}}_{{\rm - 0.03}}$ \\              
\hline
\multicolumn{7}{l}{(a) Number of galaxies in the redshift bin }\\
\multicolumn{7}{l}{(b) Number of galaxies brighter than the bias limit 
(sample used for $STY$ estimate; see the text for details)}\\

\end{tabular}
\end{flushleft}
\label{param_overd}
\end{table*}


\end{document}